\documentclass[journal,11pt,onecolumn]{IEEEtran}

\usepackage[latin1]{inputenc}
\usepackage[numbers, square, comma, compress]{natbib}
\usepackage[english]{babel}
\usepackage{dsfont}
\usepackage{setspace}
\usepackage{amsfonts}
\usepackage{amssymb}
\usepackage{amsmath}
\usepackage{mathrsfs}
\usepackage{xcolor}
\usepackage{graphicx}
\usepackage{mathdots}
\usepackage{float}
\usepackage{stmaryrd}
\usepackage[colorinlistoftodos]{todonotes}
\usepackage[ruled,vlined]{algorithm2e}
\usepackage{verbatim}
\usepackage{color}
\usepackage{lipsum}
\usepackage{mathtools}
\usepackage{paralist}   

\usepackage{longtable}
\usepackage{multirow}
\usepackage{ctable}

\usepackage{hyperref}
\hypersetup{
colorlinks,%
citecolor=black,%
filecolor=black,%
linkcolor=black,%
urlcolor=black,%
pdfauthor={Cervia, Oechtering, Skoglund},%
pdftitle={Fixed-Length Strong Coordination}%
}

\usepackage[all]{hypcap}

 \usepackage{url}

\renewcommand{\thesection}{\Roman{section}}
\renewcommand{\thesubsection}{\thesection.\Alph{subsection}}
\renewcommand{\thesubsubsection}{\thesubsection.\arabic{subsubsection}}

\pdfoutput=1
\newtheorem{theo}{Theorem}
\newtheorem{prop}{Proposition}
\newtheorem{lem}{Lemma}

\newtheorem{rem}{Remark}
\newtheorem{defi}{Definition}

\renewcommand\tabcolsep{2pt}

\newcommand{\vv}[1]{``#1''}

\renewcommand{\IEEEQED}{\IEEEQEDopen}

\makeatletter
\renewcommand*\env@matrix[1][*\c@MaxMatrixCols c]{%
  \hskip -\arraycolsep
  \let\@ifnextchar\new@ifnextchar
  \array{#1}}
\makeatother

\IEEEoverridecommandlockouts

\title{$(\epsilon,n)$ Fixed-Length Strong Coordination Capacity}

\author{ 
  \IEEEauthorblockN{Giulia~Cervia, Tobias~J.~Oechtering, and Mikael~Skoglund}

  \thanks{This work was supported in part by the Swedish foundation for strategic research, the Swedish research council, and Digital Futures.}
 \thanks{G.~Cervia is with IMT Lille Douai, Institut Mines-T\'el\'ecom, Univ. Lille, Centre for Digital Systems, F-59000 Lille, France (email: giulia.cervia@imt-lille-douai.fr), and was with the School of Electrical Engineering and Computer Science, KTH Royal Institute of Technology,
Stockholm, Sweden.
T.~J.~Oechtering, and M.~Skoglund are with the School of Electrical Engineering and Computer Science, KTH Royal Institute of Technology, Stockholm, Sweden (email: \{oech, skoglund\}@kth.se). }
}

\begin{document}
{\let\newpage\relax\maketitle}


\begin{abstract} 

This paper investigates the problem of synthesizing joint distributions in the finite-length regime.
For a fixed blocklength $n$ and an upper bound on the distribution approximation $\epsilon$, we prove a capacity result for fixed-length strong coordination. 
It is shown analytically that the rate conditions for the fixed-length regime are  lower-bounded by 
the mutual information that appears in the asymptotical condition plus $Q^{-1} \left(\epsilon  \right) \sqrt{ V/n}$, where $V$ is the channel dispersion, and $Q^{-1}$ is the inverse of the Gaussian cumulative distribution function.
\end{abstract}


\section{Introduction}

The problem of cooperation of autonomous devices in a decentralized  network, initially raised in the context of game theory by~\cite{gossner2006optimal}, with applications, for instance, to power control~\cite{larrousse2015coordination}, is concerned with  communication networks beyond the traditional problem of reliable communications. 
The goal of coordination is in fact to exceed the classical problem of reliably conveying information, and to characterize the set of  target joint probability distributions that are implementable by a choice of strategy of the agents.
Coordination is then intended as a way of enforcing a prescribed joint behavior of the devices through communication, by synthesizing joint distributions
which approximates a target behavior~\cite{cuff2009thesis,cuff2010}.
This topic, presented as \vv{channel simulation} is related to earlier work on  ``Shannon's reverse coding theorem''~\cite{bennet2002entanglement} and the compression of probability distribution sources and mixed quantum states~\cite{Soljanin2002,kramer2007communicating,winter2002compression}. 

The information-theoretic framework for coordination in networks considered in the present paper has been introduced in~\cite{cuff2009thesis,cuff2010,cuff2013distributed}. 
In~\cite{cuff2009thesis,cuff2010,cuff2013distributed} two metrics to measure the level of coordination have been defined: \emph{empirical coordination}, which requires the empirical distribution of the distributed random states to approach a target distribution with high probability, and \emph{strong coordination}, which requires the $L^1$ distance of the distribution of sequences of distributed random states to converge to an i.i.d. target distribution.
Strong and empirical coordination in the asymptotical regime have been studied in a number of works, namely~\cite{cuff2010,cuff2013distributed,haddadpour2012coordination, bloch2014strong,bloch2013strong, vellambi2015strong, vellambi2016strong,cuff2011hybrid,treust2014correlation,treust2015empirical,le2015empirical, larrousse2015coordinating,treust2017joint,haddadpour2017simulation,Cervia2017,cervia2018journal}, but until~\cite{Cervia2019Fixed} there was no attempt to tackle the problem of coordination in the finite-length regime. However, in many realistic systems of interest, non-asymptotic information-theoretic limits are of high practical interest. Originally raised by~\cite{strassen1962asymptotische}, and following~\cite{kontoyiannis1997second,baron2004quickly} and more recently~\cite{hayashi2008second,hayashi2009information,polyanskiy2010channel}, an increasingly large number of papers have  brought up finite blocklength information theory limits (see for instance~\cite{verdu2012non, jazi2012simpler, kostina2012fixed, kostina2013lossyit, tan2013dispersions,Watanabe2015Nonasymptotic,nomura2014second}),
tackling the question of whether the asymptotical results are well-suited to estimate the finite blocklength problems.

\begin{center}
\begin{figure}[t]
\centering
\includegraphics[scale=0.22]{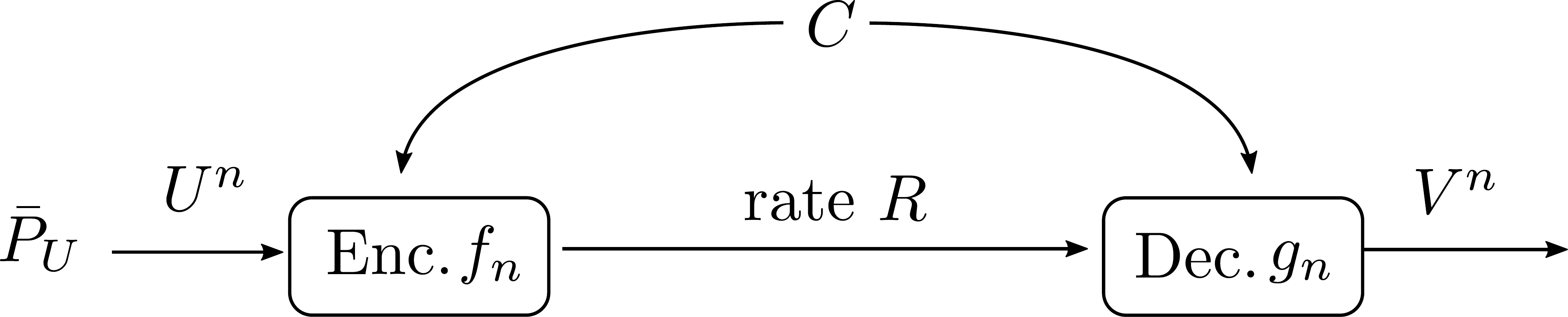}
\vspace{-1mm}
\caption{Coordination of $U^n$ and $V^n$ for a two-node network with an error-free link of rate $R$.}
\label{fig: isit2017}
\end{figure}
\end{center}

Specifically to the coordination problem,~\cite{Cervia2019Fixed} drops the simplifying assumption  that allows the blocklength to grow indefinitely, and focuses on the trade-offs achievable in the finite blocklength regime. 
The notion of $(\epsilon,n)$ \emph{fixed-length strong coordination}  demands that, for a fixed codelength $n$ and a given $\epsilon$, the $L^1$ distance between the distribution of sequences of distributed random states and an i.i.d. target distribution is upper-bounded by $\epsilon$~\cite{Cervia2019Fixed}.
In a first attempt to derive a capacity region,~\cite{Cervia2019Fixed} presents an inner bound for  a two-node network comprised of an information source and a noisy channel, in which both nodes have access to a common source of randomness. 
Even though the achievability scheme outlined in~\cite{Cervia2019Fixed} is general enough to be applied to more sophisticated  network topologies, deriving an outer bound for the region is a difficult problem 
even with the less stringent constraint of the asymptotical regime~\cite{cervia2018journal}.
Hence, to prove an outer bound as well as an inner bound, in this paper we look at the simplest setting for which   the strong coordination problem has been solved in the asymptotical regime~\cite{cuff2010}. 
Thus, we consider the point-to-point setting  comprised of an information source, a rate-limited error-free link, and a uniform source of common randomness available at the encoder and  the decoder as depicted in Figure~\ref{fig: isit2017}, and we present  both an inner and an outer bound for the $(\epsilon,n)$ \emph{fixed-length strong coordination} capacity region.
In particular,  while the inner bound exploits the achievability approach of~\cite{Cervia2019Fixed}, we delineate an outer bound proof that profits from Neyman-Pearson theory and hypothesis testing techniques~\cite{polyanskiy2010channel,kostina2013lossyit,blahut1974hypothesis,campo2012converse}, and that does not rely merely on the characteristics of the chosen network, and should therefore be well-suited for generalization to different scenarios. 

Interestingly, for fixed blocklength $n$ and bound on the $L^1$ distance $\epsilon$, we find rate constraints 
\begin{equation}\label{fund to}
\underbrace{\phantom{\!\!\!\!\!\!\!\!  \left( \frac{i}{i}\right)}\mbox{rate} \, \geq  \, \mbox{mutual information} \,\,\,}_{\mbox{constraint of the asymptotical case}}  
+\quad \underbrace{Q^{-1} \left(\epsilon +  O\left( \frac{1}{\sqrt n}\right) \right) \sqrt{ \frac{V}{n}} +  O\left( \frac{\log n}{n}\right)
}_{\mbox{approximation term}}
\end{equation}where $V$, referred to as channel dispersion, is a characteristic of the \vv{test channel} that connects the random variables of the problem ($U$ and $V$ in the setting of Fig.~\ref{fig: isit2017}) with an  auxiliary random variable representing the codebook. Then,  $Q^{-1}$ is the inverse of the Gaussian cumulative distribution function, and the approximation term is the same recovered by~\cite{polyanskiy2010channel,kostina2012fixed,kostina2013lossyit} for channel coding and compression in the finite-length regime. Since the approximation term  vanishes as $n$ increases, 
we also recover the capacity result of asymptotical strong coordination~\cite{cuff2010}, therefore answering the standard question  in information theory  of how the asymptotic limits relate to their fixed blocklength counterparts. 
We recall that the best known fixed-length capacity results for channel and source coding~\cite{polyanskiy2010channel,kostina2012fixed,kostina2013lossyit}  are also of the form 
\begin{equation*}
\mbox{rate constraint of the asymptotical case}\quad
+\quad \underbrace{Q^{-1} \left(\epsilon \right) \sqrt{ \frac{V}{n}} +  O\left( \frac{\log n}{n}\right)
}_{\mbox{approximation term}}
\end{equation*}where $\epsilon$ represents the maximal probability of error and the upper-bound on distortion respectively, meaning that the coordination problem has the same order of approximation as in~\cite{polyanskiy2010channel,kostina2012fixed,kostina2013lossyit}.
Then, even though the fundamental trade-off of~\eqref{fund to} was expected, through~\eqref{fund to} we make the trade-off more explicit, and we prove that high coordination rate and common randomness are needed to shorten the blocklength and approximate the distribution with higher precision.
Not only the better we can approximate the target distribution, the more expensive this is in terms of rate, but the shape of these rate constraints shows the direct dependence of the rate on the \vv{level of coordination} $\epsilon$ inside the approximation term. A similar relation between the rate constraint and the probability of coding error or the level of distortion has been proved in~\cite{polyanskiy2010channel,kostina2012fixed,kostina2013lossyit}.

\vspace{-1cm}
\subsection{Contributions}
The main contributions of  this paper are the following.


\paragraph*{Problem formulation} We  state the definition of $(\epsilon, n)$ \emph{fixed length strong coordination} as introduced in~\cite{Cervia2019Fixed}: for a given blocklength $n$ and bound on the $L^1$ distance $\epsilon$, an i.i.d. distribution is achievable for strong coordination in the non-asymptotic regime if we can approximate it through the coding process up to a margin $\epsilon$.
Similarly to~\cite{cuff2010}, we investigate the fixed length strong coordination region $\mathcal R(\epsilon, n)$  for the simplest point-to-point setting comprised of an i.i.d. source and a noiseless link, in which encoder and decoder share a source of common randomness.

The characterization of the coordination region involves two stages: achievability and converse, detailed in the following paragraphs.  
As in~\cite{polyanskiy2010channel,kostina2012fixed}, the key step of each stage consists in identifying a sequence of independent random variables to which we apply the Berry-Esseen Central Limit Theorem.


\paragraph*{Inner bound}  Theorem~\ref{theona_inner} presents the sufficient conditions for  achievability for non-asymptotic strong coordination. Following the approach designed by~\cite{Cervia2019Fixed}, we use the random binning approach inspired by~\cite{yassaee2013technique,yassaee2013non} to design a random binning and a random coding scheme which are close in $L^1$ distance. By combining the finite-length techniques of~\cite{polyanskiy2010channel,kostina2012fixed} and the Berry-Esseen Central Limit Theorem with the properties of random binning \cite{yassaee2013non}, we derive an inner bound on the rate conditions that guarantees coordination with given arbitrary blocklength $n$. 
Interestingly, the rate constraints, proved in Section~\ref{sec: nonasy ib}, are consistent with the inner bound for the fixed length strong coordination region of a two-nodes network with a noisy link~\cite[Theorem 1]{Cervia2019Fixed}.


\paragraph*{Outer bound} 
In Theorem~\ref{theona_outer} we derive the outer bound for the same setting of Theorem~\ref{theona_inner}. The proof 
involves a meta-converse, an approach which has proved to be optimal in several scenarios~\cite{vazquez2013meta,Vazquez2013Bayesian}.
The meta-converse exploits results from the Neyman-Pearson  hypothesis testing, similarly to~\cite{polyanskiy2010channel,kostina2013lossyit,blahut1974hypothesis,campo2012converse}. Starting from sequences which are by assumption generated with a distribution close to i.i.d., we consider a randomized test between this distribution and an i.i.d. one. Then, the probabilities of false positive and miss-detection of the test can be bounded using the Berry-Esseen Central Limit Theorem and the assumption that strong coordination holds, leading to rate conditions that asymptotically match the achievability.

Furthermore, by analyzing the two main results, we observe that we can derive a closed result on the capacity region in Remark~\ref{theona}.
 

\paragraph*{Comparison with lossy compression}  Since  coordination is conceptually related to  source coding, we compare our results with the non-asymptotic fundamental limits of lossy data compression~\cite{kostina2012fixed,kostina2013lossyit,kostina2013lossy}. Furthermore, coordination and lossy source coding involve different metrics, thus we need to adapt the  formulation of compression to derive analogies between the two problems.


\paragraph*{Discussion of the result} 
We discuss the new non-asymptotic results by looking at the trade-off between the rate required to achieve strong coordination and the threshold $\epsilon$ which measures the \vv{level of coordination}.


\subsection{Organization of the paper}
The remainder of the paper is organized as follows. 
$\mbox{Section \ref{sec: sys}}$ introduces the notation and the model, and recalls the asymptotical result for strong coordination region derived in~\cite{cuff2010}. In $\mbox{Section \ref{sec: nonasy}}$ we present the information-theoretic modelling of fixed-length strong coordination together with the main results of this paper: an inner and an outer bound for fixed-length strong coordination.
The inner bound is proved in $\mbox{Section \ref{sec: nonasy ib}}$, while the proof of the outer bound is given in  $\mbox{Section \ref{sec: nonasy ob}}$.
Finally, the result is further analyzed in $\mbox{Section \ref{sec: comp}}$ by comparing it to fixed-length lossy compression~\cite{kostina2012fixed}, and  by studying the trade-off between the rate-constraint and the level of coordination, measured by the bound on the $L^1$ distance.


\section{System model and background}\label{sec: sys}

We begin by reviewing the main concepts used in this paper.

\vspace{-5mm}
\subsection{Notation and preliminary results}\label{sec: prel}
We define the integer interval $\llbracket a,b \rrbracket$ as the set of integers from $a$ to $b.$ Given a random vector $X^{n}\coloneqq$ $(X_1, \ldots, X_{n})$, we denote $\mathbf x \in \mathcal X^n$ as a realization of $X^n$, and $x_i \in \mathcal X$ is the $i-$th component of $\mathbf x$.
We use the notation ${\lVert \cdot \rVert}_{1}$  and $\mathbb D (\cdot \Arrowvert \cdot)$ to denote the $L^1$  distance  and Kullback-Leibler (KL) divergence respectively.  
We write  $x \sim y$ if $x$ is proportional to $y$. Finally for a set $X$, we denote with $Q_X$ the uniform distribution over $\mathcal X$.

\vspace{1mm}
We recall some useful definition and results.

\begin{defi}
Given $A$ generated according to $ P_A$ and $(A,B)$ generated according to $P_{AB}$
\begin{itemize}
\item Information (or entropy density): $ h_{P_A}\coloneqq \log{\frac{1}{P_{A}(\mathbf a)}}$;
\item Conditional information:
$h_{P_{A|B}} (a|b)\coloneqq \log{\frac{1}{P_{A|B}(\mathbf a|\mathbf b)}} $;
\item Information density: $\imath_{P_{AB}}\coloneqq \log{\frac{P_{AB}(\mathbf a,\mathbf b)}{P_{A}(\mathbf a) P_{B}(\mathbf b)}}$.
\end{itemize}
Whenever the underlying distribution is clear from the context, we drop the subscript  from $h(\cdot)$ and $\imath(\cdot,\cdot)$.
\end{defi}

\vspace{1mm}

\begin{lem}[Properties of $L^1$  distance and K-L divergence]\label{tv prop}
\begin{enumerate}[(i)]
\item \label{cuff16}${\lVert P_{A}-\hat P_{A}\rVert}_{1} \leq {\lVert P_{AB}- \hat P_{AB}\rVert}_{1}$, see \cite[Lemma 16]{cuff2009thesis},
\item  \label{cuff17}${\lVert P_A- \hat P_A\rVert}_{1}={\lVert P_AP_{B|A}- \hat P_A P_{B|A}\rVert}_{1}$, see \cite[Lemma 17]{cuff2009thesis},
\item \label{lem4} If ${\lVert P_{A} P_{B|A}-P'_{A} P'_{B|A}\rVert}_{1}$ $ = \epsilon$,  then 
there exists $\mathbf a \in \mathcal A$ s.t.
${\lVert P_{B|A=\mathbf a}-P'_{B |A= \mathbf a}\rVert}_{1}\leq 2 \epsilon$, see \cite[Lemma 4]{yassaee2014achievability}.

\end{enumerate}
\end{lem}

\vspace{1mm}
\begin{defi}\label{defcoup}
A coupling of  two probability mass functions $P_A$ and $P_{A'}$ on  $\mathcal A$ is any  probability mass function
$\hat P_{AA'}$ defined on $\mathcal{A} \times \mathcal{A}$  whose marginals are $P_A$ and $P_{A'}$.
\end{defi}

\begin{prop}[Coupling property $\mbox{\cite[I.2.6]{Lindvall1992coupling}}$]\label{theocoup}
Given $A $ generated according to $ P_{A}$, $A' $ generated according to $P_{A'}$, any coupling 
$\hat P_{AA'}$ of $P_{A}$, $P_{A'}$ satisfies
\begin{equation*}
{\lVert P_A- P_{A'}\rVert}_{1}\leq 4 \, \mathbb P_{\hat P_{AA'}}\{A \neq A'\}.
\end{equation*}
\end{prop}

Now, we recall the Berry-Esseen Central Limit Theorem.
\vspace{1mm}
\begin{theo}[Berry-Esseen Central Limit Theorem $\mbox{\cite[Thm.~2]{erokhin1958varepsilon}}$]\label{berryessen}
Given $n>0$ and $Z_i$, $i=1, \ldots,n$  independent r.v.s. Then, for any real $t$,
\vspace{0.5mm}
 \begin{equation*}
  \left \lvert \mathbb{P} \left\{ \sum_{i=1}^n Z_i >n \left(\mu_n +t \sqrt{\frac{V_n}{n}}\right)\right\} -Q(t) \right \rvert \leq \frac{B_n}{\sqrt{n}},
 \end{equation*}where
 {\allowdisplaybreaks
 \begin{align*}
 \mu_n &= \frac{1}{n} \sum_{i=1}^n \mathbb E [Z_i],\\
 V_n  &= \frac{1}{n} \sum_{i=1}^n \text{Var} [Z_i],\\
 T_n &=\frac{1}{n}\sum_{i=1}^n \mathbb{E} [{\lvert Z_i - \mu_i \rvert}^3],\\
 B_n &= 6 \frac{T_n}{V_n^{3/2}},
 \end{align*}}and $Q(\cdot)$  is the tail distribution function of the standard normal distribution.
\end{theo}

\vspace{-2mm}
\subsection{Point-to-point setting}\label{subsecsec: sys}

As in~\cite{cuff2010}, we consider two nodes connected by a one-directional error-free link of rate $R$ and sharing a common source of uniformly distributed randomness $C$ defined on $\mathcal C$ of rate $R_0 = \frac{\log \lvert \mathcal C \rvert}{n}$. 
At time $i=1,\ldots,n$, the nodes perform  $U_i$ and $V_i$ respectively. 
The sequence $U^n$ is assigned by nature and behaves 
according to the fixed distribution $\bar P_{U^n}$. Then, the encoder generates a message $M$ defined on $\mathcal M$ of rate $R= \frac{\log \lvert \mathcal M \rvert}{n}$ as a  function of $U^n$ and the common randomness $ C$, via the stochastic map
$f_n : \mathcal U^n \times \mathcal C  \rightarrow \mathcal M$.
The message is sent through the error-free link of rate $R$
and the sequence  $V^n$ is generated through the map $g_n: \mathcal M \times \mathcal C \rightarrow \mathcal V^n$ as a function of the message $M$ and of the common randomness $C$.

\vspace{-2mm}
\subsection{Asymptotic case}
We recall the definitions of \emph{achievability for strong coordination}  and \emph{strong coordination region} in the asymptotic regime~\cite{cuff2010} 
for the setting of Figure \ref{fig: isit2017}.
Let $P_{U^{n}V^{n}}$ be the joint distribution induced by the code $(f_n, g_n)$,
a triplet $(\bar{P}_{UV}, R, R_0)$, composed of the target distribution, the error-free channel rate and the rate of common randomness, is achievable for strong coordination  if  
\vspace{0.5mm}
\begin{equation*}
\lim_{n \to \infty} {\lVert P_{U^{n}V^{n}}- \bar{P}_{UV}^{\otimes n} \rVert}_{1}=0.
\end{equation*}Then, the strong coordination region is the closure of the set of achievable  triplets $(\bar{P}_{UV}, R, R_0)$~\cite{cuff2010}.
In more detail, the strong coordination region is characterized in~\cite[Theorem 10]{cuff2010} as follows:  
\begin{equation}\label{region cuff1}
\mathcal R_{\text{Cuff}}:=\begin{Bmatrix}[l]
(\bar P_{UV}, R, R_0)\\
\qquad \bar P_{UV}= \bar P_{U}  \bar P_{V|U}  \\
\qquad \exists \,W\in\mathcal{W},\, W \mbox{ generated according to } \bar P_{W|UV} \textnormal{ s.t. }\\ 
\qquad \bar P_{UWV}= \bar P_{U} \bar P_{W|U} \bar P_{V|W}\\
\qquad R \geq I(U;W) \\
\qquad R+R_0 \geq I(UV;W)\\
\qquad \lvert \mathcal W \rvert \leq \lvert \mathcal U \times   {\mathcal V} \rvert+1 
\end{Bmatrix}.
\end{equation}To derive a closed result, an auxiliary random variable $W$ is introduced, which represents a \vv{description} of the source on which the encoder and the decoder have to agree on in order to generate a distribution close in $L^1$ distance to i.i.d..


\vspace{1mm}
\section{Non-asymptotic case: definition and main results}\label{sec: nonasy}

We recall  the notion of $(\epsilon,n)$ fixed-length strong coordination as introduced in~\cite{Cervia2019Fixed}.
\vspace{1mm}

\begin{defi}[$(\epsilon,n)$ Fixed-length strong coordination]\label{def strong coord fix len}
For a fixed $\epsilon>0$ and $n>0$, a triplet $(\bar{P}_{UV}, R, R_0)$ is $(\epsilon,n)$-achievable for strong coordination if 
there exists  a code $(f_n,g_n)$ with common randomness rate $R_0$, such that
$${\lVert P_{U^{n}V^{n}}- \bar{P}_{UV}^{\otimes n}  \rVert}_{1}\leq \epsilon,$$
where $P_{U^{n}V^{n}}$ is the joint distribution induced by the code.
Then, the $(\epsilon,n)$  \emph{fixed-length strong coordination region} $\mathcal R{(\epsilon,n)}$ is the closure of the set of achievable  $(\bar{P}_{UV}, R,R_0)$.
\end{defi}

\vspace{1mm}
For the setting of Figure \ref{fig: isit2017}, the main result of this paper is the following inner and outer bounds for the $(\epsilon,n) $ fixed-length strong coordination region $\mathcal R{(\epsilon,n)}$. 

\vspace{1mm}

\begin{theo}[Inner bound for $\mathcal R{(\epsilon,n)}$ -- Sufficient Conditions]\label{theona_inner}
Let $\bar P_{U}$  be the given source distribution, then the triplets $(\bar{P}_{UV}, R, R_0)$
that satisfy the following conditions are achievable for $(\epsilon,n)$  \emph{fixed-length strong coordination}:
{\allowdisplaybreaks\begin{align}
&\bar P_{UV}=\bar P_{U} \bar P_{V|U}, \nonumber \\[2mm]
&\exists \,W\in\mathcal{W},\, W \mbox{ generated according to } \bar P_{W|UV}, \textnormal{ such that } \bar P_{UWV}=\bar P_{U} \bar P_{W|U}  \bar P_{V|W},\nonumber\\[1mm]
& R \geq I(W;U)  
+ Q^{-1} \left(\epsilon + O\left( \frac{1}{\sqrt n}\right) \right) \sqrt{ \frac{V_{\bar P_{W|U}}}{n}} +  O\left( \frac{\log n}{n}\right),\nonumber \\[2mm]
&R_0+R \geq  I(W;UV) 
 + Q^{-1} \left(\epsilon + O\left( \frac{1}{\sqrt n}\right) \right)  \sqrt{  \frac{V_{\bar P_{W|UV}}}{n}}+  O\left( \frac{\log n}{n}\right),\label{region fn inner}
\end{align}}where $Q(t)= \int_{t}^{\infty} \frac{1}{\sqrt{2 \pi}} e^{-x^2/2} dx$  is the tail distribution function of the standard normal distribution, and $V_{\bar P_{W|U}} $ and $V_{\bar P_{W|UV}} $  are the dispersions of the \emph{test channels} $\bar P_{W|U}$  and $\bar P_{W|UV}$ respectively, as defined in \cite[Thm.~49]{polyanskiy2010channel}:
{\allowdisplaybreaks
\begin{align*}
V_{\bar P_{W|U}}&\coloneqq \min_{\bar P_{W|U}} {\text{Var}} \left( \imath(W; U) |W\right)= \min_{\bar P_{W|U}} {\text{Var}} \left( \imath(W; U)\right),\\
V_{\bar P_{W|UV}}&\coloneqq \min_{\bar P_{W|UV}} {\text{Var}}\left( \imath(W; UV) |W\right)=\min_{\bar P_{W|UV}} {\text{Var}}\left( \imath(W; UV)\right),
\end{align*}}where $\bar P_{W|U}$  and $\bar P_{W|UV}$ are the \emph{test channels}  that connect the random variables $U$ and $V$ of the given setting with the auxiliary random variable $W$ representing the codebook, and the last identification follows from the fact that the channels $\bar P_{W|U} $ and $ \bar P_{W|UV}$ have no cost constraint~\cite[Section 22.3]{polyanskiy2014lecture}.
\end{theo}

\vspace{3mm}

\begin{theo}[Outer bound for $\mathcal R{(\epsilon,n)}$ -- Necessary Conditions]\label{theona_outer}
Let $\bar P_{U}$  be the given source distribution, then the triplets $(\bar{P}_{UV}, R, R_0)$
that are achievable for $(\epsilon,n)$  \emph{fixed-length strong coordination} have to satisfy the following conditions:
{\allowdisplaybreaks\begin{align}
& \bar P_{UV}=\bar P_{U} \bar P_{V|U}, \nonumber  \\[2mm]
&\exists \,W\in\mathcal{W},\, W \mbox{ generated according to }\bar P_{W|UV} \textnormal{ such that }  \bar P_{UWV}=\bar P_{U} \bar P_{W|U}  \bar P_{V|W},\nonumber\\[1mm]
& R \geq I(W;U)  
+ Q^{-1} \left(\epsilon + O\left( \frac{1}{\sqrt n}\right) \right) \sqrt{ \frac{V_{\bar P_{W|U}}}{n}} +  O\left( \frac{1}{n }\right),\nonumber\\[2mm]
& R_0+R \geq  I(W;UV) 
 + Q^{-1} \left(\epsilon + O\left( \frac{1}{\sqrt n}\right) \right)  \sqrt{  \frac{V_{\bar P_{W|UV}}}{n}}+  O\left( \frac{1}{n}\right)+ O \left( \frac{\log {\frac{1}{\epsilon}}}{\frac{1}{\epsilon}}\right),\nonumber\\[2mm]
& \lvert \mathcal W \rvert \leq \lvert\, \mathcal U \times {\mathcal V} \rvert+1.\label{region fn outer}
\end{align}}
\end{theo}
\vspace{2mm}

\begin{rem}[Closed result -- Sufficient conditions are also necessary]\label{theona} By getting a closer look at the rate conditions, we note that by taking the condition of the inner bound~\eqref{region fn inner}, we can retrieve a closed result.
In fact, for a given $(\epsilon,n)$ as well as target distribution $\bar P_{UV}$, any $(R, R_0)$ that satisfies the condition in~\eqref{region fn outer}, also satisfy the conditions in~\eqref{region fn inner}, since 
{\allowdisplaybreaks
\begin{align*}
&f(n)= O\left( \frac{\log n}{n}\right) \quad \mbox{ if } \exists k_1 >0,\, \exists n_0, \forall n\geq n_0 \quad \lvert f(n) \rvert \leq k_1  \frac{\log n}{n},\\
&g(n)= O\left( \frac{1}{n}\right) \quad \mbox{ if } \exists k_2 >0,\, \exists n_0, \forall n\geq n_0  \quad \lvert g(n) \rvert \leq k_2 \, \frac{1}{n},\\
&\mbox{and} \quad \lvert g(n) \rvert \leq k_2\,  \frac{1}{n}  \leq k_2\, \frac{\log n}{n} \quad \forall n \geq 2 \, \Rightarrow \, g(n)=O\left( \frac{\log n}{n}\right).
\end{align*}}
\end{rem}

\vspace{2mm}

\begin{rem}[Comparison with the asymptotic case]
We observe the following analogies between the asymptotic and the fixed-length case:
\begin{itemize}
\item The decomposition of the target joint distribution is the same (see~\eqref{region cuff1} and~\eqref{region fn inner},~\eqref{region fn outer}).
\vspace{1mm}
\item Even though necessary and sufficient conditions for $(\epsilon, n)$ strong coordination
lead to different rate constraints, we observe that the constant terms are the same in~\eqref{region fn inner} and in~\eqref{region fn outer}, whereas the difference is 
only in the growth rate of two functions of $n$ that go to zero as $n$ increases, although with different speeds.
More precisely,  both $O \left(  \frac{\log n}{n}   \right) $ and  $O \left(  \frac{1}{n}   \right) $ vanish when $n \to \infty$, hence the following terms in the inner bound of~\eqref{region fn inner} and in the outer bound  of~\eqref{region fn outer}
{\allowdisplaybreaks
\begin{align}
& O \left(  \frac{\log n}{n}   \right) + Q^{-1} \left(\epsilon + O\left( \frac{1}{\sqrt n}\right) \right)  \sqrt{ \frac{ V_{\bar P}}{n}}\nonumber\\
& O \left(  \frac{1}{n}   \right) + Q^{-1} \left(\epsilon + O\left( \frac{1}{\sqrt n}\right) \right)  \sqrt{ \frac{ V_{\bar P}}{n}}\label{lognn}
\end{align}}coincide asymptotically to
\begin{equation}
Q^{-1} \left(\epsilon \right)  \sqrt{ \frac{ V_{\bar P}}{n}}.
\end{equation}Thus the rate conditions of both the inner bound and the outer bound are reduced to
{\allowdisplaybreaks
\begin{align}
& R \geq I(W;U)  
+ Q^{-1} \left(\epsilon  \right) \sqrt{ \frac{V_{\bar P_{W|U}}}{n}} ,\nonumber\\[2mm]
& R_0+R \geq  I(W;UV) + Q^{-1} \left(\epsilon \right)  \sqrt{  \frac{V_{\bar P_{W|UV}}}{n}}.
\end{align}}

\vspace{1mm}
\item Perhaps more interestingly, as in~\cite{polyanskiy2010channel,kostina2012fixed,kostina2013lossyit} for channel coding and compression in the finite-length regime, we observe that by letting $n$ tend to infinity, we end up with the same rate conditions of the asymptotic case~\eqref{region cuff1}, therefore reproving the results of~\cite{cuff2010} with different techniques.
In fact, in the asymptotic regime $\epsilon$ vanishes when $n \to \infty$, and 
\begin{equation}\label{lognn2}
Q^{-1} \left(\epsilon + O\left( \frac{1}{\sqrt n}\right) \right)  \sqrt{ \frac{ V_{\bar P}}{n}} \sim \log {\left(\frac{1}{O (\epsilon)}\right)} \sqrt{ \frac{ V_{\bar P}}{n}}. 
\end{equation}Then, if for example, $\epsilon ~\sim \frac{1}{\sqrt n}$,~\eqref{lognn2} becomes
{\allowdisplaybreaks
\begin{align*}
 \sqrt{ V_{\bar P}} \,\frac{Q^{-1} \left(\epsilon + O\left( \frac{1}{\sqrt n}\right) \right) }{\sqrt n} \sim   \sqrt{ V_{\bar P}} \,\frac{Q^{-1}\left( \frac{1}{\sqrt n}\right)}{\sqrt n}  \sim \sqrt{ V_{\bar P}} \, \frac{\log{\sqrt{n}}}{\sqrt n}\to 0.
\end{align*}}Hence, we can recover the   asymptotic region of \eqref{region cuff1} from  the fixed-length necessary and sufficient conditions of \eqref{region fn inner} and~\eqref{region fn outer}. 
Moreover, with this choice the bound $\epsilon_{\text{Tot}}$ on the  $L^1$ distance between the two distribution goes to zero as $1/\sqrt n$.
\end{itemize}
\end{rem}


\section{Inner bound}\label{sec: nonasy ib}


\paragraph*{Outline of the proof of Theorem \ref{theona_inner}}
The achievability proof is based on  non-asymptotic \emph{ output statics of random binning} \cite{yassaee2013non} and is decomposed into the following steps:
\vspace{1mm}
\begin{enumerate}[A.]
\item preliminary definitions and results on random binning are recalled;
\vspace{1mm}
\item two schemes are defined  for a fixed $n$, a random binning and a random coding scheme;
using the properties of random binning, it is possible to derive an upper bound on the  $L^1$ distance between the i.i.d. random binning distribution $P^{\text{RB}}$ and random coding distribution $P^{\text{RC}}$, providing a first bound on ${\lVert P^{\text{RB}} - P^{\text{RC}} \rVert}_{1} $.
Then, a second bound $\epsilon_{\text{Tot}}$ is recovered, by reducing the rate of common randomness to obtain the conditions in \eqref{region fn inner};
\vspace{1mm}
\item the term $\epsilon_{\text{Tot}}$ is analyzed;
\vspace{1mm}
\item the rate conditions are summarized.
\end{enumerate}

\vspace{0.5mm}
 \begin{rem}
Observe that, as we will see in Section \ref{subsec: reduce rate cr},  the final  bound $\epsilon_{\text{Tot}}$ on  the $L^1$ distance between $P^{\text{RB}}$ and $P^{\text{RC}}$ is worse than the one found in Section \ref{subsec: before reducing}.
However, by worsening the $L^1$ distance, we can reduce the rate of common randomness.
 \end{rem}

\vspace{-1mm}
\subsection{Preliminaries on random binning}\label{osrb properties}
Let $A$ taking values in $\mathcal A$ be partitioned into $2^R$ \emph{bins} at random, we denote with $\varphi:$ $ \mathcal A$ $\to$ $  \llbracket 1,2^{R} \rrbracket,$ $\mathbf a \mapsto \mathbf k$, the realization of such partition (or binning), and we call $K\coloneqq \varphi(A)$ a \emph{random binning} of $\mathcal A$. With a slight abuse of notation, throughout this text we may refer to the map $\varphi$ as a \emph{uniform random binning}, if $\varphi(A)$ is a binning of $A$ and the partition into bins is performed uniformly at random through $\varphi$.\\
Now, let the pair $(A,B)$ generated according to $P_{AB} $ be a discrete source, and $\varphi$ introduced above be a uniform map,
we denote the distribution induced by the binning as
\begin{equation}\label{prb1}
 P^{\text{RB}} (\mathbf a,\mathbf b,\mathbf k )\coloneqq P_{AB}(\mathbf a,\mathbf b) \mathds 1 \{\varphi(\mathbf a)=\mathbf k \}.
\end{equation}

The first objective consists in ensuring that the binning is almost uniform and almost independent from the source so that the random binning scheme and the random coding scheme generate joint distributions that have the same statistics. 
\vspace{0.5mm}
\begin{theo}[$\mbox{\cite[Thm.~1]{yassaee2013non}}$]\label{oneshot1}
 Given $P_{AB}$, for every distribution $T_B$ on $\mathcal B$ and any $\gamma \in \mathbb R^{+}$,  the marginal of $P^{\text{RB}}$ in \eqref{prb1}
 satisfies
 {\allowdisplaybreaks
 \begin{align}
  &\mathbb E {\lVert  P^{\text{RB}} (\mathbf b,\mathbf k) - Q_{K}(\mathbf k) P_B(\mathbf b) \rVert}_{1} \leq \epsilon_{\text{App}}, \nonumber\\
  & \epsilon_{\text{App}}\coloneqq P_{AB}{ \left( \mathcal S_{\gamma} (P_{AB} \| T_B )^{\mathrm{c}}  \right)}+2^{-\frac{\gamma+1}{2}},\label{eqos1}\
 \end{align}}where for a set $X$, we denote with $Q_X$ the uniform distribution over $\mathcal X$ and
\begin{equation}
 \mathcal S_{\gamma}(P_{AB} \| T_B ) \coloneqq  \left\{  (\mathbf a,\mathbf b) : h_{P_{AB}} (\mathbf a,\mathbf b) -  h_{T_B}  (\mathbf b)  -  nR  >  \gamma \right\} .
\end{equation}
\end{theo}With the previous result we measure in terms of $L^1$ distance how well we can approximate a distribution for which the binning of $A$ is independently generated from $B$ and uniform, of which we characterize the upper bound $\epsilon_{\text{App}}$.  Intuitively, the approximation error $\epsilon_{\text{App}}$ is small if 
the number of bins in which we partition $A$ is high, in particular it has to be higher than the conditional information of $A$ given $B$, with the real number $\gamma$ allowing some degree of freedom.

Before stating the second property,  we introduce the decoder that we will use, sometimes called the  \emph{mismatch stochastic likelihood coder (SLC)}~\cite{scarlett2013mismatched,martinez2009bit}.
\begin{defi}\label{dec slc}
Let $T_{AB}$ be an arbitrary probability mass function, and  $\varphi :   \mathcal A  \to  \llbracket 1,2^{R} \rrbracket$, $\mathbf a \mapsto \mathbf k$ a uniform random binning of $A$. A  mismatch SLC is defined by the following induced conditional distribution
 \begin{equation}\label{eq dec slc}
 \hat T_{\hat A|BK} (\hat {\mathbf a}|\mathbf b,\mathbf k)\coloneqq \frac{T_{A|B}(\hat {\mathbf a}|\mathbf b) \mathds 1 \{\varphi(\hat{\mathbf a})=\mathbf k \}}{\sum_{\bar{\mathbf a} \in \mathcal A} T_{A|B}(\bar{\mathbf a}|\mathbf b) \mathds 1 \{\varphi(\bar{\mathbf a})=\mathbf k \}}.
\end{equation}\end{defi}

Then, the following result is used to bound the error probability of decoding $A$ when the decoder has access to the side information $B$ as well as to the binning $\varphi(A)=K$.
\begin{theo}[$\mbox{\cite[Thm.~2]{yassaee2013non}}$]\label{oneshot2}
 Given $P_{AB}$ and any distribution $T_{AB}$, the following bound on the error probability of the decoder defined in~\eqref{eq dec slc}  holds
 \begin{equation}\label{eqos2}
  \mathbb E \left[ P[\mathcal E] \right] \leq P_{AB} {\left(\mathcal S_{\gamma} (T_{AB})^{\mathrm{c}} \right)} + 2^{-\gamma } =: \epsilon_{\text{Dec}},
 \end{equation}where $\gamma$ is an arbitrary positive number and 
 \begin{equation}
  \mathcal S_{\gamma} (T_{AB})\coloneqq \left\{ (\mathbf a,\mathbf b) : nR -h_{T_{A|B}} (\mathbf a|\mathbf b) > \gamma \right\}.
 \end{equation}
\end{theo}While we will use Theorem~\ref{oneshot1} at the encoder to ensure that the random binning probability distribution approximates  well a random coding process, the latter is used at the decoder's side to minimize the probability of error of generating the wrong sequence. In this context, the bound on the error probability $\epsilon_{\text{Dec}}$ is small if the number of bins is upper bounded by the conditional information. Since at the encoder's side we had the opposite request, by demanding that the rate  (or equivalently the number of bins) is large enough, we would have to find a compromise between two seemingly competing goals. This issue will be resolved by carefully choosing different side information at the encoder and at the decoder, and by playing with different values of $\gamma$.


\vspace{-1mm}
\subsection{Fixed-length coordination scheme}\label{fixed length scheme}

The encoder and the decoder share a source of uniform randomness $C \in \llbracket 1,2^{n R_0} \rrbracket$. 
Moreover, suppose that  the encoder and decoder have access not only to common randomness $C$ but also to extra randomness $F$, where $C$ is generated uniformly at random  in $\llbracket 1,2^{nR_0} \rrbracket$ with distribution $Q_C$ and $F$ is generated uniformly at random in $\llbracket 1,2^{n \tilde R} \rrbracket$ with distribution $Q_F$ independently of $C$. 
The encoder observes the source  $U^n $ generated according to  $\bar P_{U^n}$  and selects a message  $M$ of rate $R$,  which is then transmitted through an error-free link to the decoder.
Then, the decoder exploits the message and the common randomness to select  $V^n$. 

In the rest of this section, we introduce an auxiliary random variable $W^n$ which is not part of the setting such that the Markov chain $U-W-V$ holds. This random variable represents the \vv{description} of the source on which the encoder and the decoder have to agree  to produce the right distributions.
In order to do so, we consider the i.i.d. target distribution,  and we define three binnings of $W^n$, thus inducing a joint distribution on the random variables $(U^n, W^n, V^n)$ and on the binnings, which we call \emph{random binning distribution}.
Then, we define a \emph{random coding distribution}, and we use the binning properties of Theorem~\ref{oneshot1}  and Theorem~\ref{oneshot2}  to estimate the $L^1$ distance between this random coding distribution and the random binning distribution. Since the marginal of the random binning distribution coincides with the target distribution, we can estimate the  upper bound on the $L^1$ distance $\epsilon_{\text{Tot}}$ between the target distribution and the distribution induced by the code. This will be done in two steps: first, we derive an upper bound on the $L^1$ distance by coordinating the sequences $(U^n, W^n, V^n)$; finally we see how  to reuse Theorem~\ref{oneshot2} to reduce the amount of common randomness and coordinate $U^n$ and $V^n$ only.


\vspace{2mm}
\subsubsection{Random binning scheme}

Let $\bar P_{U}  \bar P_{V| U}$ be the target distribution, we introduce an auxiliary random variable $W$ such that 
$U^{n}$, $W^{n}$, and $V^{n}$ 
are jointly i.i.d. with distribution 
\begin{equation*}
\bar P\coloneqq \bar P_{U^{n}} \bar P_{W^{n}| U^{n}} \bar P_{V^{n}|W^{n}}.
\end{equation*}

We consider three uniform random binnings for $W^n$:
\begin{enumerate}[i)]\setlength{\itemsep}{0.2em}
\item binning $C= \varphi_C(W)$, where  $\varphi_C: \mathcal{W}^n \to \llbracket 1,2^{nR_0} \rrbracket$,  
\item binning $M= \varphi_F(W)$, where $\varphi_F: \mathcal{W}^n \to \llbracket 1,2^{n R} \rrbracket$,  
\item binning $F= \varphi_M(W)$, where $\varphi_M: \mathcal{W}^n \to \llbracket 1,2^{n \tilde R} \rrbracket$,
\end{enumerate}and, inspired by~\cite{yassaee2013technique, yassaee2013non, scarlett2013mismatched}, we consider a decoder defined according to~\eqref{eq dec slc} that reconstructs $\hat W^n$:
 {\allowdisplaybreaks
 \begin{align}
  \hat T_{\hat W^n|FCM} (\hat {\mathbf w}| \mathbf f, \mathbf c,\mathbf m)
 \coloneqq \frac{T_{W^n}(\hat {\mathbf w}) \mathds 1 \{\varphi(\hat {\mathbf w})=( \mathbf f, \mathbf c,\mathbf m) \}}{\sum_{\bar {\mathbf w} \in \mathcal W^n} T_{W^n}(\bar {\mathbf w}) \mathds 1 \{\varphi(\bar {\mathbf w})= ( \mathbf f, \mathbf c,\mathbf m)\}}\label{decoder}
 \end{align}}where $\varphi=(\varphi_C, \varphi_F, \varphi_M)$.
This induces the joint distribution
\begin{equation}\label{prb}
P^{\text{RB}}  \coloneqq   \bar P_{U^{n}} \bar P_{W^{n}| U^{n}} \bar P_{F|W^{n}}   \bar P_{C| W^{n}}  \bar P_{M|W^{n}} \bar P_{V^{n}|W^n}  \hat T_{\hat W^n|FCM}.
\end{equation}In particular, $P^{\text{RB}}_{W^n|FCU^n}$ is well defined.

\vspace{2mm}


\subsubsection{Random coding scheme}
The encoder generates $W^n$ according to $P^{\text{RB}}_{W^n|FCU^n}$ defined above.
At the decoder, $\hat W^n$ is generated via the conditional distribution  $\hat T_{\hat W^n|FCM}$. 
The decoder then generates $V^n$ according to the distribution 
$P^{\text{RC}}_{V^n|\hat W^n}(\hat{\mathbf v}|\hat{\mathbf w})\coloneqq\bar P_{V^n|\hat W^n}(\hat{\mathbf v}|\hat{\mathbf w}),$
where $\hat{\mathbf w}$ is the output of a decoder defined as in~\eqref{eq dec slc}.
This induces the joint distribution
\begin{equation}\label{prc}
P^{\text{RC}} \coloneqq  Q_F Q_C  \bar P_{U^n} P^{\text{RB}}_{W^n|FCU^n} \bar P_{M|W^n}  \hat T_{\hat W^n|FCM} P^{\text{RC}}_{V^n|\hat W^n}.
\end{equation}

Observe that the marginal of the distribution $ P^{\text{RB}}$ is by construction trivially close in $L^1$ distance to the target distribution $\bar P$. 
We use the properties of random binning to show that the random binning distribution $ P^{\text{RB}}$ and the random coding distribution $ P^{\text{RC}}$ are $\epsilon$-close in $L^1$ distance, and therefore so are the marginals of $ P^{\text{RC}}$ and  $\bar P$.


 \vspace{2mm}
\subsubsection{Strong coordination of $(U^n,  V^n , W^n )$ --- Initial bound}\label{subsec: before reducing}
By applying Theorem \ref{oneshot1} and  Theorem \ref{oneshot2} to  $P^{\text{RB}}$ and $P^{\text{RC}}$, we have  
{\allowdisplaybreaks
\begin{align*}
  &{\lVert P^{\text{RB}}_{U^n W^n CFM}  - P^{\text{RC}}_{U^n W^n CFM} \rVert}_{1} \overset{\mathclap{(a)}}{=} {\lVert \bar P_{U^n }  \bar P_{W^n|U^n}    \bar P_{C|W^n}   \bar P_{F|W^n}  -   Q_C Q_F  \bar P_{U^n}  P^{\text{RB}}_{W^n|CFU^n}    \rVert}_{1} \leq \epsilon_{\text{App}},\\[1mm]
  & \mathbb E \left[ P[\mathcal E] \right] \leq \epsilon_{\text{Dec}}, 
\end{align*}}where $(a)$ comes from Lemma \ref{tv prop}~(\ref{cuff17}), and
{\allowdisplaybreaks
\begin{subequations}
\begin{align}
 \epsilon_{\text{App}} &\coloneqq \bar P_{U^n CF}{ \left( \mathcal S_{\gamma_1}^{\mathrm{c}} \right)} +2^{-\frac{\gamma_1+1}{2}},\label{epsilonapp}\\[1mm]
 \epsilon_{\text{Dec}} &\coloneqq \bar P_{W^n} {\left(\mathcal S_{\gamma_2}^{\mathrm{c}} \right)} + 2^{-\gamma_2}, \label{epsilondec}
\end{align}\end{subequations}}with $\gamma_1$ and $\gamma_2$ arbitrary positive numbers, and 
 {\allowdisplaybreaks
 \begin{subequations}
 \begin{align}
&\mathcal S_{\gamma_1}\coloneqq  \, \mathcal S_{\gamma_1} (\bar P_{U^n CF} \|  \bar P_{U^n} )
=\{  ( \mathbf u,  \mathbf w): h_{\bar P}(\mathbf u, \mathbf w)
  -  h_{ \bar P} (\mathbf u)  -  n(\tilde R+ R_0) >  \gamma_1   \},\label{sgamma1}\\[1.5mm]
 &\mathcal S_{\gamma_2} \coloneqq \mathcal S_{\gamma_2} (\bar P_{W^n}) 
=   \{\mathbf w : n(R+R_0+\tilde R) - h_{\bar P} (\mathbf w)  > \gamma_2 \}\nonumber\\
&\phantom{\!\! \mathcal S_{\gamma_1} (\bar P_{U^n CF} \|  \bar P_{U^n} )} \overset{\mathclap{(b)}}{=} \Big\{ \mathbf w : n(R+R_0+\tilde R) - \sum_{i=1}^n h_{\bar P} (w_i)  > \gamma_2\Big\},\label{sgamma2}\vspace{-2mm}
 \end{align}\end{subequations} }where $(b)$ comes from the choice of the decoder \eqref{decoder}.
Then, we have
{\allowdisplaybreaks
\begin{align*}
 {\lVert P^{\text{RB}}_{U^n W^n CF}   \bar P_{M|W^n} \hat T_{\hat W^n|CFM}  -    P^{\text{RC}}_{U^n W^n CF}  \bar P_{M|W^n}   \hat T_{\hat W^n|CFM}   \rVert}_{1} 
&=  {\lVert P^{\text{RB}}_{U^n W^n CFM \hat W^n}    -    P^{\text{RC}}_{U^n W^n CFM  \hat W^n}   \rVert}_{1} \\& \leq  \epsilon_{\text{App}}+ \epsilon_{\text{Dec}}.
\end{align*}}
To conclude, observe that in the random binning scheme we have $V^n $ generated according to $  \bar P_{V^n|W^n} $, $W^n$ generated according to $ \bar P_{W^n|U^n}$, while in the random coding scheme we have $V^n$ generated according to  $P^{\text{RC}}_{V^n|\hat W^n}$, $\hat W^n$ generated according to $ \hat T_{\hat W^n|CFM}$. Then, by applying the coupling result of Proposition \ref{theocoup}, we have
\begin{equation*}
 {\lVert P^{\text{RB}} - P^{\text{RC}} \rVert}_{1}\leq  \epsilon_{\text{App}}+5 \, \epsilon_{\text{Dec}}.
\end{equation*}


\vspace{1mm}

\subsubsection{Reducing the rate of common randomness --- Final bound}\label{subsec: reduce rate cr}

Although in a first instance we have exploited the extra randomness $F$  to coordinate the whole triplet $(U^n, V^n, W^n)$ with maximal $L^1$ distance $\epsilon_{\text{App}}+5 \, \epsilon_{\text{Dec}}$, we now show that we do not need it in order to coordinate only $(U^n, V^n)$.
As in~\cite{yassaee2014achievability} we can reduce the required amount of common randomness by having the two nodes agree on a suitable realization of the extra randomness $F$. By fixing $F = \mathbf f$, we will reduce the rate requirements at the expense of the upper bound on the $L^1$ distance by introducing a third approximation term, ending up with a new upper bound $\epsilon_{\text{Tot}}$. Now, we  detail under which circumstance such suitable realization of $F$ exists, and we derive the final upper bound on the $L^1$ distance $\epsilon_{\text{Tot}}$.
To do so, first we apply 
Theorem \ref{oneshot1} to $A=W^n$, $B=(U^n, V^n)$, $P_{B}=  P^{\text{RB}}_{U^nV^n}$, $P_{AB}= P^{\text{RB}}_{U^nV^n W^n}$ and $K=F$. Then, we have 
 \begin{equation}\label{eq3}
\lVert  P^{\text{RB}}_{U^n V^n F }   {- Q_{F}P^{\text{RB}}_{U^n V^n  }  \rVert}_{1} \leq \epsilon_{\text{App},2},
 \end{equation}where
 {\allowdisplaybreaks
 \begin{subequations}
 \begin{align}
   \epsilon_{\text{App},2} & \coloneqq
  P^{\text{RB}}{ \left( \mathcal S_{\gamma_3}^{\mathrm{c}} \right)}+2^{-\frac{\gamma_3+1}{2}},\label{epsilonapp2}\\[1mm]
    \mathcal S_{\gamma_3} &\coloneqq\mathcal S_{\gamma_3} (P^{\text{RB}}_{U^nV^n W^n} \| P^{\text{RB}}_{U^nV^n} )
     =   \{ (\mathbf u, \mathbf v,  \mathbf w)  : h_{P^{\text{RB}}}(\mathbf u,  \mathbf v,  \mathbf w)   -  h_{P^{\text{RB}}} (\mathbf u, \mathbf v) - n\tilde R  >  \gamma_3 \}. \label{sgamma3}
 \end{align}\end{subequations}}Now, we recall that  by  Lemma \ref{tv prop}~(\ref{cuff16}), we have
{\allowdisplaybreaks
\begin{align}
{\lVert P^{\text{RB}}_{U^n V^n  F}-P^{\text{RC}}_{U^n  V^n F} \rVert}_{1} &  \leq  {\lVert P^{\text{RB} }- P^{\text{RC}}  \rVert}_{1}  \leq   \epsilon_{\text{App}}+5\, \epsilon_{\text{Dec}},\label{eqtv1}
\end{align}}and combining  \eqref{eq3} and \eqref{eqtv1} with  the triangle inequality, we have
 {\allowdisplaybreaks
 \begin{align*}
 { \lVert Q_F P^{\text{RB}}_{U^n V^n}-Q_F P^{\text{RC}}_{U^n V^n} \rVert}_{1}  
  & \leq {\lVert P^{\text{RB}}_{U^n V^n F}- Q_F P^{\text{RB}}_{U^n V^n} \rVert}_{1}  + {\lVert P^{\text{RB}}_{U^n V^n F}- P^{\text{RC}}_{U^n V^n F} \rVert}_{1}  \\[1mm] 
& \leq \epsilon_{\text{App},2} +\epsilon_{\text{App}}+5\,\epsilon_{\text{Dec}}.
 \end{align*}}Finally by  Lemma \ref{tv prop}~(\ref{lem4}), there exists an instance $F= \mathbf f$, such that 
 {\allowdisplaybreaks
 \begin{subequations}
 \begin{align}
  & {\lVert P^{\text{RB}}_{U^n V^n| F=\mathbf f }- P^{\text{RC}}_{U^n V^n| F=\mathbf f } \rVert}_{1}   \leq \epsilon_{\text{Tot}},\label{var tot}\\[1mm]
  & \epsilon_{\text{Tot}}\coloneqq2 \,(\epsilon_{\text{App},2} +  \epsilon_{\text{App}}+5\, \epsilon_{\text{Dec}}).\label{epsilon tot}
 \end{align}\end{subequations}}

\vspace{-1mm}
\subsection{Analysis of the $L^1$ distance $\epsilon_{\text{Tot}}$}\label{tv analysis}

Here, we take a closer look at the overall $L^1$ distance between the i.i.d. distribution and the random coding one, denoted with $\epsilon_{\text{Tot}}$.
First, substituting the explicit expression for $(\epsilon_{\text{App}}, \epsilon_{\text{Dec}}, \epsilon_{\text{App},2} )$ of \eqref{epsilonapp}, \eqref{epsilondec}, and \eqref{epsilonapp2} into \eqref{epsilon tot}, the bound in \eqref{var tot} becomes
{\allowdisplaybreaks
\begin{align}
 & \epsilon_{\text{Tot}} =    2 \,\bar P_{U^n CF}{ \left( \mathcal S_{\gamma_1}^{\mathrm{c}}\right)}+ 10\, \bar P_{W^n} {\left(\mathcal S_{\gamma_2}^{\mathrm{c}}   \right)}+   2\, P^{\text{RB}}{ \left( \mathcal S_{\gamma_3}^{\mathrm{c}}  \right)}
 + 2 \, \Big[2^{-\frac{\gamma_1+1}{2}} +5 \cdot 2^{-\gamma_2}+ 2^{-\frac{\gamma_3+1}{2}} \Big]. \label{epsilon tot2}
\end{align}}By the union bound and De Morgan's law, we have
{\allowdisplaybreaks
\begin{align}
  \epsilon_{\text{Tot}}& \leq     10 \left[ \,\bar P_{U^n CF}{ \left( \mathcal S_{\gamma_1}^{\mathrm{c}}\right)}+ \, \bar P_{W^n} {\left(\mathcal S_{\gamma_2}^{\mathrm{c}}   \right)}+   \, P^{\text{RB}}{ \left( \mathcal S_{\gamma_3}^{\mathrm{c}}  \right)} \right]
+ 2 \, \Big[2^{-\frac{\gamma_1+1}{2}} +5 \cdot 2^{-\gamma_2}+ 2^{-\frac{\gamma_3+1}{2}} \Big]\nonumber\\
  & =     10 \left[ \,\bar P{  \left(  \mathcal S_{\gamma_1}^{\mathrm{c}} \cup \mathcal S_{\gamma_2}^{\mathrm{c}} \cup \mathcal S_{\gamma_3}^{\mathrm{c}}  \right)}\right]
 + 2 \, \Big[2^{-\frac{\gamma_1+1}{2}} +5 \cdot 2^{-\gamma_2}+ 2^{-\frac{\gamma_3+1}{2}} \Big]\nonumber\\
  &  \leq     10 \left[ \,\bar P  \left( {  \left(  \mathcal S_{\gamma_1} \cap \mathcal S_{\gamma_2} \cap \mathcal S_{\gamma_3}\right)}^{\mathrm{c}}\right)  \right]
 + 2 \, \Big[2^{-\frac{\gamma_1+1}{2}} +5 \cdot 2^{-\gamma_2}+ 2^{-\frac{\gamma_3+1}{2}} \Big].\end{align}}In the next paragraph, we investigate  $ \left(  \mathcal S_{\gamma_1} \cap \mathcal S_{\gamma_2} \cap \mathcal S_{\gamma_3}\right)^{\mathrm{c}}$, to  understand  which rate conditions are dominant to minimize the measure of the set as a function of  $\gamma_i $, $i=1,2,3$. In a second instance, we choose the parameters $(\gamma_2, \gamma_2, \gamma_3)$ such that $\epsilon_{\text{Tot}} $  defined above is small.


\vspace{2mm}

\subsubsection{Analysis of $ ( \mathcal S_{\gamma_1} \cap \mathcal S_{\gamma_2} \cap \mathcal S_{\gamma_3})^{\mathrm{c}}$}

First, we write explicitly the set:
{\allowdisplaybreaks
\begin{align}
\MoveEqLeft[1]
\left(  \mathcal S_{\gamma_1} \cap \mathcal S_{\gamma_2} \cap \mathcal S_{\gamma_3}\right)^{\mathrm{c}}\nonumber\\
&= \begin{Bmatrix}[l]
(\mathbf u,\mathbf v,\mathbf w)  \,  : \\
 \,  h(\mathbf u,\mathbf v,\mathbf w)-h(\mathbf u,\mathbf v)-n \tilde R > \gamma_3\\
 \,   h(\mathbf u,\mathbf w)-h(\mathbf u)-n (R_0 + \tilde R) > \gamma_1\\
 \,   n (R_0 + \tilde R +R) - h(\mathbf w)> \gamma_2
\end{Bmatrix}^{\mathrm{c}}\nonumber\\[1mm]
& = \begin{Bmatrix}[l]
(\mathbf u,\mathbf v,\mathbf w)  \,  : \\
 \,  h(\mathbf w|\mathbf u\mathbf v)-n \tilde R > \gamma_3\\
 \,   h(\mathbf w|\mathbf u)-n (R_0 + \tilde R) > \gamma_1\\
 \,   n (R_0 + \tilde R +R) - h(\mathbf w)> \gamma_2
\end{Bmatrix}^{\mathrm{c}}\nonumber\\[1mm]
& = \begin{Bmatrix}[l]
(\mathbf u,\mathbf v,\mathbf w) \, : \\
 \,  n \tilde R < h(\mathbf w|\mathbf u\mathbf v)-\gamma_3\\
 \,   n (R_0 +  R) >  \imath(\mathbf w;\mathbf u\mathbf v) +\gamma_2+\gamma_3\\
 \,  n R >\imath(\mathbf w;\mathbf u)+ \gamma_2+\gamma_1
\end{Bmatrix}^{\mathrm{c}}\nonumber\\[1mm]
& = \begin{Bmatrix}[l]
(\mathbf u,\mathbf v,\mathbf w)  \,  : \\
 \,   n \tilde R < h(\mathbf w|\mathbf u\mathbf v)-\gamma_3
\end{Bmatrix}^{\mathrm{c}} \cup  \begin{Bmatrix}[l]
(\mathbf u,\mathbf v,\mathbf w)  \, : \\
 \,  n (R_0 +  R) >  \imath(\mathbf w;\mathbf u\mathbf v) +\gamma_2+\gamma_3
\end{Bmatrix}^{\mathrm{c}} \cup \begin{Bmatrix}[l]
(\mathbf u,\mathbf v,\mathbf w)  \,  : \\
 \,   n R >\imath(\mathbf w;\mathbf u)+ \gamma_2+\gamma_1
\end{Bmatrix}^{\mathrm{c}}\nonumber\\[1mm]
& = \begin{Bmatrix}[l]
(\mathbf u,\mathbf v,\mathbf w) \,  : \\
 \,   n \tilde R \geq h(\mathbf w|\mathbf u\mathbf v)-\gamma_3
\end{Bmatrix} \cup  \begin{Bmatrix}[l]
(\mathbf u,\mathbf v,\mathbf w)  \,  : \\
 \,   n (R_0 +  R) \leq \imath(\mathbf w;\mathbf u\mathbf v) +\gamma_2+\gamma_3
\end{Bmatrix} \cup \begin{Bmatrix}[l]
(\mathbf u,\mathbf v,\mathbf w)  \,  : \\
 \,   n R \leq \imath(\mathbf w;\mathbf u)+ \gamma_2+\gamma_1
\end{Bmatrix}\label{demorgan}.
\end{align}}Now, recall that in Section~\ref{subsec: reduce rate cr} the extra common randomness $F$ of rate $\tilde R$ has been fixed to an instance $F=\mathbf f$. Hence, to minimize the measure of the set $\left(  \mathcal S_{\gamma_1} \cap \mathcal S_{\gamma_2} \cap \mathcal S_{\gamma_3}\right)^{\mathrm{c}} $, we only need to minimize the second and third terms of~\eqref{demorgan}. We define the sets 
{\allowdisplaybreaks
\begin{align}
 \mathcal S_{\imath(\mathbf w;\mathbf u\mathbf v)} &\coloneqq \begin{Bmatrix}[l]
(\mathbf u,\mathbf v,\mathbf w)  \,  : \\
 \,   n (R_0 + R) \leq \imath(\mathbf w;\mathbf u\mathbf v) +\gamma_2+\gamma_3
\end{Bmatrix}, \\[2mm]
  \mathcal S_{\imath(\mathbf w;\mathbf u)} & \coloneqq \begin{Bmatrix}[l]
(\mathbf u,\mathbf v,\mathbf w)  \,  : \\
 \,   n R \leq \imath(\mathbf w;\mathbf u)+ \gamma_2+\gamma_1
\end{Bmatrix},
\end{align}}and we treat them separately in the following.


\vspace{2mm}

\paragraph{Analysis of $ \mathcal S_{\imath(\mathbf w;\mathbf u \mathbf v)}$}\label{siwuv}

We observe that, since the distribution $\bar P$ is i.i.d., 
the terms
$Z_i= \imath_{\bar P}(w_i, u_i v_i)$, 
are mutually independent for $i=1, \ldots n$. 
 Then, we consider the following inequality
 {\allowdisplaybreaks
 \begin{align}
 n(R+R_0)>\underbrace{\sum_{i=1}^n { \mathbb E}_{\bar P_{W UV}} [\imath_{\bar P}(w_i; u_i v_i)]}_{n \mu_n }+ Q^{-1} (\epsilon_1)  \underbrace{\sqrt{  \sum_{i=1}^n   {\text{Var}}_{\bar P_{WUV}} (\imath_{\bar P}(w_i;u_i v_i))}}_{n \, \sqrt{ V_n/n}} +\, \gamma_2, \label{be rate}
 \end{align}}where $ \mu_n = \frac{1}{n} \sum_{i=1}^n \mathbb E [Z_i]$, $V_n  = \frac{1}{n} \sum_{i=1}^n \text{Var} [Z_i] $  and $Q(\cdot)$  is the tail distribution function of the standard normal distribution. We prove that, assuming that~\eqref{be rate} holds, we can successfully bound $ \mathcal S_{\imath(\mathbf w;\mathbf u \mathbf v)}$. In fact, 
 the chain of inequalities
  {\allowdisplaybreaks
 \begin{align*}
  \sum_{i=1}^n i_{\bar P}(W;UV) >n(R+R_0)-(\gamma_2 + \gamma_3)> n\left(\mu_n + t\sqrt{\frac{V_n}{n}}\right)
  \end{align*}}implies that, if~\eqref{be rate} holds, $ \mathcal S_{\imath(\mathbf w;\mathbf u \mathbf v)}$ is contained in 
   {\allowdisplaybreaks
 \begin{align*}
 \left\{  (\mathbf u, \mathbf v, \mathbf w) : \sum_{i=1}^n   \imath_{\bar P}(w_i;u_i, v_i)> n\mu_n + n \, Q^{-1}(\epsilon_1) \sqrt{\frac{V_n}{n}}\right\} .\stepcounter{equation}\tag{\theequation}\label{be2}
 \end{align*}}Therefore, if we find an upper bound on~\eqref{be2}, we have an upper bound on $ \mathcal S_{\imath(\mathbf w;\mathbf u \mathbf v)}$ as well. To obtain that, we apply Theorem \ref{berryessen} (Berry-Esseen CLT) to the right-hand side of \eqref{be2}, and we choose 
{\allowdisplaybreaks \begin{align}
&Q(t)=\epsilon_1,\nonumber\\
& \epsilon_1^{*} =\epsilon_1+ \frac{B_n}{\sqrt{n}}, \label{epsilon4},
 \end{align}}where, as in the statement of Theorem \ref{berryessen} (Berry-Esseen CLT),
 $B_n = 6 \frac{T_n}{V_n^{3/2}}$, and  $T_n =\frac{1}{n}\sum_{i=1}^n \mathbb{E} [{\lvert Z_i - \mu_i \rvert}^3].$
Then, we have
    {\allowdisplaybreaks
 \begin{align}
 &\left\lvert  \mathbb P \left\{ \sum_{i=1}^n    \imath_{\bar P}(w_i, u_i, v_i)>n\mu_n  +  n\, Q^{-1}(\epsilon_1) \sqrt{\frac{V_n}{n}} \right\}  -  \epsilon_1  \right\rvert \leq   \frac{B_n}{\sqrt{n}},\nonumber\\[1mm]
&\Rightarrow \mathbb P \left\{ \sum_{i=1}^n    \imath_{\bar P}(w_i, u_i, v_i)>n\mu_n 
+ n\, Q^{-1}(\epsilon_1) \sqrt{\frac{V_n}{n}}\right\} \leq \epsilon_1^{*}.\label{be4}
 \end{align}}Finally, \eqref{be4} combined with \eqref{be2} implies $\bar P {\left(\mathcal S_{\imath(w:uv)}  \right)} \leq \epsilon_1^{*}.$

Moreover, we can simplify~\eqref{be rate} with the following identifications.
\begin{rem}[Mutual Information and Channel Dispersion]
Similarly to~\cite[Section IV.A]{yassaee2013nonarxiv}, observe that, since $(u_i, w_i, v_i)$ are generated i.i.d. according to the same distribution $\bar P_{WUV}$, we have
{\allowdisplaybreaks
 \begin{subequations}
 \begin{align}
 \mu_n &\coloneqq\frac{1}{n} \sum_{i=1}^n { \mathbb E}_{\bar P_{WUV}} [\imath_{\bar P}(w_i; u_i ,v_i)]\nonumber\\[0.5mm]
 &= { \mathbb E}_{\bar P_{WUV}} [\imath_{\bar P}(w; u ,v)]\nonumber\\[0.5mm]
&=I_{\bar P}(W;UV),\label{rate entropy}\\[2mm]
 V_n&\coloneqq  \frac{1}{n} \sum_{i=1}^n {\text{Var}}_{\bar P_{WUV}} (\imath_{\bar P}(w_i;u_iv_i))\nonumber\\
&=  {\text{Var}}_{\bar P_{WUV}} (\imath_{\bar P}(W;UV))
\label{dispersion term}
\end{align} \end{subequations}}and $V_{\bar P_{W|UV}}  =\min_{\bar P_{W|UV}}  \left[ {\text{Var}}_{\bar P_{WUV}} (\imath_{\bar P}(W;UV))\right]=\min_{\bar P_{W|UV}}  \left[  {\text{Var}}_{\bar P_{WUV}} (\imath_{\bar P}(W;UV)|W) \right] $ is the dispersion of the channel $\bar P_{W|UV}$ as defined in \cite[Thm.~49]{polyanskiy2010channel}.
\end{rem}

Then, \eqref{be rate} can be rewritten as
 \begin{equation}\label{be rate3}
 n(R+R_0)>n I_{\bar P}(W;UV)  + n\, Q^{-1}(\epsilon_1) \sqrt{\frac{V_{\bar P_{W|UV}}}{n}} +( \gamma_2+\gamma_3).
 \end{equation}
 
 
\vspace{2mm}

\paragraph{Analysis of $ \mathcal S_{\imath(\mathbf w;\mathbf u)}$}\label{siwu} Similarly, we use Theorem \ref{berryessen} (Berry-Esseen CLT) to estimate $\bar P( \mathcal S_{\imath(\mathbf w;\mathbf u)})$. If we apply the same reasoning to  $Z'_i= \imath_{\bar P}(w_i, U_i)$ for $i=1, \ldots n$ , and  
$\mu'_n = \frac{1}{n} \sum_{i=1}^n \mathbb E [Z'_i]=I(W;U)$,
$V'_n  = \frac{1}{n} \sum_{i=1}^n \text{Var} [Z'_i]=V_{\bar P_{W|U}}$,
$ T'_n =\frac{1}{n}\sum_{i=1}^n \mathbb{E} [{\lvert Z'_i - \mu'_i \rvert}^3]$,
$B'_n = 6 \frac{T_n}{V_n^{3/2}}$, we find that  $\bar P {\left(\mathcal S_{\imath(w:u)}  \right)} \leq \epsilon_2^{*}=\epsilon_2+\frac{B'_n}{n}$ if 
 \begin{equation}\label{be rate4}
 nR>n I_{\bar P}(W;U)  + n\, Q^{-1}(\epsilon_2) \sqrt{\frac{V_{\bar P_{W|U}}}{n}} +( \gamma_2+\gamma_1).
 \end{equation}
 A more detailed proof can be found in Appendix~\ref{appendix_siwu}.
 

\vspace{2mm}

\subsubsection{Choice of $(\gamma_1, \gamma_2, \gamma_3)$ and rate conditions} \label{section rate}

If we choose $(\gamma_1, \gamma_2, \gamma_3)=( \log{n},  \frac{1}{2} \log{n},  \log{n})$, then the bound \eqref{epsilon tot2}  on the $L^1$ distance becomes
\vspace{0.5mm}
{\allowdisplaybreaks
\begin{align}
 &{\lVert  P^{\text{RB}}_{U^n V^n}- P^{\text{RC}}_{U^n V^n} \rVert}_{1} \leq \epsilon_{\text{Tot}} ,\nonumber\\[1mm]
 & \epsilon_{\text{Tot}} = 10 \bar P{\left ( (\mathcal S_{\gamma_1} \cap \mathcal S_{\gamma_2} \cap \mathcal S_{\gamma_3})^{\mathrm{c}} \right)}
  + 2 \, \Big(2^{-\frac{\gamma_1+1}{2}} +5 \cdot 2^{-\gamma_2}+ 2^{-\frac{\gamma_3+1}{2}} \Big)\nonumber\\[1mm]
& \phantom{ \epsilon_{\text{Tot}} =} = 10 \bar P{\left ( (\mathcal S_{\gamma_1} \cap \mathcal S_{\gamma_2} \cap \mathcal S_{\gamma_3})^{\mathrm{c}} \right)}+  \frac{10  +  2\sqrt{2}}{\sqrt{n}}\nonumber\\[1mm]
& \phantom{ \epsilon_{\text{Tot}} =} \leq  10\,  \bar P{\left( \mathcal S_{\imath(\mathbf w;\mathbf u \mathbf v)} \right)}+ 10\,  \bar P{\left( \mathcal S_{\imath(\mathbf w;\mathbf u)} \right)} +  \frac{10  +  2\sqrt{2}}{\sqrt{n}}\nonumber\\[1mm]
& \phantom{ \epsilon_{\text{Tot}} =} \leq  10\,( \epsilon_1^{*} + \epsilon_2^{*}) +\frac{10+2\sqrt{2}}{\sqrt{n}}\nonumber\\[1mm]
& \phantom{ \epsilon_{\text{Tot}} =} =  10\,( \epsilon_1 + \epsilon_2 ) +\frac{10(1+B_n+B'_n)+2\sqrt{2} }{\sqrt{n}}\nonumber\\[1mm]
& \phantom{ \epsilon_{\text{Tot}} =} =  10\,( \epsilon_1 + \epsilon_2 ) + O\left(\frac{1 }{\sqrt{n}}\right).\label{speed conv}
\end{align}}

With this choice for $\gamma_i$, the rate conditions become
\vspace{-1mm}
{\allowdisplaybreaks
\begin{align}
 & R+R_0 >  I_{\bar P}(W;UV) + Q^{-1}(\epsilon_1) \sqrt{ \frac{V_{\bar P_{W|UV}}}{n}}+ \frac{3\log{n}}{2 \, n},\nonumber\\[1mm] 
 & R > I_{\bar P}(W;U) + Q^{-1}(\epsilon_2) \sqrt{ \frac{V_{\bar P_{W|U}}}{n}}+  \frac{3\log{n}}{2 \, n}.\label{final rate}
\end{align}}From now on, we drop the subscript  $ \bar P$ from  $I(\cdot,\cdot)$ to simplify the notation.


\section{Outer bound}\label{sec: nonasy ob}

For a fixed $\epsilon>0$ and $n>0$, we consider a triplet $(\bar{P}_{UV}, R, R_0)$ which is $(\epsilon,n)$-achievable for strong coordination, and we want to prove that it is contained in~\eqref{region fn outer}.
Since $(\bar{P}_{UV}, R, R_0)$ is achievable, there exist a code $(f_n,g_n)$ that induces a distribution $P_{U^{n}V^{n}}$ such that
\vspace{-1mm}
{\allowdisplaybreaks
\begin{align}
&{\lVert P_{U^nV^n} - \bar P_{U}^{\otimes n} \bar P_{V|U}^{\otimes n}  \rVert }_1 \leq \epsilon.\label{esc2}
\vspace{-1mm}
 \end{align}}
Now, we consider $n$ i.i.d. pairs $(C,M)$ of the message and the common randomness, generated by the stochastic  code via   $P_{(C,M)|U^nV^n}$, and  a time index $T$ uniformly  distributed from 1 to $n$. Observe that  since $U_T$ and $V_T$ are conditionally independent through $(C,M,T)$, we have:
\vspace{-1.5mm}
\begin{align}
& P_{(C,M, T)^n|U_T^nV_T^n} \bar P_{U_T}^{\otimes n} \bar P_{V_T|U_T}^{\otimes n}\!\!\!=\!\bar P_{V_T^n| (C,M, T)^n} \bar P_{(C,M,T)^n|U_T^n} \bar P_{U_T}^{\otimes n} \nonumber  \\ &=  \bar P_{V_T| (C,M, T)}^{\otimes n} \bar P_{(C,M,T)|U_T}^{\otimes n} \bar P_{U_T}^{\otimes n}. \label{mcwt}
\end{align}
Then, using~\eqref{mcwt} and  the properties of the $L_1$ distance~\cite[Lemma 17]{cuff2009thesis},~\eqref{esc2} becomes
\begin{align}
\vspace{-1.5mm}
 \epsilon&\geq{\lVert P_{U_T^nV_T^n} \!-\! \bar P_{U_T}^{\otimes n} \!\bar P_{V_T|U_T}^{\otimes n}  \rVert }_1\nonumber\\
&= {\lVert P_{U_T^n(M,C,T)^nV_T^n}   \!-\! P_{(C,M, T)^n|U_T^nV_T^n} \bar P_{U_T}^{\otimes n} \bar P_{V_T|U_T}^{\otimes n} \rVert }_1 \nonumber\\
&={\lVert P_{U_T^n (M,C,T)^n V_T^n} - \bar P_{V_T| (C,M, T)}^{\otimes n} \bar P_{(C,M,T)|U_T}^{\otimes n} \bar P_{U_T}^{\otimes n} \rVert }_1 \label{esc3}
 \end{align}with $P_{U_T^n (M,C,T)^n V_T^n} \coloneqq P_{U^nV^n} P_{(M,C,T)^n|U_T^nV_T^n}$.
 
\vspace{1mm}
We denote  $(C,M,T)$ as $(W_T, T)$, $W_t$ as $(C,M)$ for each $t \in \llbracket 1,n\rrbracket$ and $W$ as $(W_T,T)=(C,M, T)$, and we will see  from the following that this identification satisfies the conditions in~\eqref{region fn outer}.
Before proceeding, we observe that with this notation we find the following equivalent to~\eqref{esc3} 
\vspace{-1mm}
 \begin{align}
&{\lVert P_{U^nV^n} \!-\! \bar P_{U}^{\otimes n} \!\bar P_{V|U}^{\otimes n}  \rVert }_1
 \!\!=\! {\lVert P_{U^nW^nV^n}   \!-\! \bar P_{U}^{\otimes n}\! \bar P_{V|U}^{\otimes n}  \bar P^{\otimes n}_{(W_T,T)|UV} \rVert }_1 \nonumber\\
&\!=\!{\lVert P_{U^nW^nV^n}\! - \!\bar P_{U}^{\otimes n} \bar P_{W|U}^{\otimes n}  \bar P_{V|W}^{\otimes n} \rVert }_1 \! \leq \! \epsilon\label{esc}
\vspace{-1mm}
 \end{align}and  we  later refer to~\eqref{esc} as the $(\epsilon,n)$-\emph{strong coordination assumption}. 
Note that  we have yet to prove that this is the appropriate choice for $W$, that is that $W$ verifies all the constraints of~\eqref{region fn outer}. First, observe that by definition of encoder and decoder the following Markov chains hold: 
\vspace{-1mm}
{\allowdisplaybreaks\begin{align*}
&U_t - (C,M) - V_t  \Leftrightarrow U_t - W_t - V_t,\\
&U_T - (C,M, T) - V_T \Leftrightarrow  U - W - V.
\end{align*}
In the next section, we present a proof of the rate constraints based on a meta-converse, following the approach of~\cite{polyanskiy2010channel}. More precisely, we show the bound on the rate $R$, while the proof for the bound on $R+R_0$ and the cardinality bound are developed in  Appendix~\ref{appendix_ssec_R_R0} and Appendix~\ref{appendix bounds} respectively.


\subsection{First bound --  $R$}\label{ssec_R}

Similarly to~\cite{polyanskiy2010channel}, for an observation $ \mathbf w= (m,c)$ composed of a message and an instance of the common randomness, we define the hypothesis:
{\allowdisplaybreaks
\begin{align*}
\mathcal H_0)&\quad \mathbf w \mbox{ generated according to } \sum_{\mathbf u}  P_{U^nW^{n}}(\mathbf u, \mathbf w),\\
\mathcal H_1)&\quad  \mathbf w \mbox{ generated according to }  \sum_{\mathbf u} \bar P_{U}^{\otimes n} \bar{P}_{W}^{\otimes n} (\mathbf u, \mathbf w),
\end{align*}}where $P_{U^nW^n}$ is the coding distribution $\epsilon$-close to i.i.d. by assumption, and $\bar P^{\otimes n}_{UW}$ is the i.i.d. target distribution. We now carry out a thought experiment: we consider a  test with a (possibly) stochastic decision rule between the distributions $P_{U^{n}W^{n}}$ and $ \bar P^{\otimes n}_U \bar P^{\otimes n}_W$: a test is defined by a random transformation 
\begin{equation*}P_{Z|U^nW^n}: \mathcal U^n \times \mathcal W^n \to {\{\mathcal H_0,\mathcal H_1\}},\end{equation*} where $\mathcal H_0$ indicates that the test chooses $ P_{U^{n}W^{n}}$, and  $\mathcal H_1$ indicates that the test chooses $ \bar P^{\otimes n}_U \bar P^{\otimes n}_W$. The purpose of this randomized test is to bound the error probability of such test declaring that a pair message-common randomness, here identified with the
sequence $\mathbf w$, is generated according to the product distribution $\bar P^{\otimes n}_{UW}$, while $\mathbf w$ is by assumption generated according to $P_{U^nW^n}$. Later on, the bound on this error probability will be reduced to the rate conditions of the outer bound~\eqref{region fn outer} by applying the Theorem \ref{berryessen} (Berry-Esseen CLT) and the $(\epsilon,n)$-strong coordination assumption~\eqref{esc}.
Hence, we start by defining the probability of type-I error (probability of choosing $\mathcal H_1$ when the true hypothesis is $\mathcal H_0$)  and type-II error (probability of choosing $\mathcal H_0$ when the true hypothesis is $\mathcal H_1$) as 
{\allowdisplaybreaks
\begin{subequations}
\begin{align}
  &
  P_{e}^{\text{I}}(P_{Z|U^nW^n})\,\coloneqq \mathbb P\{\hat{\mathcal H}_1|\mathcal H_0 \}=\sum_{\mathbf u} \bar P^{\otimes n}_U  (\mathbf u) \bar P^{\otimes n}_W (\mathbf w) P_{Z|U^nW^n}(\mathcal H_0|\mathbf u,\mathbf w),
  \label{ty2}\\
   &
P_{e}^{\text{II}}(P_{Z|U^nW^n})\coloneqq \mathbb P\{\hat{\mathcal H}_0|\mathcal H_1 \}= \sum_{\mathbf u} P_{U^{n}W^{n}} (\mathbf u, \mathbf w) P_{Z|U^nW^n}(\mathcal H_1|\mathbf u,\mathbf w)  \label{ty1}.
 \end{align} \end{subequations}}Similar to~\cite{polyanskiy2010channel}, we denote
the minimum type-I error for a maximum type-II error $1-\alpha$:
{\allowdisplaybreaks\begin{align}
\beta_{\alpha}&\coloneqq \min_{\substack{P_{Z|U^nW^n}: \\[1mm] 
P_{e}^{\text{II}}(P_{Z|U^nW^n}) \leq 1-\alpha }} P_{e}^{\text{I}}(P_{Z|U^nW^n})\nonumber\\[2mm]
&=
\min_{\substack{P_{Z|U^nW^n}: \\[1mm] 
\sum_{\mathbf u} P_{U^{n}W^{n}} (\mathbf u,\mathbf w) P_{Z|U^nW^n}(\mathcal H_1|\mathbf u, \mathbf w) \leq 1-\alpha }} 
\sum_{\mathbf u} \bar P^{\otimes n}_U  (\mathbf u) \bar P^{\otimes n}_W (\mathbf w) P_{Z|U^nW^n}(\mathcal H_0|\mathbf u,\mathbf w), \label{defbeta}
\end{align}}where the error probability $\alpha$ will be defined later.
For the error probabilities $\alpha$ and  $\beta_{\alpha}$,~\cite[Section 12.4]{polyanskiy2014lecture} proves the following relations:
{\allowdisplaybreaks
\begin{subequations}\begin{align}
& \bullet \quad \text{upper bound on $\min P_{e}^{\text{I}}(P_{Z|U^nW^n})$}\quad \beta_{\alpha} \leq \frac{1}{\gamma_0}, \mbox{ with $\gamma_0$ s.t. } \mathbb P_{P_{U^{n}W^{n}}} \left\{ \log \frac{P_{U^{n}W^{n}}}{\bar P^{\otimes n}_U \bar P^{\otimes n}_W} > \log \gamma_0 \right\}\geq \alpha,\label{rel2}\\[1mm]
&\bullet \quad \text{lower bound on $\min P_{e}^{\text{I}}(P_{Z|U^nW^n})$}\quad\alpha \leq \mathbb P_{P_{U^{n}W^{n}}} \left\{ \log \frac{P_{U^{n}W^{n}}}{\bar P^{\otimes n}_U \bar P^{\otimes n}_W} > \log \gamma\right\} +\gamma \, \beta_{\alpha}  \quad \forall \gamma >0.\label{rel1}
\end{align}\end{subequations}}
Then, our goal is to use the  inequalities~\eqref{rel2} and~\eqref{rel1} to prove the rate constraint 
\begin{equation}\label{rate converse nr}
nR \geq n I(W;U) + Q^{-1}(\epsilon) \sqrt{ n V_{\hat P_{W|U}}} -x
\end{equation}where $\epsilon$  is the approximation term,  and  $x\in\mathbb R$ is a parameter which will be defined later.  First, we split~\eqref{rate converse nr} into  an upper and a lower bound on the logarithm of the error probability $\beta_{\alpha}$:
{\allowdisplaybreaks\begin{subequations}
\begin{align}
\text{upper bound on $\log{\beta_{\alpha}}$}\qquad &\phantom{nR+x \geq } \log{\frac{1}{\beta_{\alpha}}}\geq n I(W;U) + Q^{-1}(\epsilon) \sqrt{ n V_{\hat P_{W|U}}},
\label{mcpartb}\\[1mm]
\text{lower bound on $\log{\beta_{\alpha}}$}\qquad &nR+x \geq  \log{\frac{1}{\beta_{\alpha}}}.\label{mcparta}
\end{align}\end{subequations}}Then, the proof of~\eqref{rate converse nr} is divided in the following steps, detailed in the next sections:
\begin{enumerate}[(i)]
\item \textit{Proof of the upper bound on $\log{\beta_{\alpha}}$:} we use the upper bound on $\min P_{e}^{\text{I}}(P_{Z|U^nW^n})$ of~\eqref{rel2} combined with the $(\epsilon,n)$-strong coordination assumption~\eqref{esc} and Theorem \ref{berryessen} (Berry-Esseen CLT) to derive the following upper bound on the logarithm $\beta_{\alpha}$ of \eqref{mcpartb} by choosing the parameter $\gamma_0$;
\item \textit{Proof of the lower bound on $\log{\beta_{\alpha}}$:} we use the lower bound on $\min P_{e}^{\text{I}}(P_{Z|U^nW^n})$ of~\eqref{rel1} combined with the $(\epsilon,n)$-strong coordination assumption~\eqref{esc} and classical information theory properties  to derive the lower bound on $\beta_{\alpha}$ of \eqref{mcparta} by choosing the parameter $\gamma$;
\item \textit{Proof of the rate constraint:} we combine \eqref{mcpartb} and \eqref{mcparta} proved in the previous steps and we derive~\eqref{rate converse nr}. 
\end{enumerate}

Before proceeding, observe that by the  $(\epsilon,n)$-strong coordination assumption~\eqref{esc} and the properties of $L^1$ distance~\cite[Lemma 16]{cuff2009thesis}, we have   
{\allowdisplaybreaks\begin{align*}
&{\lVert P_{U^nW^n}\! -\! \bar P_{U}^{\otimes n} \bar P_{W|U}^{\otimes n} \rVert }_1 \! \leq \!{\lVert P_{U^nW^nV^n} \!-\! \bar P_{U}^{\otimes n} \bar P_{W|U}^{\otimes n} \bar P_{V|W}^{\otimes n} \rVert }_1\!\!= \!\epsilon, \\
& \Rightarrow  \forall (\mathbf u, \mathbf w) \quad  \lvert P_{U^nW^n}(\mathbf u, \mathbf w) - \bar P_{U}^{\otimes n} \bar P_{W|U}^{\otimes n}(\mathbf u, \mathbf w)  \rvert \leq \epsilon\\
& \Rightarrow  \forall (\mathbf u, \mathbf w)  ~ \exists \epsilon_{uw} \leq \epsilon ~\text{such that }  \lvert P_{U^nW^n}(\mathbf u, \mathbf w) - \bar P_{U}^{\otimes n} \bar P_{W|U}^{\otimes n}(\mathbf u, \mathbf w)  \rvert = \epsilon_{uw}
\end{align*}}which we can distinguish into two cases:
\begin{itemize}
\item \textit{Case 1} \quad $P_{U^nW^n}(\mathbf u, \mathbf w) =\bar P_{U}^{\otimes n} \bar P_{W|U}^{\otimes n}(\mathbf u, \mathbf w) +\epsilon_{uw}$;
\item \textit{Case 2} \quad $P_{U^nW^n}(\mathbf u, \mathbf w) =\bar P_{U}^{\otimes n} \bar P_{W|U}^{\otimes n}(\mathbf u, \mathbf w) -\epsilon_{uw}$.
\end{itemize}Thus, we prove steps (i)--(iii) separately for both cases.  With a slight abuse of notation from now on we will drop the index from $\epsilon_{uw}$ and use $\epsilon$ instead, since $\epsilon_{uw}$ just has to be smaller that $\epsilon$. Similarly, we omit the pairs $ (\mathbf u, \mathbf w)$ in order to simplify the notation.


\vspace{1mm}

\subsubsection{Proof of the upper bound on $\log{\beta_{\alpha}}$~\eqref{mcpartb} -- Case 1 ($P_{U^nW^n}=\bar P_{U}^{\otimes n} \bar P_{W|U}^{\otimes n}+\epsilon$)}\label{sssec_mcpartb}

By the $(\epsilon,n)$-strong coordination assumption~\eqref{esc}, we have
{\allowdisplaybreaks
\begin{align}
\log{ \left(P_{U^nW^n} (\mathbf u,\mathbf w ) \right)}
&= \log { \left( \bar P_{U}^{\otimes n} \bar P_{W|U}^{\otimes n} (\mathbf u,\mathbf w ) + \epsilon \right)} \nonumber\\
&= \log { \left( \bar P_{U}^{\otimes n} \bar P_{W|U}^{\otimes n}  (\mathbf u,\mathbf w ) \right)} + \log { \left(1 + \frac{ \epsilon} {\bar P_{U}^{\otimes n} \bar P_{W|U}^{\otimes n}  (\mathbf u,\mathbf w )} \right)}.
 \label{ineq_tv2}
\end{align}}Thus, we rewrite the left-hand side of the upper bound on $\min P_{e}^{\text{I}}(P_{Z|U^nW^n})$ in~\eqref{rel2} as
{\allowdisplaybreaks
\begin{align}
&\mathbb P \left\{ \log{ P_{U^nW^n} }\geq \log{\bar P^{\otimes n}_U \bar P^{\otimes n}_W} + \log {\gamma_0}\right\} \nonumber\\[1.5mm]
&=  \mathbb P \left\{ \log { \bar P_{U}^{\otimes n} \bar P_{W|U}^{\otimes n}} + \log { \left(1 + \frac{ \epsilon} {\bar P_{U}^{\otimes n} \bar P_{W|U}^{\otimes n}  } \right)} \geq \log{\bar P^{\otimes n}_U \bar P^{\otimes n}_W} + \log {\gamma_0}\right\}  \nonumber\\[1.5mm]
& =\mathbb P \left\{ \log { \frac {\bar P_{U}^{\otimes n} \bar P_{W|U}^{\otimes n}}{\bar P^{\otimes n}_U \bar P^{\otimes n}_W} }  \geq  \log {\gamma_0}  -\log { \left(1 + \frac{ \epsilon} {\bar P_{U}^{\otimes n} \bar P_{W|U}^{\otimes n}  } \right)}\right\}.
\label{probgamma}
\end{align}}Given the parameter  $\epsilon$, we choose the parameter $\gamma_0$ as:
{\allowdisplaybreaks
\begin{align}
\log{\gamma_0} 
&\coloneqq n \mu_n + Q^{-1}\left(\epsilon\right) \sqrt{ n V_{n} } + \log { \left(1 + \frac{ \epsilon} {\bar P_{U}^{\otimes n} \bar P_{W|U}^{\otimes n}  } \right)}\label{log}
\end{align}}where, as in the statement of Theorem \ref{berryessen} (Berry-Esseen CLT), $ \sum_{i=1}^n \log (\bar P_{U} \bar P_{W|U}/ \bar P_{U} \bar P_{W})=:  \sum_{i=1}^n Z_i$ is the sum of $n$ i.i.d. random variables, and 
 $\mu_n = \frac{1}{n} \sum_{i=1}^n \mathbb E [Z_i],$
 $V_n  = \frac{1}{n} \sum_{i=1}^n \text{Var} [Z_i]$,
 $T_n =\frac{1}{n}\sum_{i=1}^n \mathbb{E} [{\lvert Z_i - \mu_i \rvert}^3]$,
 $B_n = 6 \frac{T_n}{V_n^{3/2}}$,
and $Q(\cdot)$  is the tail distribution function of the standard normal distribution. Moreover, we can rewrite these terms by using the identification of the following remark:
\vspace{2mm}

\begin{rem}[Mutual Information and Channel Dispersion]\label{rem mi cd}
Similarly to~\cite{polyanskiy2010channel}, we observe that  for the discrete i.i.d. distributions $\bar P^{\otimes n}_U \bar P^{\otimes n}_{W|U}$ and $\bar P^{\otimes n}_U \bar P^{\otimes n}_{W}$, we have:
{\allowdisplaybreaks
\begin{align}
&\mu_n =  \mathbb D(\bar P_U \bar P_{W|U} \Arrowvert \bar P_U \bar P_{W})=I(W;U),\nonumber\\
&V_n= \sum_{u,w} \bar P_U(u) \bar P_{W|U}(w|u) {\left[ \log{\frac{\bar P_U(u) \bar P_{W|U}(w|u) }{\bar P_U(u) \bar P_{W}(w)}}\right]}^2- {\mathbb D(\bar P_U \bar P_{W|U} \Arrowvert \bar P_U \bar P_{W})}^2= V_{\bar P_{W|U}},\nonumber\\
&T_n= \sum_{u,w} \bar P_U(u) \bar P_{W|U}(w|u) {\left\lvert \log{\frac{\bar P_U(u) \bar P_{W|U}(w|u) }{\bar P_U(u) \bar P_{W}(w)}}- \mathbb D(\bar P_U\bar P_{W|U}\Arrowvert \bar P_U \bar P_{W})\right\rvert}^3,\nonumber\\
& B_n = 6 \frac{T_n}{V_n^{3/2}}.\label{eqbn}\end{align}}\end{rem}
Now, observe that we have chosen parameter $\gamma_0$ in~\eqref{log} appropriately such, that with the identifications of Remark~\ref{rem mi cd}, the probability of error~\eqref{probgamma} becomes
{\allowdisplaybreaks
\begin{align}
& \mathbb P  \left\{ \sum_{i=1}^n \log \frac{ \bar P_{U} \bar P_{W|U}}{ \bar P_{U} \bar P_{W}} \geq   n \mu_n + Q^{-1}\left(\epsilon \right) \sqrt{ n V_{n} } \right\}\nonumber\\
&= \mathbb P  \left\{ \sum_{i=1}^n \log \frac{ \bar P_{U} \bar P_{W|U}}{ \bar P_{U} \bar P_{W}} \geq   n I(U;W) + Q^{-1}\left(\epsilon \right) \sqrt{ n V_{\bar P_{W|U}} }
\right\}.\label{prob be}
\end{align}}Since $ \sum_{i=1}^n \log (\bar P_{U} \bar P_{W|U}/ \bar P_{U} \bar P_{W})=  \sum_{i=1}^n Z_i$ is the sum of $n$ i.i.d. random variables, the next step is to bound the probability in~\eqref{prob be} using  Theorem \ref{berryessen} (Berry-Esseen CLT): 

{\allowdisplaybreaks
\begin{align}
& \mathbb P \left\{ \log \prod_{i=1}^n \frac{ \bar P_{U} \bar P_{W|U}}{ \bar P_{U} \bar P_{W}} \geq n I(U;W) + Q^{-1}\left(\epsilon \right) \sqrt{ n V_{\bar P_{W|U}} }\right\} - \epsilon   \geq - \frac{B_n}{\sqrt n}\nonumber\\[2mm]
&\Longleftrightarrow \,\,\mathbb P \left\{ \log \prod_{i=1}^n \frac{ \bar P_{U} \bar P_{W|U}}{ \bar P_{U} \bar P_{W}} \geq n I(U;W) + Q^{-1}\left(\epsilon \right) \sqrt{ n V_{\bar P_{W|U}} }\right\}  \geq \left(\epsilon - \frac{B_n}{\sqrt n}\right).\label{deltanew}
 \end{align}}Then, we identify 
 \begin{equation}
 \alpha \coloneqq \epsilon - \frac{B_n}{\sqrt n},
 \end{equation}and by combining the upper bound on $\min P_{e}^{\text{I}}(P_{Z|U^nW^n})$ of~\eqref{rel2} with the parameter $\gamma_0$ as chosen in~\eqref{log}, we obtain:
{\allowdisplaybreaks
\begin{align}
\log{\frac{1}{\beta_{\alpha}}} & \geq  \log{ \gamma_0} 
=  n \mu_n + Q^{-1}\left(\epsilon \right) \sqrt{ n V_{n} } + \log { \left(1 + \frac{ \epsilon} {\bar P_{U}^{\otimes n} \bar P_{W|U}^{\otimes n}  } \right)}. \label{beta2part1}\end{align}} 


\vspace{1mm}
 
\subsubsection{Proof of the lower bound on $\log{\beta_{\alpha}}$~\eqref{mcparta} -- Case 1 $(P_{U^nW^n}=\bar P_{U}^{\otimes n} \bar P_{W|U}^{\otimes n}+\epsilon$)}\label{sssec_mcparta}

First, we observe that 
{\allowdisplaybreaks\begin{align}
&nR=H(M)\overset{\mathclap{(a)}}{\geq}  n I(U;W)  \overset{\mathclap{(b)}}{\geq}  n I(U;W) + Q^{-1}(y) \sqrt{n V_n}, \qquad \frac{1}{2}<y<1, \label{eqy1}
\end{align}}where $(a)$  is proved in Appendix~\ref{appendix_ssec_HM} and 
$(b)$ comes from the fact that $Q^{-1}(y) \sqrt{n V_n}\leq 0$ for every  $\frac{1}{2}<y<1 $.

\vspace{1mm}
Now, we recall that by  the lower bound on $\min P_{e}^{\text{I}}(P_{Z|U^nW^n})$ of~\eqref{rel1}, for every $\gamma >0$ we have
{\allowdisplaybreaks
\begin{align}
&\beta_{\alpha}  \geq \frac{1}{\gamma} \left[
\alpha - \mathbb P_{P_{U^{n}W^{n}}} \left\{ \log \frac{P_{U^{n}W^{n}}}{\bar P^{\otimes n}_U \bar P^{\otimes n}_W} > \log \gamma\right\} \right]  \nonumber\\[2mm]
&\phantom{\beta_{\alpha}} \overset{\mathclap{(c)}}{=} \frac{1}{\gamma} \left[
\alpha - \mathbb P_{P_{U^{n}W^{n}}} \left\{ \log \frac{\bar P_{U}^{\otimes n} \bar P_{W|U}^{\otimes n}+\epsilon}{\bar P^{\otimes n}_U \bar P^{\otimes n}_W} > \log \gamma\right\} \right] \nonumber\\[2mm]
&\phantom{\beta_{\alpha}} =\frac{1}{\gamma} \left[
\alpha - \mathbb P_{P_{U^{n}W^{n}}} \left\{ \log \frac{\bar P_{U}^{\otimes n} \bar P_{W|U}^{\otimes n}}{\bar P^{\otimes n}_U \bar P^{\otimes n}_W} > \log \gamma - \log { \left(1 + \frac{ \epsilon} {\bar P_{U}^{\otimes n} \bar P_{W|U}^{\otimes n}  } \right)} \right\} \right],\nonumber\\[2mm]
&\Longleftrightarrow \,\log{\beta_{\alpha}}  \geq \log{\frac{1}{\gamma} \left[
\alpha - \mathbb P_{P_{U^{n}W^{n}}} \left\{ \log \frac{\bar P_{U}^{\otimes n} \bar P_{W|U}^{\otimes n}}{\bar P^{\otimes n}_U \bar P^{\otimes n}_W} > \log \gamma - \log { \left(1 + \frac{ \epsilon} {\bar P_{U}^{\otimes n} \bar P_{W|U}^{\otimes n}  } \right)} \right\} \right]},
 \label{eq beta+sc}
\end{align}}where in  $(c)$ we have used the $(\epsilon,n)$-strong coordination assumption~\eqref{esc}. Then, similarly to Section~\ref{sssec_mcpartb}, we choose appropriately the parameter $\gamma$: 
\begin{equation}
\log \gamma = H(M)+\log { \left(1 + \frac{ \epsilon} {\bar P_{U}^{\otimes n} \bar P_{W|U}^{\otimes n}  } \right)}. \end{equation}With this choice of $\gamma$, we plug~\eqref{eqy1} into~\eqref{eq beta+sc}, and we have
{\allowdisplaybreaks
\begin{align}
\log{\beta_{\alpha} }&
\geq \log{\frac{1}{\gamma}} +\log{ \left[
\alpha - \mathbb P_{P_{U^{n}W^{n}}} \left\{ \log \frac{\bar P_{U}^{\otimes n} \bar P_{W|U}^{\otimes n}}{\bar P^{\otimes n}_U \bar P^{\otimes n}_W} > \log \gamma - \log { \left(1 + \frac{ \epsilon} {\bar P_{U}^{\otimes n} \bar P_{W|U}^{\otimes n}  } \right)} \right\} \right]}\nonumber\\[2mm]
&=- \left[H(M)+\log { \left(1 + \frac{ \epsilon} {\bar P_{U}^{\otimes n} \bar P_{W|U}^{\otimes n}  } \right)}\right] +\log{ \left[
\alpha - \mathbb P_{P_{U^{n}W^{n}}} \left\{ \log \frac{\bar P_{U}^{\otimes n} \bar P_{W|U}^{\otimes n}}{\bar P^{\otimes n}_U \bar P^{\otimes n}_W} >  H(M) \right\} \right]}\nonumber\\[2mm]
&\geq- \left[H(M)+\log { \left(1 + \frac{ \epsilon} {\bar P_{U}^{\otimes n} \bar P_{W|U}^{\otimes n}  } \right)}\right] \nonumber\\[2mm]
&\qquad +\log{ \left[
\alpha - \mathbb P_{P_{U^{n}W^{n}}} \left\{ \log \frac{\bar P_{U}^{\otimes n} \bar P_{W|U}^{\otimes n}}{\bar P^{\otimes n}_U \bar P^{\otimes n}_W} >  n I(U;W) + Q^{-1}(y) \sqrt{n V_n} \right\} \right]}\nonumber\\[2mm]
&= - H(M) -\log { \left(1 + \frac{ \epsilon} {\bar P_{U}^{\otimes n} \bar P_{W|U}^{\otimes n}  } \right)} + \log{\left(\alpha -y - \frac{B_n}{\sqrt n}\right)}\label{partbcase1}
\end{align}}which is equivalent to the lower bound on $\log{\beta_{\alpha}}$ of~\eqref{mcparta} if we identify 
\begin{equation}
x=\log { \left(1 + \frac{ \epsilon} {\bar P_{U}^{\otimes n} \bar P_{W|U}^{\otimes n}  } \right)} - \log{\left(\alpha -y - \frac{B_n}{\sqrt n}\right)},
\end{equation}since
{\allowdisplaybreaks
\begin{align}
nR+x&=n R + \log { \left(1 + \frac{ \epsilon} {\bar P_{U}^{\otimes n} \bar P_{W|U}^{\otimes n}  } \right)} - \log{\left(\alpha -y - \frac{B_n}{\sqrt n}\right)} \nonumber\\
& = H(M)+ \log { \left(1 + \frac{ \epsilon} {\bar P_{U}^{\otimes n} \bar P_{W|U}^{\otimes n}  } \right)} - \log{\left(\alpha -y - \frac{B_n}{\sqrt n}\right)} \geq 
\log{\frac{1}{\beta_{\alpha} }}.\label{partbcase1fin}
\end{align}}


\subsubsection{Proof of the rate constraint -- Case 1 $(P_{U^nW^n}=\bar P_{U}^{\otimes n} \bar P_{W|U}^{\otimes n}+\epsilon$)}\label{sssec_fin_mcparta}
Now, we can conclude the proof of this part of the outer bound by combining~\eqref{beta2part1} and~\eqref{partbcase1fin}. In fact, for $1/2<y<1$ we have
{\allowdisplaybreaks\begin{align*}
\underbrace{H(M)}_{nR} +\log { \left(1 + \frac{ \epsilon} {\bar P_{U}^{\otimes n} \bar P_{W|U}^{\otimes n}  } \right)} - \log{\left(\alpha -y - \frac{B_n}{\sqrt n}\right)} \geq  n \mu_n + Q^{-1}\left(\epsilon \right) \sqrt{ n V_{n} } + \log { \left(1 + \frac{ \epsilon} {\bar P_{U}^{\otimes n} \bar P_{W|U}^{\otimes n}  } \right)}
\end{align*}}which is equivalent to
{\allowdisplaybreaks\begin{align}
R  \geq   \mu_n + Q^{-1}\left(\epsilon \right) \sqrt{ \frac{ V_{n}}{n} }+ \frac{\log{\left(\alpha -y - \frac{B_n}{\sqrt n}\right)}}{n}.\label{r+log1}
\end{align}}
  

\vspace{1mm}

\subsubsection{Proof of the rate constraint -- Case 2 $P_{U^nW^n}=\bar P_{U}^{\otimes n} \bar P_{W|U}^{\otimes n}-\epsilon$}\label{sssec_mcpartba case 2}  
The proofs of the upper bound  and of the lower bound on $\log{\beta_{\alpha}}$ of~\eqref{mcpartb} and~\eqref{mcparta}  are similar to the one of Section~\ref{sssec_mcpartb} and Section~\ref{sssec_mcparta}, and are therefore deferred to Appendix~\ref{appendix_ssec_R_Case2}.

\vspace{1mm}
\begin{rem}[Speed of convergence]
Note that for both case 1 and case 2 we retrieve the same rate condition as in~\eqref{r+log1}, and the term 
{\allowdisplaybreaks
\begin{align*}
& \frac{\log{\left(\alpha -y - \frac{B_n}{\sqrt n}\right)} }{n}\leq  \frac{\alpha -y - \frac{B_n}{\sqrt n} }{n}=\frac{\alpha -y}{n}-\frac{ B_n }{n \, \sqrt n} =O\left( \frac{1}{n}\right).
\end{align*}}which goes to zero faster than the term $\log n/n$ in the achievability.
\end{rem}

\subsection{Second bound --  $R+R_0$}\label{ssec_R_R0}
The proof is similar to the one of Section~\ref{ssec_R}, and it is deferred to Appendix~\ref{appendix_ssec_R_R0}.


\section{Discussion on the result}\label{sec: comp}


\subsection{Comparison with fixed-length lossy compression}\label{sec: comp kostina}

In~\cite{cuff2010,treust2017joint} the authors show  that \emph{empirical coordination} in the asymptotic regime yields the rate-distortion result of Shannon~\cite{shannon1959coding}.
\emph{Empirical coordination} is the weaker form of coordination which requires the joint histogram of the devices' distributed random states to approach a target distribution in $L^1$ distance with high probability~\cite{cuff2010}, thus capturing an  ``average behavior''  of the agents. 
This metric of choice can be specialized to the probability of distortion, therefore connecting empirical coordination with source coding~\cite{cuff2010,treust2017joint}.
In this paper  however we have considered the \emph{strong coordination} metric, which requires  the joint distribution of sequences of distributed random states to converge to an i.i.d. target distribution in $L^1$ distance instead~\cite{cuff2010}, hence dealing with a different and more stringent constraint which demands a positive rate of common randomness. 
Nonetheless, by looking at the known results for fixed-length rate-distortion~\cite{kostina2012fixed}, we can derive similarities with our case of study.

First, we recall the setting and notation of fixed-length lossy compression~\cite{kostina2012fixed}. The output of a source $S$ generated according to $ \bar P_S$ with alphabet $\mathcal M$ is mapped to one of the $M$ codewords from $\hat{\mathcal M}$, and a lossy code  consists of a pair of mappings $f : \mathcal M \mapsto \{1,\ldots ,M\}$ and $c: \{1, \ldots ,M\} \mapsto \hat{\mathcal M}$. Then, a distortion measure $d: \mathcal M \times  \hat{\mathcal M} \mapsto [0,\infty)$ is used to quantify the performance of a lossy code.
Given the decoder $c$, the best encoder maps the source output to the codeword which minimises the distortion. Then,  $\epsilon$ is the  excess-distortion probability if
\begin{equation}\label{condr-d}
\mathbb P \{ d\left(S, c(f(S))\right)>d\} \leq \epsilon.
\end{equation}
The minimum achievable code size at excess-distortion probability  $\epsilon$ and distortion $d$ is defined by 
{\allowdisplaybreaks\begin{align*}
M^{*}(d, \epsilon) &=\min \{ M \, : \, \exists \, (M,d, \epsilon) \,\, \mbox{code} \}, \\
R(n,d, \epsilon)&= \frac{1}{n} \log M^{*}(M,d, \epsilon).
\end{align*}}

Then, the following achievability results are presented in~\cite{kostina2012fixed,kostina2013lossy}.
\begin{theo}[Achievability for fixed-length lossy compression~$\mbox{\cite[Thm.~2.21]{kostina2013lossy}}$]
There exists an $(M, d,\epsilon)$ code with
{\allowdisplaybreaks
\begin{align}
\epsilon \leq \inf_{P_{Z|S}}  \left\{ \mathbb P \{d(S,Z)>d\}+ \inf_{\gamma>0} \left\{ \sup_{z\in \hat{\mathcal M}} \mathbb P \{\imath_{S;Z}(S;z) \geq \log M -\gamma \}\right\} + \exp (-\gamma)\right\}
\end{align}}
\end{theo}

\begin{theo}[Gaussian approximation for fixed-length lossy compression~$\mbox{\cite[Thm.~12]{kostina2012fixed}}$]\label{theo_kostina}
When the source is memoryless, 
{\allowdisplaybreaks
\begin{align}
R(n,d, \epsilon)&= R(d)+ \sqrt{\frac{V}{n}} Q^{-1}(\epsilon)+ \Theta \left(\frac{\log n}{n} \right)
\nonumber\\
&\geq \min I(S;Z)+ \sqrt{\frac{V}{n}} Q^{-1}(\epsilon)+ \Theta \left(\frac{\log n}{n} \right)
\end{align}}where $V$ is the dispersion term, and  $f(n)=\Theta \left(\frac{\log n}{n} \right)$ indicates that $f$ is bounded both above and below by $\frac{\log n}{n}  $ asymptotically: $\exists k_1>0, \, \exists k_2>0,
, \exists n_0$ such that $\forall n >n_0$, $ k_1 \frac{\log n}{n}  \leq f(n) \leq k_2 \frac{\log n}{n} $.
\end{theo}

\vspace{2mm}

Now, if we choose as a distance $d(\cdot)$ the $L^1$ distance,  the strong coordination condition
\begin{equation}
{\lVert P_{U^nV^n} - \bar P_{UV}^{\otimes n}\rVert}_{1} \leq d 
\end{equation}implies the rate-distortion condition~\eqref{condr-d}.
Then, we can  interpret the strong coordination problem outlined in Section~\ref{sec: sys} as two connected \vv{stronger} rate-distortion problems depicted in Figure~\ref{fig: sourceprob1} and Figure~\ref{fig: sourceprob2} respectively:
\begin{itemize}
\item \emph{Rate-distortion Problem 1:} first, at the encoder we have to generate a pair $(U,V)$ which is close in $L^1$ distance to the one generated according to the fixed i.i.d. distribution $\bar P_{UV}$;
\item \emph{Rate-distortion Problem 2:} in a second instance, the decoder has to reconstruct the source $U$ to produce $V$ via the conditional distribution $\bar P_{V|U}$. 
\end{itemize}
\begin{center}
\begin{figure}[h]
\centering
\includegraphics[scale=0.22]{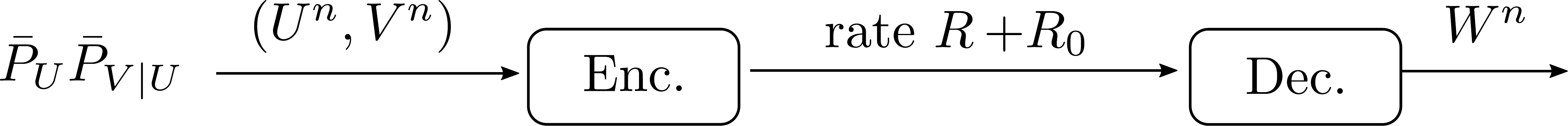}
\vspace{-1mm}
\caption{Rate-distortion Problem 1: Compression of the source $(U^n,V^n)$  with a link of rate $R+R_0$.}
\label{fig: sourceprob1}
\end{figure}

\begin{figure}[h]
\centering
\includegraphics[scale=0.22]{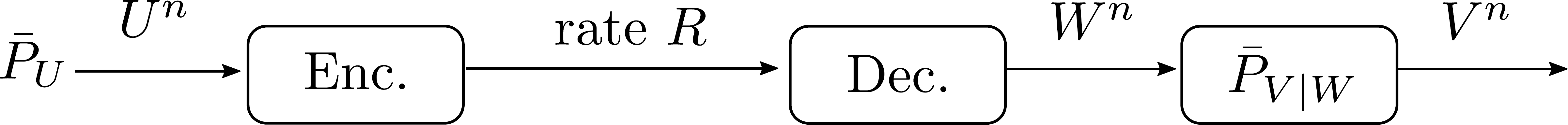}
\vspace{-1mm}
\caption{Rate-distortion Problem 2: Compression of the source $U^n$  with a link of rate $R$ and reconstruction of $V^n$.}
\label{fig: sourceprob2}
\end{figure}
\end{center} 

\vspace{-5mm}
\noindent
If the two goals are fulfilled the strong coordination requirements are met, and each one implies lossy compression of a source.
Then, at the encoder, we can interpret the strong coordination problem as \vv{distorting} a source of i.i.d. distribution $\bar P_{UV}$, by exploiting a link of rate $R+R_0$ as in Figure~\ref{fig: sourceprob1}. Then, the constraint on rate of the coordination problem
{\allowdisplaybreaks
\begin{align}
R+R_0&\geq \min I(UV;W)+ \sqrt{\frac{V}{n}} Q^{-1}(\epsilon)+O \left(\frac{\log n}{n} \right)\label{concoord1}
\end{align}}is similar  and implies the  rate-distortion condition of  Theorem~\ref{theo_kostina}: 
{\allowdisplaybreaks
\begin{align}
R+R_0&\geq \min I(UV;W)+ \sqrt{\frac{V}{n}} Q^{-1}(\epsilon)+ \Theta \left(\frac{\log n}{n} \right).\label{con1}
\end{align}}Moreover, as in Figure~\ref{fig: sourceprob2}, if the decoder is able to reconstruct $U^n$ reliably, then it can generate $V^n$ by using the i.i.d. distribution, and strong coordination would be achieved. By allowing $W^n$ to be the reliable reconstruction of $U^n$ at the decoder, we can also reformulate this in terms of \vv{stronger} rate-distortion, and we find
{\allowdisplaybreaks
\begin{align}
R&\geq \min I(U;W)+ \sqrt{\frac{V}{n}} Q^{-1}(\epsilon)+O \left(\frac{\log n}{n} \right),\label{concoord2}
\end{align}}which implies the  rate-distortion condition of  Theorem~\ref{theo_kostina}: 
{\allowdisplaybreaks
\begin{align}
R&\geq 
 \min I(U;W)+ \sqrt{\frac{V}{n}} Q^{-1}(\epsilon)+ \Theta \left(\frac{\log n}{n} \right).\label{con2}
\end{align}}Then, the constraints~\eqref{con1} and~\eqref{con2} which ensure lossy compression are similar to the rate conditions~\eqref{concoord1} and~\eqref{concoord2} in Theorem~\ref{theona_inner} for achievability in fixed-length strong coordination, with the only difference being the order of approximation. However, as we will see in Section~\ref{sec: nonasy ib}, we prove the achievability with rate constraints
\begin{equation*}
\mbox{rate} \geq \mbox{mutual information} + \sqrt{\frac{V}{n}} Q^{-1}(\epsilon) + \underbrace{\mbox{constant} \, \cdot \, \frac{\log n}{n}}_{\Theta\left(\frac{\log n}{n}\right)}
\end{equation*}which is consistent with~\eqref{con1} and~\eqref{con2}.
We keep the term $O\left(\frac{\log n}{n}\right)$ in Theorem~\ref{theona_inner} because, being less restrictive, it allows us to derive the closed result for the fixed-length coordination region of Corollary~\ref{theona}.

Notice that there is no rate of common randomness in this second constraint. This is because the two problems have to be solved together, and once that all the possible distributions $\bar P_{UV}$ are generated at the \vv{encoder side} in Rate-distortion Problem 1, the decoder's role merely relies on generating the correct random variable. This concept is better explained in the following remarks.

\vspace{2mm}

\begin{rem}[Stochasticity at the decoder does not help]\label{stoch}
Note that we can always represent discrete stochastic decoders as discrete deterministic decoders with auxiliary randomness $S$ that takes value in $ \llbracket 1, 2^{n R^*}\rrbracket$. Then, instead of the stochastic decoder function $\mbox{dec}$, we can consider the deterministic decoder $\mbox{dec}'$, that exploits external randomness $S$. Then, when focusing on the probability of error $p_e$, we have
{\allowdisplaybreaks
\begin{align}
p_e \! &=  \! \mathbb P\{ d(U^n, W^n)>d\} \nonumber\\&= \mathbb E_{S} \left[ \, \mathbb P \{ d(U^n, W^n)>d\,  |\, S\}\right]\nonumber\\&=  \mathbb E_{S} \left[ \, p_e(S)\right].\label{perroru}
\end{align}}Since each realization $\mathbf s $ of $S$ gives a deterministic decoder, and the average over all $\mathbf s $ is equal to $p_e$ by \eqref{perroru}, there exists at least one choice  $\mathbf s^{\star}$ for which $p_e(\mathbf s^{\star}) \leq p_e$. Because of this, and because the choice of the deterministic decoder only concerns reliable reconstruction and not approximating the target distribution, we can assume that the decoder is deterministic without loss of generality.
\end{rem}

\vspace{2mm}

\begin{rem}[The encoder has to be stochastic]
While we can suppose that the decoder is deterministic, the encoding function should be stochastic to achieve the whole coordination region. This is because we not only want to characterize the rates such that the rate  condition holds, but also the target distributions $\bar P_{UV}$. When restricting the case to deterministic functions,  we would restrict the choice of distributions $\bar P_{UV}$ that can be coordinated. 
More in details, $W^n$ generated according to $ \bar P_{W|UV}^{\otimes n}$ comes from:
{\allowdisplaybreaks
\begin{align*}
\llbracket 1, 2^{nR} \rrbracket    \! \times\!  \llbracket 1, 2^{nR_{0}} \rrbracket \! \times\!  {\mathcal U}^n  \times\!  {\mathcal V}^n &\!\xrightarrow{\, \phantom{iiiiiiiiiiiii} \mbox{ enc. } \phantom{iiiiiiiiiiiii}\,}  \! {\mathcal W}^n,\\
(\mathbf c,\mathbf m, \mathbf u,  \mathbf v) &\!\xmapsto{\, \phantom{iiiiiiiiiiiii} \phantom{\mbox{ enc. }} \phantom{iiiiiiiiiiiii}\,} \, \mathbf w,
\end{align*}}and if the encoder is a deterministic function, $\mbox{enc}(\mathbf c, \mathbf m, \mathbf u, \mathbf v)= \mathbf w$ with probability 1, whereas if the encoder is  stochastic, $\mbox{enc}(\mathbf c, \mathbf m, \mathbf u, \mathbf v)= \mathbf w$ with probability $ \bar P_{V|U}^{\otimes n}(\mathbf v| \mathbf u) $. Thus, the \vv{deterministic encoder choice} restricts the possibilities for $\bar P_{U^nW^nV^n}$ and therefore for the target distributions $\bar P_{U^nV^n}$, since the realization $(\mathbf u, \mathbf w, \mathbf v)$ would be generated with probability
{\allowdisplaybreaks
\begin{align*}
\frac{1}{2^{nR_{C}}} \frac{1}{2^{nR_{M}}}  \bar P_{U}^{\otimes n}(\mathbf u) \bar P_{V|U}^{\otimes n}(\mathbf v| \mathbf u) 
\end{align*}}instead of 
{\allowdisplaybreaks
\begin{align*}
\frac{1}{2^{nR_{C}}} \frac{1}{2^{nR_{M}}}  \bar P_{U}^{\otimes n}(\mathbf u) \bar P_{V|U}^{\otimes n}(\mathbf v| \mathbf u)  \bar P_{W|UV}^{\otimes n}(\mathbf v| \mathbf u, \mathbf v).
\end{align*}}
\end{rem}


\subsection{Trade-off between  $\epsilon_{\text{Tot}}$ and rate}\label{sec: trade-off}

Observe that in order to minimize  $\epsilon_{\text{Tot}}$, in the achievability proof we can choose $\epsilon_1^{*}$ and $\epsilon_2^{*}$ equal to zero. On the other hand,  this would require more common randomness since $Q^{-1}(\cdot)$ increases as its argument approaches zero.
Note that one can minimize $ \epsilon_{\text{Tot}}$ (for example, we can have $ \epsilon_{\text{Tot}}= \text{constant} \cdot 2^{-n}$) simply by choosing different $(\gamma_1, \gamma_2, \gamma_3)$ in   Section~\ref{section rate} in the achievability,  but this increases the rate conditions~\eqref{final rate}. If, for example we choose $(\gamma_1, \gamma_2, \gamma_3)=\left(2cn, cn, 2cn\right)$ for every constant $c$, the rate conditions become
\vspace{-1mm}
{\allowdisplaybreaks
\begin{align}
 & R+R_0 >  I(W;UV) + Q^{-1}(\epsilon_1) \sqrt{ \frac{V_{\bar P_{W|UV}}}{n}}+ 3c,\nonumber\\[1mm] 
 & R > I(W;U) + Q^{-1}(\epsilon_2) \sqrt{ \frac{V_{\bar P_{W|U}}}{n}}+ 3c,\label{final rate nv3}
\end{align}}and therefore match the rate constraints of the outer bound~\eqref{region fn outer}.
With this choice, the bound \eqref{epsilon tot2}  on the $L^1$ distance decreases exponentially:
\vspace{0.5mm}
{\allowdisplaybreaks
\begin{align}
 &{\lVert  P^{\text{RB}}_{U^n V^n}- P^{\text{RC}}_{U^n V^n} \rVert}_{1} \leq \epsilon_{\text{Tot}} ,\nonumber\\[1mm]
 & \epsilon_{\text{Tot}} = 10 \bar P{\left ( \mathcal S_{\gamma_1} \cap \mathcal S_{\gamma_2} \cap \mathcal S_{\gamma_3})^{\mathrm{c}} \right)}
  + 2 \, \Big(2^{-\frac{\gamma_1+1}{2}} +5 \cdot 2^{-\gamma_2}+ 2^{-\frac{\gamma_3+1}{2}} \Big)\nonumber\\[1mm]
& \phantom{ \epsilon_{\text{Tot}} }\leq 10\,( \epsilon_1 + \epsilon_2 ) +  2^{1-c}\, \left(2^{\frac{1}{2}} +5 \right) 2^{-n}.\label{speed conv c2v3}
\end{align}}

Suppose instead that we want to recover the same conditions of the outer bound~\eqref{region fn outer}. Then  we can choose $(\gamma_1, \gamma_2, \gamma_3)=\left(2, 1, 2\right)$. With this choice for $\gamma_i$, the rate conditions become
\vspace{-1mm}
{\allowdisplaybreaks
\begin{align}
 & R+R_0 >  I(W;UV) + Q^{-1}(\epsilon_1) \sqrt{ \frac{V_{\bar P_{W|UV}}}{n}}+ \frac{3}{n},\nonumber\\[1mm] 
 & R > I(W;U) + Q^{-1}(\epsilon_2) \sqrt{ \frac{V_{\bar P_{W|U}}}{n}}+  \frac{3}{n},\label{final rate nv2}
\end{align}}and therefore match the rate constraints of the outer bound~\eqref{region fn outer}.
With this choice, the bound \eqref{epsilon tot2}  on the $L^1$ distance becomes
\vspace{0.5mm}
{\allowdisplaybreaks
\begin{align}
 & \epsilon_{\text{Tot}} 
\leq 10\,( \epsilon_1 + \epsilon_2 ) +   \, \left(2^{\frac{1}{2}} +5 \right).\label{speed conv c2v2}
\end{align}}More to this point, in the achievability we can choose $(\gamma_1, \gamma_2, \gamma_3)=\left(\frac{2c}{n^{k}},\frac{c}{n^{k}}, \frac{2c}{n^{k}}\right)$ for any constant $c$ and any $k\geq 0$. With this choice for $\gamma_i$, the rate conditions become
\vspace{-1mm}
{\allowdisplaybreaks
\begin{align}
 & R+R_0 >  I(W;UV) + Q^{-1}(\epsilon_1) \sqrt{ \frac{V_{\bar P_{W|UV}}}{n}}+ \frac{3c}{n^{k+1}},\nonumber\\[1mm] 
 &\phantom{R+R_0}=I(W;UV) + Q^{-1}(\epsilon_1) \sqrt{ \frac{V_{\bar P_{W|UV}}}{n}}+O\left( \frac{1}{n}\right),\nonumber\\[1mm] 
 & R > I(W;U) + Q^{-1}(\epsilon_2) \sqrt{ \frac{V_{\bar P_{W|U}}}{n}}+  \frac{3}{n\sqrt{n}},\nonumber\\[1mm] 
& \phantom{R} = I(W;U) + Q^{-1}(\epsilon_2) \sqrt{ \frac{V_{\bar P_{W|U}}}{n}}+O\left( \frac{1}{n}\right)
 \label{final rate v4}
\end{align}}because $ \frac{3c}{n^{k+1}}= O\left( \frac{1}{n}\right) $, since $ \forall k \geq 0$ we have $\frac{3c}{n^{k+1}} \leq  \, \frac{3c}{n}$,  and the bound on the $L^1$ distance becomes
\vspace{0.5mm}
{\allowdisplaybreaks
\begin{align}
 & \epsilon_{\text{Tot}} 
 \leq  10\,( \epsilon_1 + \epsilon_2 ) + 2 \, \left(2^{\frac{1}{2}} +5 \right) \, 2^{- \frac{c}{n^k} }.\label{speed conv c2v4}
\end{align}}

\begin{center}
\begin{figure}[h]
\centering
\includegraphics[scale=0.8]{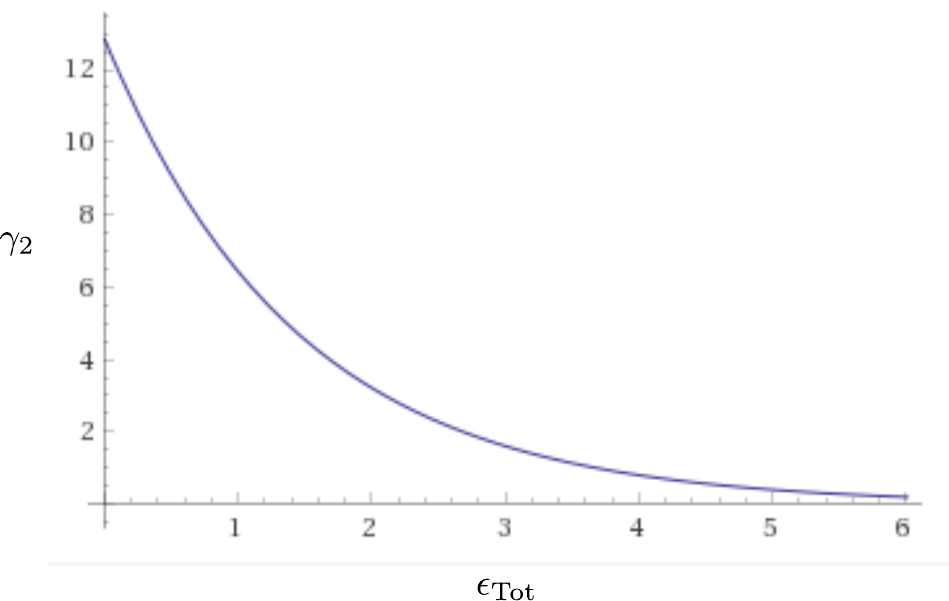}
\vspace{-1mm}
\caption{Trade-off between $(\gamma_1, \gamma_2, \gamma_3)$ and $\epsilon_{\text{Tot}}$ for $\epsilon_1 = \epsilon_2=0$}
\label{fig: graph}
\end{figure}
\end{center}

We can generalize this by fixing $(\gamma_1, \gamma_2, \gamma_3)=\left(2x,x, 2x\right)$, the rate conditions and the bound on the $L^1$ distance become
\vspace{-1mm}
{\allowdisplaybreaks
\begin{align}
 & R+R_0 >  I(W;UV) + Q^{-1}(\epsilon_1) \sqrt{ \frac{V_{\bar P_{W|UV}}}{n}}+ \frac{3x}{n},\nonumber\\[1mm] 
 & R > I(W;U) + Q^{-1}(\epsilon_2) \sqrt{ \frac{V_{\bar P_{W|U}}}{n}}+  \frac{3x}{n},\nonumber\\[1mm] 
 & \epsilon_{\text{Tot}} 
 \leq  10\,( \epsilon_1 + \epsilon_2 ) + 2 \, \left(2^{\frac{1}{2}} +5 \right) \, 2^{- x }.\label{speed conv c2v5}
\end{align}} and the trade-off between $(\gamma_1, \gamma_2, \gamma_3)$, and therefore the rate conditions, and $\epsilon_{\text{Tot}}$ is depicted in Figure~\ref{fig: graph}.

\appendices


\section{Detailed analysis of $ \mathcal S_{\imath(\mathbf w;\mathbf u )}$}\label{appendix_siwu}
We observe that, since the distribution $\bar P$ is i.i.d., 
the terms
$Z'_i= \imath_{\bar P}(w_i, u_i)$, 
are mutually independent for $i=1, \ldots n$
 Then, we consider the following inequality
 {\allowdisplaybreaks
 \begin{align}
 nR>\underbrace{\sum_{i=1}^n { \mathbb E}_{\bar P_{WU}} [\imath_{\bar P}(w_i; u_i )]}_{n \mu'_n }+ Q^{-1} (\epsilon_2)  \underbrace{\sqrt{  \sum_{i=1}^n   {\text{Var}}_{\bar P_{WU}} (\imath_{\bar P}(w_i;u_i ))}}_{n \, \sqrt{V'_n/n}} +\, \gamma_2, \label{be rate_a}
 \end{align}}where $ \mu'_n = \frac{1}{n} \sum_{i=1}^n \mathbb E [Z_i]$, ${V'}_n  = \frac{1}{n} \sum_{i=1}^n \text{Var} [Z_i] $  and $Q(\cdot)$  is the tail distribution function of the standard normal distribution. We prove that, assuming that~\eqref{be rate_a} holds, we can successfully bound $ \mathcal S_{\imath(\mathbf w;\mathbf u)}$. In fact, 
 the chain of inequalities
  {\allowdisplaybreaks
 \begin{align*}
  \sum_{i=1}^n i_{\bar P}(W;U) >nR-(\gamma_1 + \gamma_2)> n\left(\mu'_n + t\sqrt{\frac{{V'}_n}{n}}\right)
  \end{align*}}implies that, if~\eqref{be rate_a} holds, $ \mathcal S_{\imath(\mathbf w;\mathbf u )}$ is contained in 
   {\allowdisplaybreaks
 \begin{align*}
 \left\{  (\mathbf u, \mathbf w) : \sum_{i=1}^n   \imath_{\bar P}(w_i;u_i,)> n\mu'_n + n \, Q^{-1}(\epsilon_2) \sqrt{\frac{V_n}{n}}\right\} .\stepcounter{equation}\tag{\theequation}\label{be2_a}
 \end{align*}}Therefore, if we find an upper bound on~\eqref{be2_a}, we have an upper bound on $ \mathcal S_{\imath(\mathbf w;\mathbf u)}$ as well. To obtain that, we apply Theorem \ref{berryessen} (Berry-Esseen CLT) to the right-hand side of \eqref{be2_a}, and we choose 
{\allowdisplaybreaks \begin{align}
&Q(t)=\epsilon_2,\nonumber\\
& \epsilon_2^{*} =\epsilon_2+ \frac{B'_n}{\sqrt{n}}, \label{epsilon4_a}
 \end{align}}where, as in the statement of Theorem \ref{berryessen} (Berry-Esseen CLT),
 $B'_n = 6 \frac{T'_n}{{V'}_n^{3/2}}$, and  $T'_n =\frac{1}{n}\sum_{i=1}^n \mathbb{E} [{\lvert Z'_i - \mu_i \rvert}^3].$
Then, we have
    {\allowdisplaybreaks
 \begin{align}
 &\left\lvert  \mathbb P \left\{ \sum_{i=1}^n    \imath_{\bar P}(w_i, u_i)>n\mu'_n  +  n\, Q^{-1}(\epsilon_2) \sqrt{\frac{{V'}_n}{n}} \right\}  -  \epsilon_2  \right\rvert \leq   \frac{B'_n}{\sqrt{n}},\nonumber\\[1mm]
&\Rightarrow \mathbb P \left\{ \sum_{i=1}^n    \imath_{\bar P}(w_i, u_i)>n\mu'_n 
+ n\, Q^{-1}(\epsilon_2) \sqrt{\frac{{V'}_n}{n}}\right\} \leq \epsilon_2^{*}.\label{be4_a}
 \end{align}}Finally, \eqref{be4_a} combined with \eqref{be2_a} implies $\bar P {\left(\mathcal S_{\imath(w:uv)}  \right)} \leq \epsilon_1^{*}.$
Moreover, we can simplify~\eqref{be rate_a} with the following identifications:
similarly to~\cite{yassaee2013non}, observe that
{\allowdisplaybreaks
 \begin{subequations}
 \begin{align}
 \mu'_n &\coloneqq\frac{1}{n} \sum_{i=1}^n { \mathbb E}_{\bar P_{WU }} [\imath_{\bar P}(w_i; u_i )]\nonumber\\[0.5mm]
%
&={ \mathbb E}_{\bar P_{WU }} [\imath_{\bar P}(w; u )]= I(W;U),\label{rate entropy_a}\\[2mm]
V'_n&\coloneqq  \frac{1}{n} \sum_{i=1}^n {\text{Var}}_{\bar P_{WU}} (\imath_{\bar P}(W;U))\nonumber\\
&= {\text{Var}}_{\bar P_{WU}} (\imath_{\bar P}(W;U)),\label{dispersion term_a} 
\end{align} \end{subequations}}and $V_{\bar P_{W|U}}=\min_{\bar P_{W|U}} \left[  {\text{Var}}_{\bar P_{WU}} (\imath_{\bar P}(W;U))\right] =\min_{\bar P_{W|U}} \left[  {\text{Var}}_{\bar P_{WU}} (\imath_{\bar P}(W;U)|W)\right] $ is the dispersion of the channel $\bar P_{W|U}$ as defined in \cite[Thm.~49]{polyanskiy2010channel}.
Hence, \eqref{be rate_a} can be rewritten as
 \begin{equation}\label{be rate3_a}
 nR>n I(W;U)  + n\, Q^{-1}(\epsilon_2) \sqrt{\frac{V_{\bar P_{W|U}}}{n}} +( \gamma_1+\gamma_2).
 \end{equation}


\section{Proof of $H(M) \geq nI(U;W)$}\label{appendix_ssec_HM}
We have 
{\allowdisplaybreaks\begin{align}
H(M)&\geq H(M|C) \geq I(U^n;M|C)=\sum_{i=1}^n I(U_t;M|U^{t-1}C)\nonumber\\
&=\sum_{i=1}^n I(U_t;MC|U^{t-1}) - \sum_{i=1}^n I(U_t;C|U^{t-1}) \nonumber\\
&\geq \sum_{i=1}^n I(U_t;MC|U^{t-1}) - \sum_{i=1}^n I(U^t;C)\overset{\mathclap{(a)}}{=} \sum_{i=1}^n I(U_t;MC|U^{t-1}) \nonumber\\
& =\sum_{i=1}^n I(U_t;MCU^{t-1}) - \sum_{i=1}^n I(U_t;U^{t-1})\nonumber\\
&\overset{\mathclap{(d)}}{=}  \sum_{i=1}^n I(U_t;MCU^{t-1}) \geq  \sum_{i=1}^n I(U_t;MC) \nonumber\\
& \overset{\mathclap{(c)}}{=} \sum_{i=1}^n I(U_t;W_t)\overset{\mathclap{(f)}}{=}  n I(U;W) 
\end{align}}where $(a)$ and $(b)$ follow from the i.i.d. nature of the channel, and $(c)$ and $(d)$ from the identifications $W_t =(C,M)$ for each $t \in \llbracket 1,n\rrbracket$ and $W=(W_T,T)=(C,M, T)$.


\section{Proof of the bound on $R$: Case 2}\label{appendix_ssec_R_Case2}
\vspace{1mm}


\subsection{Proof of the upper bound on $\log{\beta_{\alpha}}$~\eqref{mcpartb} -- Case 2 ($P_{U^nW^n}=\bar P_{U}^{\otimes n} \bar P_{W|U}^{\otimes n}-\epsilon$)}\label{sssec_mcpartb case 2}  
We have
{\allowdisplaybreaks
\begin{align}
\log { \left( P_{U^nW^n} (\mathbf u,\mathbf w )  \right)}  
&= \log{\left(\bar P_{U}^{\otimes n} \bar P_{W|U}^{\otimes n} (\mathbf u,\mathbf w )  - \epsilon\right)}\nonumber\\[1.5mm]
&= \log { \left( \bar P_{U}^{\otimes n} \bar P_{W|U}^{\otimes n} (\mathbf u,\mathbf w )  \right)} + \log { \left(1 - \frac{ \epsilon}{\bar P_{U}^{\otimes n} \bar P_{W|U}^{\otimes n} (\mathbf u,\mathbf w )} \right)}\nonumber\\[2mm]
&= \log { \left( \bar P_{U}^{\otimes n} \bar P_{W|U}^{\otimes n} (\mathbf u,\mathbf w )  \right)} - \log { \left( \frac{\bar P_{U}^{\otimes n} \bar P_{W|U}^{\otimes n} (\mathbf u,\mathbf w )}{\bar P_{U}^{\otimes n} \bar P_{W|U}^{\otimes n} (\mathbf u,\mathbf w )- \epsilon} \right)}.\label{ineq_tv22}
\end{align}}Thus, the following holds
{\allowdisplaybreaks
\begin{align}
&\mathbb P \left\{ \log{ P_{U^nW^n} }\geq \log{\bar P^{\otimes n}_U \bar P^{\otimes n}_W} + \log {\gamma_0}\right\} \nonumber\\[1.5mm]
&=  \mathbb P \left\{ \log { \bar P_{U}^{\otimes n} \bar P_{W|U}^{\otimes n}} -\log { \left( \frac{\bar P_{U}^{\otimes n} \bar P_{W|U}^{\otimes n} }{\bar P_{U}^{\otimes n} \bar P_{W|U}^{\otimes n} - \epsilon} \right)} \geq \log{\bar P^{\otimes n}_U \bar P^{\otimes n}_W} + \log {\gamma_0} \right\}  \nonumber\\[1.5mm]
& =\mathbb P \left\{ \log { \bar P_{U}^{\otimes n} \bar P_{W|U}^{\otimes n}}  \geq \log{\bar P^{\otimes n}_U \bar P^{\otimes n}_W} +\log {\gamma_0}+\log { \left( \frac{\bar P_{U}^{\otimes n} \bar P_{W|U}^{\otimes n} }{\bar P_{U}^{\otimes n} \bar P_{W|U}^{\otimes n} - \epsilon} \right)}\right\} \nonumber\\[1.5mm]
& =\mathbb P \left\{ \log { \frac {\bar P_{U}^{\otimes n} \bar P_{W|U}^{\otimes n}}{\bar P^{\otimes n}_U \bar P^{\otimes n}_W} }  \geq  \log {\gamma_0}+\log { \left( \frac{\bar P_{U}^{\otimes n} \bar P_{W|U}^{\otimes n} }{\bar P_{U}^{\otimes n} \bar P_{W|U}^{\otimes n} - \epsilon} \right)}\right\} \nonumber\\[1.5mm]
&=\mathbb P  \left\{ \sum_{i=1}^n \log \frac{ \bar P_{U} \bar P_{W|U}}{ \bar P_{U} \bar P_{W}} \geq  \log {\gamma_0}+\log { \left( \frac{\bar P_{U}^{\otimes n} \bar P_{W|U}^{\otimes n} }{\bar P_{U}^{\otimes n} \bar P_{W|U}^{\otimes n} - \epsilon} \right)}\right\}.\label{probgamma2}
\end{align}}Since $ \sum_{i=1}^n \log (\bar P_{U} \bar P_{W|U}/ \bar P_{U} \bar P_{W})=:  \sum_{i=1}^n Z_i$ is the sum of $n$ i.i.d. random variables, the next step is to evaluate the probability in~\eqref{probgamma} using Theorem \ref{berryessen} (Berry-Esseen CLT). To accomplish that, we choose appropriately the parameter $\gamma$:
{\allowdisplaybreaks
\begin{align}
\log {\gamma_0}
&\coloneqq n \mu_n + Q^{-1}\left(\epsilon \right) \sqrt{ n V_{n} } -\log { \left( \frac{\bar P_{U}^{\otimes n} \bar P_{W|U}^{\otimes n} }{\bar P_{U}^{\otimes n} \bar P_{W|U}^{\otimes n} - \epsilon} \right)}\label{log2}
\end{align}}where, as in Theorem \ref{berryessen},
 $\mu_n = \frac{1}{n} \sum_{i=1}^n \mathbb E [Z_i],$
 $V_n  = \frac{1}{n} \sum_{i=1}^n \text{Var} [Z_i]$,
 $T_n =\frac{1}{n}\sum_{i=1}^n \mathbb{E} [{\lvert Z_i - \mu_i \rvert}^3]$,
 $B_n = 6 \frac{T_n}{V_n^{3/2}}$,
and $Q(\cdot)$  is the tail distribution function of the standard normal distribution.
Observe that by~\eqref{log2} and Remark~\ref{rem mi cd},~\eqref{probgamma2} becomes
{\allowdisplaybreaks
\begin{align}
& \mathbb P  \left\{ \sum_{i=1}^n \log \frac{ \bar P_{U} \bar P_{W|U}}{ \bar P_{U} \bar P_{W}} \geq   n \mu_n + Q^{-1}\left(\epsilon \right) \sqrt{ n V_{n} } \right\}\nonumber\\[2mm]
&= \mathbb P  \left\{ \sum_{i=1}^n \log \frac{ \bar P_{U} \bar P_{W|U}}{ \bar P_{U} \bar P_{W}} \geq   n I(U;W) + Q^{-1}\left(\epsilon \right) \sqrt{ n V_{\bar P_{W|U}} }
\right\}\nonumber\\[2mm]
& \geq \epsilon- \frac{B_n}{\sqrt n} =: \alpha.\label{prob be2}
\end{align}}

Then, we have
{\allowdisplaybreaks
\begin{align}
\log{\frac{1}{\beta_{\alpha}}}  &\geq \log{\gamma_0} = n \mu_n + Q^{-1}\left(\epsilon \right) \sqrt{ n V_{n} } -\log { \left( \frac{\bar P_{U}^{\otimes n} \bar P_{W|U}^{\otimes n} }{\bar P_{U}^{\otimes n} \bar P_{W|U}^{\otimes n} - \epsilon} \right)}.
\label{beta3}
\end{align}}


\subsection{Proof of the lower bound on $\log{\beta_{\alpha}}$~\eqref{mcparta} -- Case 2 ($P_{U^nW^n}=\bar P_{U}^{\otimes n} \bar P_{W|U}^{\otimes n}-\epsilon$)}\label{sssec_mcparta case 2}   
By~\eqref{rel1} for every $\gamma >0$ we have
{\allowdisplaybreaks\begin{align*}
\beta_{\alpha} &\geq \frac{1}{\gamma} \left[
\alpha - \mathbb P_{P_{U^{n}W^{n}}} \left\{ \log \frac{P_{U^{n}W^{n}}}{\bar P^{\otimes n}_U \bar P^{\otimes n}_W} > \log \gamma\right\} \right]  \\[2mm]
&=\frac{1}{\gamma} \left[
\alpha - \mathbb P_{P_{U^{n}W^{n}}} \left\{ \log \frac{\bar P_{U}^{\otimes n} \bar P_{W|U}^{\otimes n}-\epsilon}{\bar P^{\otimes n}_U \bar P^{\otimes n}_W} > \log \gamma\right\} \right] \\[2mm]
&=\frac{1}{\gamma} \left[
\alpha - \mathbb P_{P_{U^{n}W^{n}}} \left\{ \log \frac{\bar P_{U}^{\otimes n} \bar P_{W|U}^{\otimes n}}{\bar P^{\otimes n}_U \bar P^{\otimes n}_W} > \log \gamma + \log { \left( \frac{\bar P_{U}^{\otimes n} \bar P_{W|U}^{\otimes n} }{\bar P_{U}^{\otimes n} \bar P_{W|U}^{\otimes n} - \epsilon} \right)} \right\} \right].
\end{align*}}Then, we set $$ \log \gamma = H(M)-\log { \left( \frac{\bar P_{U}^{\otimes n} \bar P_{W|U}^{\otimes n} }{\bar P_{U}^{\otimes n} \bar P_{W|U}^{\otimes n} - \epsilon} \right)}$$ and, as proved in~\eqref{eqy1},
{\allowdisplaybreaks\begin{align*}
&nR=H(M)\geq n I(U;W) \geq  n I(U;W) + Q^{-1}(y) \sqrt{n V_n} \qquad \frac{1}{2}<y<1
\end{align*}}which implies
{\allowdisplaybreaks
\begin{align}
\log{\beta_{\alpha} }&
\geq \log{\frac{1}{\gamma}} +\log{ \left[
\alpha - \mathbb P_{P_{U^{n}W^{n}}} \left\{ \log \frac{\bar P_{U}^{\otimes n} \bar P_{W|U}^{\otimes n}}{\bar P^{\otimes n}_U \bar P^{\otimes n}_W} > \log \gamma + \log { \left( \frac{\bar P_{U}^{\otimes n} \bar P_{W|U}^{\otimes n} }{\bar P_{U}^{\otimes n} \bar P_{W|U}^{\otimes n} - \epsilon} \right)} \right\} \right]}\nonumber\\[2mm]
&\geq- H(M)+\log { \left( \frac{\bar P_{U}^{\otimes n} \bar P_{W|U}^{\otimes n} }{\bar P_{U}^{\otimes n} \bar P_{W|U}^{\otimes n} - \epsilon} \right)}+\log{ \left[
\alpha - \mathbb P_{P_{U^{n}W^{n}}} \left\{ \log \frac{\bar P_{U}^{\otimes n} \bar P_{W|U}^{\otimes n}}{\bar P^{\otimes n}_U \bar P^{\otimes n}_W} >  n I(U;W) + Q^{-1}(y) \sqrt{n V_n} \right\} \right]}\nonumber\\[2mm]
&= - H(M)+\underbrace{\left[\log { \left( \frac{\bar P_{U}^{\otimes n} \bar P_{W|U}^{\otimes n} }{\bar P_{U}^{\otimes n} \bar P_{W|U}^{\otimes n} - \epsilon} \right)}+ \log{\left(\alpha -y - \frac{B_n}{\sqrt n}\right)}\right]}_{-x}\label{partbcase2}
\end{align}}and similarly to~\eqref{partbcase1fin},~\eqref{mcparta} holds for this case as well.


\subsection{Proof of the rate constraint -- Case 2 ($P_{U^nW^n}=\bar P_{U}^{\otimes n} \bar P_{W|U}^{\otimes n}-\epsilon$)}
Now, we conclude the proof by combining ~\eqref{beta3} and~\eqref{partbcase2}. For $1/2<y<1$, we have
{\allowdisplaybreaks
\begin{align*}
 \underbrace{H(M)}_{nR} -\left[\log { \left( \frac{\bar P_{U}^{\otimes n} \bar P_{W|U}^{\otimes n} }{\bar P_{U}^{\otimes n} \bar P_{W|U}^{\otimes n} - \epsilon} \right)} +\log{\left(\alpha -y - \frac{B_n}{\sqrt n}\right)}\right] \geq  n \mu_n + Q^{-1}\left(\epsilon\right) \sqrt{ n V_{n} } -\log { \left( \frac{\bar P_{U}^{\otimes n} \bar P_{W|U}^{\otimes n} }{\bar P_{U}^{\otimes n} \bar P_{W|U}^{\otimes n} - \epsilon} \right)} \end{align*}}which is equivalent to
{\allowdisplaybreaks\begin{align}
 R  \geq   \mu_n + Q^{-1}\left(\epsilon \right) \sqrt{ \frac{ V_{n}}{n} }+ 
\frac{\log{\left(\alpha -y - \frac{B_n}{\sqrt n}\right)} }{n}.\label{r+log}
\end{align}}  


\section{Proof of the bound on $R+R_0$}\label{appendix_ssec_R_R0}

As in Section~\ref{ssec_R},  for an observation $ \mathbf w= (m,c)$, we define the hypothesis:
{\allowdisplaybreaks
\begin{align*}
\mathcal H_0:&\quad \mathbf w \mbox{ generated according to } P_{W^{n}|U^n V^n}(\mathbf w)= \sum_{\mathbf u,  \mathbf v}  P_{U^n} (\mathbf{w}) P_{U^nW^{n} V^n} (\mathbf u, \mathbf w,  \mathbf v) =\sum_{\mathbf u, \mathbf v}  P_{U^nW^{n}}(\mathbf u, \mathbf w, \mathbf v),\\
\mathcal H_1:&\quad  \mathbf w \mbox{ generated according to }   \bar{P}_{W}^{\otimes n} (\mathbf w)=\sum_{\mathbf u,\mathbf v } \bar P_{UV}^{\otimes n} \bar{P}_{W}^{\otimes n} (\mathbf u, \mathbf w, \mathbf v),
\end{align*}}where $\bar P$ is the i.i.d. target distribution. Then, we consider a randomized test between the distributions $P_{U^{n}W^{n}V^n}$ and $ \bar P^{\otimes n}_{UV} \bar P^{\otimes n}_W$: 
$$P_{Z|U^nW^nV^n}: \mathcal U^n \times \mathcal W^n \times V^n \to {\{\mathcal H_0,\mathcal H_1\}},$$ 
where $\mathcal H_0$ indicates that the test chooses $ P_{U^{n}W^{n}V^n}$, and  $\mathcal H_1$ indicates that the test chooses $ \bar P^{\otimes n}_{UV} \bar P^{\otimes n}_W$.
The probability of type-I error (probability of choosing $\mathcal H_1$ when the true hypothesis is $\mathcal H_0$)  and type-II error (probability of choosing $\mathcal H_0$ when the true hypothesis is $\mathcal H_1$) are 
{\allowdisplaybreaks
\begin{subequations}
\begin{align}
  &
 P_{e}^{\text{I}}(P_{Z|U^nW^nV^n})\,\coloneqq \mathbb P\{\hat{\mathcal H}_1|\mathcal H_0 \}=\ \sum_{\mathbf u, \mathbf v} \bar P^{\otimes n}_{UV}  (\mathbf u, \mathbf v) \bar P^{\otimes n}_W (\mathbf w) P_{Z|U^nW^nV^n}(\mathcal H_0|\mathbf u,\mathbf w, \mathbf v),
  \label{ty20}\\
   &
P_{e}^{\text{II}}(P_{Z|U^nW^nV^n})\,\coloneqq \mathbb P\{\hat{\mathcal H}_0|\mathcal H_1 \}=\ \sum_{\mathbf u, \mathbf v} P_{U^{n}W^{n}V^n} (\mathbf u, \mathbf w, \mathbf v) P_{Z|U^nW^nV^n}(\mathcal H_1|\mathbf u,\mathbf w,\mathbf v)  \label{ty10}.
 \end{align} \end{subequations}}Similar to~\eqref{defbeta}, we denote with $\beta'_{\alpha'}$
the minimum type-I error for a maximum type-II error $1-\alpha'$:
{\allowdisplaybreaks\begin{align}
\beta'_{\alpha'}&\coloneqq \min_{\substack{P_{Z|U^nW^nV^n}: \\[1mm] 
P_{e}^{\text{II}}(P_{Z|U^nW^nV^n}) \leq 1-\alpha' }} P_{e}^{\text{I}}(P_{Z|U^nW^nV^n})\label{defbeta0}
\end{align}}where the error probability $\alpha'$ will be defined later.
The following relations between $\alpha'$ and  $\beta'_{\alpha'}$, proved in~\cite[Section 12.4]{polyanskiy2014lecture}, hold:
{\allowdisplaybreaks
\begin{subequations}\begin{align}
& \beta'_{\alpha'} \leq \frac{1}{\gamma_0}, \mbox{ if $\gamma_0$ is such that } \mathbb P_{P_{U^{n}W^{n}V^n}} \left\{ \log \frac{P_{U^{n}W^{n}V^n}}{\bar P^{\otimes n}_{UV} \bar P^{\otimes n}_W} > \log \gamma_0 \right\}\geq \alpha',\label{rel20}\\
&\alpha' \leq \mathbb P_{P_{U^{n}W^{n}V^n}} \left\{ \log \frac{P_{U^{n}W^{n}V^n}}{\bar P^{\otimes n}_{UV} \bar P^{\otimes n}_W} > \log \gamma\right\} +\gamma \, \beta'_{\alpha'}  \quad \forall \gamma >0. \label{rel10}
\end{align}\end{subequations}}Similarly to Section~\ref{ssec_R}, we prove the rate constraint by separately deriving an upper and a lower bound on $\log{\beta'_{\alpha'}}$:
{\allowdisplaybreaks\begin{subequations}
\begin{align}
\text{upper bound on $\log{\beta'_{\alpha'}}$}\qquad & \phantom{n(R+R_0)+x \geq}\log{\frac{1}{\beta'_{\alpha'}}}\geq n I(W;UV) + Q^{-1}(\epsilon) \sqrt{ n V_{\hat P_{W|UV}}}, 
\label{mcpartb0}\\
\text{lower bound on $\log{\beta'_{\alpha'}}$}\qquad  &n(R+R_0)+x \geq  \log{\frac{1}{\beta'_{\alpha'}}}\label{mcparta0}
\end{align}\end{subequations}}for a certain $x \in \mathbb R$ which will be defined later. 
Then, the proof of the rate constraint is divided in the following steps, detailed in the next sections:
\begin{enumerate}[(i)]
\item \textit{Proof of the upper bound on $\log{\beta'_{\alpha'}}$:} we use the upper bound on $\min P_{e}^{\text{I}}(P_{Z|U^nW^nV^n})$ of~\eqref{rel20} combined with the $(\epsilon,n)$-strong coordination assumption~\eqref{esc} and Theorem \ref{berryessen} (Berry-Esseen CLT) to derive the following upper bound on the logarithm $\beta'_{\alpha'}$ of \eqref{mcpartb0} by choosing the parameter $\gamma_0$;
\item \textit{Proof of the lower bound on $\log{\beta'_{\alpha'}}$:} we use the lower bound on $\min P_{e}^{\text{I}}(P_{Z|U^nW^nV^n})$ of~\eqref{rel10} combined with the $(\epsilon,n)$-strong coordination assumption~\eqref{esc} and classical information theory properties  to derive the lower bound on $\beta'_{\alpha'}$ of \eqref{mcparta0} by choosing the parameter $\gamma$;
\item we combine \eqref{mcpartb0} and \eqref{mcparta0} proved in the previous steps and we derive the rate constraint. 
\end{enumerate}

Moreover, as before by the  $(\epsilon,n)$-strong coordination assumption~\eqref{esc}  we have   
{\allowdisplaybreaks\begin{align*}
&{\lVert P_{U^nW^nV^n} \!-\! \bar P_{U}^{\otimes n} \bar P_{W|U}^{\otimes n} \bar P_{V|W}^{\otimes n} \rVert }_1\!\! \leq \!\epsilon, \\
& \!\! \Rightarrow  \! \forall (\mathbf u, \mathbf w, \mathbf v) \quad  \lvert P_{U^nW^nV^n}(\mathbf u, \mathbf w, \mathbf v) - \bar P_{U}^{\otimes n} \bar P_{W|U}^{\otimes n} \bar P_{V|W}^{\otimes n}(\mathbf u, \mathbf w, \mathbf v)  \rvert \leq \epsilon\\
& \!\! \Rightarrow \!  \forall (\mathbf u, \mathbf w, \mathbf v) ~  \exists \epsilon_{uwv} \leq \epsilon \text{ such that }   \lvert P_{U^nW^nV^n}(\mathbf u, \mathbf w, \mathbf v) - \bar P_{U}^{\otimes n} \bar P_{W|U}^{\otimes n} \bar P_{V|W}^{\otimes n}(\mathbf u, \mathbf w, \mathbf v)  \rvert= \! \epsilon_{uwv}
\end{align*}}which we can distinguish into two cases:
\begin{itemize}
\item \textit{Case 1} \quad $P_{U^nW^nV^n}(\mathbf u, \mathbf w, \mathbf v) = \bar P_{V|W}^{\otimes n}(\mathbf u, \mathbf w, \mathbf v) +\epsilon_{uwv}$;
\item \textit{Case 2} \quad $P_{U^nW^nV^n}(\mathbf u, \mathbf w, \mathbf v) = \bar P_{V|W}^{\otimes n}(\mathbf u, \mathbf w, \mathbf v) -\epsilon_{uwv}$.
\end{itemize}Thus, we prove steps (i)--(iii) separately for both cases. With a slight abuse of notation from now on we will drop the index from $\epsilon_{uwv}$ and use $\epsilon$ instead, since $\epsilon_{uwv}$ just has to be smaller that $\epsilon$. Similarly, we omit the pairs $ (\mathbf u, \mathbf w, \mathbf v)$ in order to simplify the notation.


\vspace{1mm}
\subsection{Proof of the upper bound on $\log{\beta'_{\alpha'}}$~\eqref{mcpartb0} -- Case 1 ($P_{U^nW^nV^n}=\bar P_{U}^{\otimes n} \bar P_{W|U}^{\otimes n} \bar P_{V|W}^{\otimes n}+\epsilon$)}\label{sssec_mcpartb0}
By the $(\epsilon,n)$-strong coordination assumption~\eqref{esc}, we have
{\allowdisplaybreaks
\begin{align}
\log{ \left(P_{U^nW^nV^n} (\mathbf u,\mathbf w, \mathbf v ) \right)}
&= \log { \left( \bar P_{U}^{\otimes n} \bar P_{W|U}^{\otimes n} \bar P_{V|W}^{\otimes n}  (\mathbf u,\mathbf w, \mathbf v ) \right)} + \log { \left(1 + \frac{ \epsilon} {\bar P_{U}^{\otimes n} \bar P_{W|U}^{\otimes n}  \bar P_{V|W}^{\otimes n}(\mathbf u,\mathbf w, \mathbf v )} \right)}.
 \label{ineq_tv20}
\end{align}}Then, similarly to~\eqref{prob be}, the following holds
{\allowdisplaybreaks
\begin{align}
&\mathbb P \left\{ \log{ P_{U^nW^nV^n} }\geq \log{\bar P^{\otimes n}_{UV} \bar P^{\otimes n}_W} + \log {\gamma_0}\right\} \nonumber\\[2mm]
& =\mathbb P \left\{ \log { \frac {\bar P_{U}^{\otimes n} \bar P_{W|U}^{\otimes n}  \bar P_{V|W}^{\otimes n} }{\bar P^{\otimes n}_{UV} \bar P^{\otimes n}_W} }  \geq  \log {\gamma_0}  -\log { \left(1 + \frac{ \epsilon} {\bar P_{U}^{\otimes n} \bar P_{W|U}^{\otimes n}  \bar P_{V|W}^{\otimes n}  } \right)}\right\}\nonumber\\
& = \mathbb P  \left\{ \sum_{i=1}^n \log \frac{ \bar P_{U} \bar P_{W|U} \bar P_{V|W}}{ \bar P_{UV} \bar P_{W}} \geq   n I(UV;W) + Q^{-1}\left(\epsilon \right) \sqrt{ n V_{\bar P_{W|UV}} }\right\}
\label{prob be0}
\end{align}}where the last equality follows from  choosing the parameter $\gamma_0$ as
{\allowdisplaybreaks
\begin{align}
\log{\gamma_0} 
&\coloneqq n \mu_n + Q^{-1}\left(\epsilon \right) \sqrt{ n V_{n} } + \log { \left(1 + \frac{ \epsilon} {\bar P_{U}^{\otimes n} \bar P_{W|U}^{\otimes n} \bar P_{V|W}^{\otimes n}   } \right)}\label{log0}
\end{align}}and, as in the statement of Theorem \ref{berryessen} and Remark~\ref{rem mi cd}, 
from the identifications:
{\allowdisplaybreaks
\begin{align*}
 &\sum_{i=1}^n Z_i=\sum_{i=1}^n \log (\bar P_{U} \bar P_{W|U}  \bar P_{V|W}/ \bar P_{UV} \bar P_{W})  ,\\[1.5mm]
 &\mu_n = \frac{1}{n} \sum_{i=1}^n \mathbb E [Z_i]=\mathbb D(\bar P_U \bar P_{W|U}\bar P_{V|W} \Arrowvert \bar P_{UV} \bar P_{W})=I(W;UV),\\[1.5mm]
 &V_n  = \frac{1}{n} \sum_{i=1}^n \text{Var} [Z_i]= V_{\bar P_{W|UV}}\\
 &\phantom{V_n}= \sum_{u,w, v} \bar P_U(u) \bar P_{W|U}(w|u) \bar P_{V|W} (v|w) {\left[ \log{\frac{\bar P_U(u) \bar P_{W|U}(w|u) \bar P_{V|W} (v|w) }{\bar P_{UV}(u,v) \bar P_{W}(w)}}\right]}^2- {\mathbb D(\bar P_U \bar P_{W|U} \bar P_{V|W} \Arrowvert \bar P_{UV} \bar P_{W})}^2\\
 &T_n =\frac{1}{n}\sum_{i=1}^n \mathbb{E} [{\lvert Z_i - \mu_i \rvert}^3]\\
 &\phantom{T_n}=\sum_{u,w,v} \bar P_U(u) \bar P_{W|U}(w|u)  \bar P_{V|W} (v|w)  {\left\lvert \log{\frac{\bar P_U(u) \bar P_{W|U}(w|u) \bar P_{V|W} (v|w)  }{\bar P_{UV}(u,v) \bar P_{W}(w)}}- \mathbb D(\bar P_U\bar P_{W|U}  \bar P_{V|W} \Arrowvert \bar P_{UV} \bar P_{W})\right\rvert}^3,\\[1.5mm]
 &B_n = 6 \frac{T_n}{V_n^{3/2}},
 \end{align*}}and $Q(\cdot)$  is the tail distribution function of the standard normal distribution.

Now, since $ \sum_{i=1}^n \log (\bar P_{U} \bar P_{W|U}  \bar P_{V|W}/ \bar P_{UV} \bar P_{W})= \sum_{i=1}^n Z_i$ is the sum of $n$ i.i.d. random variables, we bound~\eqref{prob be0} using Theorem \ref{berryessen} (Berry-Esseen CLT):
{\allowdisplaybreaks
\begin{align}
 \mathbb P \left\{ \log \prod_{i=1}^n \frac{ \bar P_{U} \bar P_{W|U}  \bar P_{V|W}}{ \bar P_{UV} \bar P_{W}} \geq n I(UV;W) + Q^{-1}\left(\epsilon \right) \sqrt{ n V_{\bar P_{W|UV}} }\right\}  \geq \left(\epsilon - \frac{B_n}{\sqrt n}\right)\label{deltanew0}
 \end{align}}and we identify 
 \begin{equation}
 \alpha' \coloneqq \epsilon - \frac{B_n}{\sqrt n}
 \end{equation}and by combining~\eqref{rel20} with~\eqref{log0}, we obtain:
{\allowdisplaybreaks
\begin{align}
\log{\frac{1}{\beta'_{\alpha'}}} & \geq  \log{ \gamma_0} 
=  n \mu_n + Q^{-1}\left(\epsilon \right) \sqrt{ n V_{n} } + \log { \left(1 + \frac{ \epsilon} {\bar P_{U}^{\otimes n} \bar P_{W|U}^{\otimes n} \bar P_{V|W}^{\otimes n} } \right)} .\label{beta2part1.1}\end{align}} 

 
\subsection{Proof of the lower bound on $\log{\beta'_{\alpha'}}$~\eqref{mcparta0} -- Case 1 ($P_{U^nW^nV^n}=\bar P_{U}^{\otimes n} \bar P_{W|U}^{\otimes n} \bar P_{V|W}^{\otimes n}+\epsilon$)}\label{sssec_mcparta0}
First, we prove that 
{\allowdisplaybreaks\begin{align}
n(R+R_0)=H(M,C) &\overset{\mathclap{(a)}}{\geq}  n I(UV;W) -4 n \epsilon \left(\log {\lvert \mathcal U \times \mathcal V \rvert} + \log {\frac{1}{\epsilon}}\right)\nonumber\\
&\overset{\mathclap{(b)}}{\geq}   n I(UV;W)-4 n \epsilon \left(\log {\lvert \mathcal U \times \mathcal V \rvert} + \log {\frac{1}{\epsilon}}\right)
 + Q^{-1}(y) \sqrt{n V_n}, \qquad \frac{1}{2}<y<1 \label{eqy10}.
\end{align}}To prove $(a)$, observe that
{\allowdisplaybreaks\begin{align}
H(M,C)& \geq I(U^n V^n ;MC)=\sum_{i=1}^n I(U_t V_t;MC|U^{t-1}V^{t-1})\nonumber\\
&=\sum_{i=1}^n I(U_t V_t;MC U^{t-1}V^{t-1}) - \sum_{i=1}^n I(U_t V_t; U^{t-1}V^{t-1}) \nonumber\\
& \overset{\mathclap{(c)}}{\geq}  \sum_{i=1}^n I(U_t V_t;MC U^{t-1}V^{t-1}) -n g(\epsilon) \nonumber\\ 
&\geq \sum_{i=1}^n I(U_t V_t;MC ) -n g(\epsilon) \nonumber\\ 
& \overset{\mathclap{(d)}}{=} \sum_{i=1}^n I(U_t V_t;W_t)  -n g(\epsilon)\overset{\mathclap{(e)}}{\geq}  n I(UV;W) -2n g(\epsilon) \nonumber\\ 
& =n I(UV;W) -4 n \epsilon \left(\log {\lvert \mathcal U \times \mathcal V \rvert} + \log {\frac{1}{\epsilon}}\right)
\end{align}}where, as in~\cite{cuff2013distributed}, the term $g(\epsilon)$ in $(c)$ and $(e)$ is defined as
\begin{equation}
g(\epsilon) \coloneqq 2 \epsilon \left(\log {\lvert \mathcal U \times \mathcal V \rvert} + \log {\frac{1}{\epsilon}}\right),
\end{equation} and the inequalities $(c)$ and $(e)$ are proved in~\cite[Lemma VI.3]{cuff2013distributed}. Moreover $(d)$ and $(e)$ use the identifications $W_t =(C,M)$ for each $t \in \llbracket 1,n\rrbracket$ and $W=(W_T,T)=(C,M, T)$. Finally,~\eqref{eqy10} is proved in~\cite[Lemma VI.3]{cuff2013distributed}
 since
$(b)$ comes from the fact that $Q^{-1}(y) \sqrt{n V_n}\leq 0$ for every  $\frac{1}{2}<y<1 $.

\vspace{1mm}
Now, we recall that by~\eqref{rel10} for every $\gamma >0$, we have
{\allowdisplaybreaks
\begin{align}
\beta'_{\alpha'} &\geq \frac{1}{\gamma} \left[
\alpha' - \mathbb P_{P_{U^{n}W^{n}V^n}} \left\{ \log \frac{P_{U^{n}W^{n}V^n}}{\bar P^{\otimes n}_{UV} \bar P^{\otimes n}_W} > \log \gamma\right\} \right]  \nonumber\\[2mm]
&=\frac{1}{\gamma} \left[
\alpha '- \mathbb P_{P_{U^{n}W^{n}V^n}} \left\{ \log \frac{\bar P_{U}^{\otimes n} \bar P_{W|U}^{\otimes n} \bar P_{V|W}^{\otimes n}}{\bar P^{\otimes n}_{UV} \bar P^{\otimes n}_W} > \log \gamma - \log { \left(1 + \frac{ \epsilon} {\bar P_{U}^{\otimes n} \bar P_{W|U}^{\otimes n} \bar P_{V|W}^{\otimes n} } \right)} \right\} \right].\label{beta2part10}
\end{align}}Then, we set 
\begin{equation} 
\log \gamma = H(M,C)+2n g(\epsilon)+ \log { \left(1 + \frac{ \epsilon} {\bar P_{U}^{\otimes n} \bar P_{W|U}^{\otimes n} \bar P_{V|W}^{\otimes n} } \right)}
\end{equation}and~\eqref{eqy10} implies
{\allowdisplaybreaks
\begin{align}
\log{\beta'_{\alpha'} }&
\geq \log{\frac{1}{\gamma}} +\log{ \left[
\alpha' - \mathbb P_{P_{U^{n}W^{n}V^n}} \left\{ \log \frac{\bar P_{U}^{\otimes n} \bar P_{W|U}^{\otimes n} \bar P_{V|W}^{\otimes n}}{\bar P^{\otimes n}_{UV} \bar P^{\otimes n}_W} > \log \gamma - \log { \left(1 + \frac{ \epsilon} {\bar P_{U}^{\otimes n} \bar P_{W|U}^{\otimes n}  \bar P_{V|W}^{\otimes n}} \right)} \right\} \right]}\nonumber\\[2mm]
&=- \left[H(M,C)+2n g(\epsilon)+\log { \left(1 + \frac{ \epsilon} {\bar P_{U}^{\otimes n} \bar P_{W|U}^{\otimes n} \bar P_{V|W}^{\otimes n} } \right)}\right] \nonumber\\[2mm]
& \quad +\log{ \left[
\alpha' - \mathbb P_{P_{U^{n}W^{n}V^n}} \left\{ \log \frac{\bar P_{U}^{\otimes n} \bar P_{W|U}^{\otimes n} \bar P_{V|W}^{\otimes n}}{\bar P^{\otimes n}_{UV} \bar P^{\otimes n}_W} >  H(M,C)+2n g(\epsilon)  \right\} \right]}\nonumber\\[2mm]
&\geq- \left[H(M,C)+2n g(\epsilon)+\log { \left(1 + \frac{ \epsilon} {\bar P_{U}^{\otimes n} \bar P_{W|U}^{\otimes n} \bar P_{V|W}^{\otimes n} } \right)}\right] \nonumber\\[2mm]
& \quad +\log{ \left[
\alpha' - \mathbb P_{P_{U^{n}W^{n}V^n}} \left\{ \log \frac{\bar P_{U}^{\otimes n} \bar P_{W|U}^{\otimes n} \bar P_{V|W}^{\otimes n}}{\bar P^{\otimes n}_{UV} \bar P^{\otimes n}_W} >  n I(UV;W) + Q^{-1}(y) \sqrt{n V_n} \right\} \right]}\nonumber\\[2mm]
&= - H(M,C) -2n g(\epsilon) -\log { \left(1 + \frac{ \epsilon} {\bar P_{U}^{\otimes n} \bar P_{W|U}^{\otimes n} \bar P_{V|W}^{\otimes n} } \right)} + \log{\left(\alpha' -y - \frac{B_n}{\sqrt n}\right)}\label{partbcase10}.\end{align}}
Similarly to~\eqref{partbcase1fin}, we identify 
\begin{equation*}
x=2n g(\epsilon) +\log { \left(1 + \frac{ \epsilon} {\bar P_{U}^{\otimes n} \bar P_{W|U}^{\otimes n} \bar P_{V|W}^{\otimes n} } \right)} - \log{\left(\alpha' -y - \frac{B_n}{\sqrt n}\right)}
\end{equation*}
and we observe that~\eqref{partbcase10} is equivalent to
{\allowdisplaybreaks
\begin{align}
n(R+R_0)+x &= n(R+R_0) + 2n g(\epsilon) +\log { \left(1 + \frac{ \epsilon} {\bar P_{U}^{\otimes n} \bar P_{W|U}^{\otimes n} \bar P_{V|W}^{\otimes n} } \right)} - \log{\left(\alpha' -y - \frac{B_n}{\sqrt n}\right)} \nonumber\\[2mm]
& = H(M,C) + 2n g(\epsilon) +\log { \left(1 + \frac{ \epsilon} {\bar P_{U}^{\otimes n} \bar P_{W|U}^{\otimes n} \bar P_{V|W}^{\otimes n} } \right)} - \log{\left(\alpha' -y - \frac{B_n}{\sqrt n}\right)} \geq \log{\frac{1}{\beta'_{\alpha'}}}.\label{partbcase1fin0}
\end{align}}


\subsection{Proof of the rate constraint -- Case 1 ($P_{U^nW^nV^n}=\bar P_{U}^{\otimes n} \bar P_{W|U}^{\otimes n} \bar P_{V|W}^{\otimes n}+\epsilon$)}

Now, we conclude the proof of this part by combining~\eqref{beta2part1.1} and~\eqref{partbcase1fin0}. For $1/2<y<1$ we have
{\allowdisplaybreaks
\begin{align*}
\MoveEqLeft[3]
\underbrace{H(M,C)}_{n(R+R_0) } + 2n g(\epsilon)+\log { \left(1 + \frac{ \epsilon} {\bar P_{U}^{\otimes n} \bar P_{W|U}^{\otimes n}  \bar P_{V|W}^{\otimes n}} \right)} - \log{\left(\alpha' -y - \frac{B_n}{\sqrt n}\right)}\nonumber\\[2mm]
 &\geq  n \mu_n + Q^{-1}\left(\epsilon \right) \sqrt{ n V_{n} } + \log { \left(1 + \frac{ \epsilon} {\bar P_{U}^{\otimes n} \bar P_{W|U}^{\otimes n} \bar P_{V|W}^{\otimes n} } \right)} 
\end{align*}}which is equivalent to
{\allowdisplaybreaks
\begin{align}
R+R_0 & \geq \mu_n + Q^{-1}\left(\epsilon \right) \sqrt{ \frac{ V_{n} }{n}}+ \frac{\log{\left(\alpha' -y - \frac{B_n}{\sqrt n}\right)}}{n} - 2 g(\epsilon)\nonumber\\
&= \mu_n + Q^{-1}\left(\epsilon \right) \sqrt{ \frac{ V_{n} }{n}}+ \frac{\log{\left(\alpha' -y - \frac{B_n}{\sqrt n}\right)}}{n}- 4 \epsilon \left(\log {\lvert \mathcal U \times \mathcal V \rvert} + \log {\frac{1}{\epsilon}}\right).
\end{align}  }


\vspace{1mm}
\subsection{Proof of the upper bound on $\log{\beta'_{\alpha'}}$~\eqref{mcpartb0} -- Case 2 ($P_{U^nW^nV^n}=\bar P_{U}^{\otimes n} \bar P_{W|U}^{\otimes n}\bar P_{V|W}^{\otimes n} -\epsilon$)}\label{sssec_mcpartb case 20}  
We have
{\allowdisplaybreaks
\begin{align}
\log { \left( P_{U^nW^nV^n} (\mathbf u,\mathbf w,\mathbf v )  \right)}  
&= \log { \left( \bar P_{U}^{\otimes n} \bar P_{W|U}^{\otimes n} \bar P_{V|W}^{\otimes n} (\mathbf u,\mathbf w, \mathbf v )  \right)} - \log { \left( \frac{\bar P_{U}^{\otimes n} \bar P_{W|U}^{\otimes n}  \bar P_{V|W}^{\otimes n} (\mathbf u,\mathbf w, \mathbf v)}{\bar P_{U}^{\otimes n} \bar P_{W|U}^{\otimes n} \bar P_{V|W}^{\otimes n} (\mathbf u,\mathbf w,\mathbf v )- \epsilon} \right)}.\label{ineq_tv220}
\end{align}}Thus, the following holds
{\allowdisplaybreaks
\begin{align}
&\mathbb P \left\{ \log{ P_{U^nW^nV^n} }\geq \log{\bar P^{\otimes n}_{UV} \bar P^{\otimes n}_W} + \log {\gamma_0}\right\} \nonumber\\[2mm]
&=\mathbb P  \left\{ \sum_{i=1}^n \log \frac{ \bar P_{U} \bar P_{W|U} \bar P_{V|W} }{ \bar P_{U} \bar P_{W}} \geq  \log {\gamma_0}+\log { \left( \frac{\bar P_{U}^{\otimes n} \bar P_{W|U}^{\otimes n} \bar P_{V|W}^{\otimes n} }{\bar P_{U}^{\otimes n} \bar P_{W|U}^{\otimes n} \bar P_{V|W}^{\otimes n} - \epsilon} \right)}\right\}\nonumber\\[2mm]
&= \mathbb P  \left\{ \sum_{i=1}^n \log \frac{ \bar P_{U} \bar P_{W|U} \bar P_{V|W}}{ \bar P_{UV} \bar P_{W}} \geq   n I(UV;W) + Q^{-1}\left(\epsilon \right) \sqrt{ n V_{\bar P_{W|UV}} }
\right\}\nonumber\\[2mm]
& \geq \epsilon- \frac{B_n}{\sqrt n} =: \alpha'.\label{prob be20}
\end{align}}by Theorem \ref{berryessen} (Berry-Esseen CLT) and
{\allowdisplaybreaks
\begin{align}
\log {\gamma_0}
&\coloneqq n \mu_n + Q^{-1}\left(\epsilon \right) \sqrt{ n V_{n} } -\log { \left( \frac{\bar P_{U}^{\otimes n} \bar P_{W|U}^{\otimes n} \bar P_{V|W}^{\otimes n}}{\bar P_{U}^{\otimes n} \bar P_{W|U}^{\otimes n}\bar P_{V|W}^{\otimes n} - \epsilon} \right)}.\label{log20}
\end{align}}Then, we have
{\allowdisplaybreaks
\begin{align}
\log{\frac{1}{\beta'_{\alpha'}}}  &\geq \log{\gamma_0}
= n \mu_n + Q^{-1}\left(\epsilon \right) \sqrt{ n V_{n} } -\log { \left( \frac{\bar P_{U}^{\otimes n} \bar P_{W|U}^{\otimes n}\bar P_{V|W}^{\otimes n} }{\bar P_{U}^{\otimes n} \bar P_{W|U}^{\otimes n} \bar P_{V|W}^{\otimes n} - \epsilon} \right)}.
\label{beta30}
\end{align}}

 
\subsection{Proof of the lower bound on $\log{\beta'_{\alpha'}}$~\eqref{mcparta0} -- Case 2 ($P_{U^nW^nV^n}=\bar P_{U}^{\otimes n} \bar P_{W|U}^{\otimes n}\bar P_{V|W}^{\otimes n}-\epsilon$)}\label{sssec_mcparta case 20}   
By~\eqref{rel10} $\forall \gamma >0$ we have
{\allowdisplaybreaks\begin{align*}
\beta'_{\alpha'} &\geq \frac{1}{\gamma} \left[
\alpha' - \mathbb P_{P_{U^{n}W^{n}V^n}} \left\{ \log \frac{P_{U^{n}W^{n}V^n}}{\bar P^{\otimes n}_{UV} \bar P^{\otimes n}_W} > \log \gamma\right\} \right]  \nonumber\\[2mm]
&=\frac{1}{\gamma} \left[
\alpha' - \mathbb P_{P_{U^{n}W^{n}V^n}} \left\{ \log \frac{\bar P_{U}^{\otimes n} \bar P_{W|U}^{\otimes n} \bar P_{V|W}^{\otimes n}}{\bar P^{\otimes n}_{UV} \bar P^{\otimes n}_W} > \log \gamma + \log { \left( \frac{\bar P_{U}^{\otimes n} \bar P_{W|U}^{\otimes n} \bar P_{V|W}^{\otimes n}}{\bar P_{U}^{\otimes n} \bar P_{W|U}^{\otimes n} \bar P_{V|W}^{\otimes n}- \epsilon} \right)} \right\} \right].
\end{align*}}Then, we set $$ \log \gamma = H(M,C)-\log { \left( \frac{\bar P_{U}^{\otimes n} \bar P_{W|U}^{\otimes n} \bar P_{V|W}^{\otimes n}}{\bar P_{U}^{\otimes n} \bar P_{W|U}^{\otimes n} \bar P_{V|W}^{\otimes n}- \epsilon} \right)}$$ and we recall the  following chain of inequalities, proved in~\eqref{eqy10}
{\allowdisplaybreaks\begin{align}
n(R+R_0)=H(M,C) &\geq n I(UV;W)-4 n \epsilon \left(\log {\lvert \mathcal U \times \mathcal V \rvert} + \log {\frac{1}{\epsilon}}\right)
 + Q^{-1}(y) \sqrt{n V_n}, \qquad \frac{1}{2}<y<1 
\end{align}}which implies
{\allowdisplaybreaks\begin{align}
\log{\beta'_{\alpha'} }&
\geq \log{\frac{1}{\gamma}} +\log{ \left[
\alpha' - \mathbb P_{P_{U^{n}W^{n}V^n}} \left\{ \log \frac{\bar P_{U}^{\otimes n} \bar P_{W|U}^{\otimes n}\bar P_{V|W}^{\otimes n}}{\bar P^{\otimes n}_{UV} \bar P^{\otimes n}_W} > \log \gamma + \log { \left( \frac{\bar P_{U}^{\otimes n} \bar P_{W|U}^{\otimes n} \bar P_{V|W}^{\otimes n}}{\bar P_{U}^{\otimes n} \bar P_{W|U}^{\otimes n} \bar P_{V|W}^{\otimes n}- \epsilon} \right)} \right\} \right]}\nonumber\\[2mm]
&\geq - H(M,C)-\underbrace{\left[4n \epsilon \left(\log {\lvert \mathcal U \times \mathcal V \rvert} + \log {\frac{1}{\epsilon}}\right)-\log { \left( \frac{\bar P_{U}^{\otimes n} \bar P_{W|U}^{\otimes n} \bar P_{V|W}^{\otimes n}}{\bar P_{U}^{\otimes n} \bar P_{W|U}^{\otimes n} \bar P_{V|W}^{\otimes n}- \epsilon} \right)}- \log{\left(\alpha' -y - \frac{B_n}{\sqrt n}\right)}\right]}_{x}.\label{partbcase20}
\end{align}}


\subsection{Proof of the rate constraint -- Case 2 ($P_{U^nW^nV^n}=\bar P_{U}^{\otimes n} \bar P_{W|U}^{\otimes n} \bar P_{V|W}^{\otimes n}-\epsilon$)}
Now, by combining~\eqref{partbcase20} with~\eqref{beta30}, for $1/2<y<1$ we have
{\allowdisplaybreaks
\begin{align*}
\MoveEqLeft[3]
 \underbrace{H(M,C)}_{n(R+R_0) }+4n \epsilon \left(\log {\lvert \mathcal U \times \mathcal V \rvert} + \log {\frac{1}{\epsilon}}\right) -\log { \left( \frac{\bar P_{U}^{\otimes n} \bar P_{W|U}^{\otimes n} \bar P_{V|W}^{\otimes n}}{\bar P_{U}^{\otimes n} \bar P_{W|U}^{\otimes n} \bar P_{V|W}^{\otimes n}- \epsilon} \right)} -\log{\left(\alpha -y - \frac{B_n}{\sqrt n}\right)} \\[2mm]
 & \geq  n \mu_n + Q^{-1}\left(\epsilon \right) \sqrt{ n V_{n} } -\log { \left( \frac{\bar P_{U}^{\otimes n} \bar P_{W|U}^{\otimes n} \bar P_{V|W}^{\otimes n}}{\bar P_{U}^{\otimes n} \bar P_{W|U}^{\otimes n} \bar P_{V|W}^{\otimes n}- \epsilon} \right)} \end{align*}}which is equivalent to
{\allowdisplaybreaks\begin{align}
 R +R_0 \geq   \mu_n + Q^{-1}\left(\epsilon \right) \sqrt{ \frac{ V_{n}}{n} }+ 
\frac{\log{\left(\alpha' -y - \frac{B_n}{\sqrt n}\right)} }{n}-4 \epsilon \left(\log {\lvert \mathcal U \times \mathcal V \rvert} + \log {\frac{1}{\epsilon}}\right).\label{r+log0}
\end{align}}


\section{Proof of the cardinality bound}\label{appendix bounds}
Here we prove the cardinality bound for  the outer bound in Theorem~\ref{theona_outer}. 

First, we state the Support Lemma \cite[Appendix C]{elgamal2011nit}.
\begin{lem}\label{support lemma}
Let $\mathcal{A}$ a finite set and $\mathcal W$ be an arbitrary set. Let $\mathcal P$ be a connected compact subset of probability mass functions on $\mathcal A$ and $P_{A|W}$ be a collection of conditional probability mass functions on $\mathcal A$. Suppose that $h_i(\pi)$, $i=1, \ldots, d$, are real-valued continuous functions of $\pi \in \mathcal P$. Then for every $W$ defined on $\mathcal W$ there exists a random variable $W'$ with $\lvert \mathcal W' \rvert \leq d$ and a collection of conditional probability mass functions $P_{A|W'} \in \mathcal P$ such that
{\allowdisplaybreaks
 \begin{align*}
 & \sum_{w \in \mathcal W} P_W(w) h_i(P_{A|W}(a|w)) =\sum_{w \in \mathcal W'} P_{W'}(w)h_i(P_{A|W'}(a|w)) \quad i=1, \ldots, d.
 \end{align*}}
\end{lem}

Now, we consider the probability distribution $\bar P_{U}  \bar P_{W|U}  \bar P_{V|W}$ that is $\epsilon$-close in $L^1$ distance to the i.i.d. distribution. 
We identify $\mathcal{A}$ with $\{1,\ldots,\lvert \mathcal{A}\rvert \}$ and we consider $\mathcal P$  a connected compact subset of probability mass functions on $\mathcal A=\mathcal U \times  \mathcal V$.
Similarly to \cite{treust2017joint}, suppose that $h_i(\pi)$, $i=1, \ldots, \lvert \mathcal A \rvert +1$,  are real-valued continuous functions of $\pi \in \mathcal P$ such that:
{\allowdisplaybreaks
\begin{align*}
 h_i (\pi) =
 \begin{cases}
   \pi(i) &\mbox{for } i= 1, \ldots, \lvert \mathcal A \rvert -1, \\
   H(U) &\mbox{for } i= \lvert \mathcal A  \rvert,  \\
   H(V|U)& \mbox{for } i= \lvert \mathcal  A \rvert +1.\\
 \end{cases}
\end{align*}}Then by Lemma \ref{support lemma} there exists an auxiliary random variable $W'$ taking at most 
$ \lvert \mathcal U \times \mathcal X \times \mathcal Y \times \mathcal V  \rvert +1$ values such that:
{\allowdisplaybreaks
\begin{align*}
& H(U|W)= \sum_{w \in \mathcal W} P_W(w) H(U|W\!=\!w)= \sum_{w \in \mathcal W'} P_{W'}(w) H(U|W'\!=\!w)=H(U|W'),\\[2mm]
& H(V|UW)= \sum_{w \in \mathcal W} P_W(w) H(V|UW\!=\!w)
= \sum_{w \in \mathcal W'} P_{W'}(w) H(V|UW'\!=\!w)=H(V|UW').
\end{align*}}The constraints on the conditional distributions, the rate constraints and the Markov chain  $U-W-V$ are therefore still verified since we can write 
{\allowdisplaybreaks
\begin{align*}
& I(U;W) =H(U)-H(U|W),\\
& I(UV;W)= H(UV)-H(UV|W)= H(U)+H(V|U)- H(U|W)+H(V|UW),\\
& I(U;V|W)= H(U|W)- H(U|VW)=0.
\end{align*}}Note that we are not forgetting any constraints:  once the distribution $\bar P_{UV}$ and the Markov chain $U-W-V$ are preserved, the dispersion term of the channels $\bar P_{W|U}$ and $\bar P_{W|UV}$ are fixed as well.

\begin{small}
\bibliographystyle{IEEEtran}
\bibliography{mybib2021}

\begin{thebibliography}{10}
\providecommand{\url}[1]{#1}
\csname url@samestyle\endcsname
\providecommand{\newblock}{\relax}
\providecommand{\bibinfo}[2]{#2}
\providecommand{\BIBentrySTDinterwordspacing}{\spaceskip=0pt\relax}
\providecommand{\BIBentryALTinterwordstretchfactor}{4}
\providecommand{\BIBentryALTinterwordspacing}{\spaceskip=\fontdimen2\font plus
\BIBentryALTinterwordstretchfactor\fontdimen3\font minus
  \fontdimen4\font\relax}
\providecommand{\BIBforeignlanguage}[2]{{%
\expandafter\ifx\csname l@#1\endcsname\relax
\typeout{** WARNING: IEEEtran.bst: No hyphenation pattern has been}%
\typeout{** loaded for the language `#1'. Using the pattern for}%
\typeout{** the default language instead.}%
\else
\language=\csname l@#1\endcsname
\fi
#2}}
\providecommand{\BIBdecl}{\relax}
\BIBdecl

\bibitem{gossner2006optimal}
O.~Gossner, P.~Hernandez, and A.~Neyman, ``Optimal use of communication
  resources,'' \emph{Econometrica}, pp. 1603--1636, 2006.

\bibitem{larrousse2015coordination}
B.~Larrousse, S.~Lasaulce, and M.~Bloch, ``Coordination in distributed networks
  via coded actions with application to power control,'' \emph{IEEE
  Transactions on Information Theory}, vol.~64, no.~5, pp. 3633--3654, May
  2018.

\bibitem{cuff2009thesis}
P.~Cuff, ``Communication in networks for coordinating behavior,'' Ph.D.
  dissertation, Stanford University, 2009.

\bibitem{cuff2010}
P.~Cuff, H.~H. Permuter, and T.~M. Cover, ``{Coordination Capacity},''
  \emph{{IEEE Trans. Inf. Theory}}, vol.~56, no.~9, pp. 4181--4206, 2010.

\bibitem{bennet2002entanglement}
C.~H. Bennett, P.~W. Shor, J.~A. Smolin, and A.~V. Thapliyal,
  ``Entanglement-assisted capacity of a quantum channel and the reverse
  {S}hannon theorem,'' \emph{IEEE Transactions on Information Theory}, vol.~48,
  no.~10, pp. 2637--2655, 2002.

\bibitem{Soljanin2002}
E.~Soljanin, ``Compressing quantum mixed-state sources by sending classical
  information,'' \emph{IEEE Transactions on Information Theory}, vol.~4, no.~8,
  pp. 2263--2275, 2002.

\bibitem{kramer2007communicating}
G.~Kramer and S.~A. Savari, ``Communicating probability distributions,''
  \emph{IEEE Transactions on Information Theory}, vol.~53, no.~2, pp. 518--525,
  2007.

\bibitem{winter2002compression}
\BIBentryALTinterwordspacing
A.~Winter, ``Compression of sources of probability distributions and density
  operators,'' 2002. [Online]. Available:
  \url{http://arxiv.org/abs/quant-ph/0208131}
\BIBentrySTDinterwordspacing

\bibitem{cuff2013distributed}
P.~Cuff, ``{Distributed Channel Synthesis},'' \emph{{IEEE Trans. Inf. Theory}},
  vol.~59, no.~11, pp. 7071--7096, Nov. 2013.

\bibitem{haddadpour2012coordination}
F.~Haddadpour, M.~H. Yassaee, A.~Gohari, and M.~R. Aref, ``Coordination via a
  relay,'' in \emph{Proc. of IEEE International Symposium on Information Theory
  (ISIT)}, 2012, pp. 3048--3052.

\bibitem{bloch2014strong}
M.~R. Bloch and J.~Kliewer, ``Strong coordination over a three-terminal relay
  network,'' in \emph{Information Theory Workshop (ITW), 2014 IEEE}.\hskip 1em
  plus 0.5em minus 0.4em\relax IEEE, 2014, pp. 646--650.

\bibitem{bloch2013strong}
------, ``Strong coordination over a line network,'' in \emph{Proc. of IEEE
  International Symposium on Information Theory (ISIT)}, 2013, pp. 2319--2323.

\bibitem{vellambi2015strong}
B.~N. Vellambi, J.~Kliewer, and M.~R. Bloch, ``Strong coordination over
  multi-hop line networks,'' in \emph{Proc. of IEEE Information Theory
  Workshop-Fall (ITW)}, 2015, pp. 192--196.

\bibitem{vellambi2016strong}
------, ``Strong coordination over a line when actions are markovian,'' in
  \emph{Proc. of Annual Conference on Information Science and Systems (CISS)},
  2016, pp. 412--417.

\bibitem{cuff2011hybrid}
P.~Cuff and C.~Schieler, ``Hybrid codes needed for coordination over the
  point-to-point channel,'' in \emph{Proc. of Allerton Conference on
  Communication, Control and Computing}, 2011, pp. 235--239.

\bibitem{treust2014correlation}
M.~Le~Treust, ``Correlation between channel state and information source with
  empirical coordination constraint,'' in \emph{Proc. of IEEE Information
  Theory Workshop (ITW)}, 2014, pp. 272--276.

\bibitem{treust2015empirical}
------, ``Empirical coordination with two-sided state information and
  correlated source and state,'' in \emph{Proc. of IEEE International Symposium
  on Information Theory (ISIT)}, 2015, pp. 466--470.

\bibitem{le2015empirical}
------, ``Empirical coordination with channel feedback and strictly causal or
  causal encoding,'' in \emph{Proc. of IEEE International Symposium on
  Information Theory (ISIT)}.\hskip 1em plus 0.5em minus 0.4em\relax IEEE,
  2015, pp. 471--475.

\bibitem{larrousse2015coordinating}
B.~Larrousse, S.~Lasaulce, and M.~Wigger, ``Coordinating partially-informed
  agents over state-dependent networks,'' in \emph{Proc. of IEEE Information
  Theory Workshop (ITW)}, 2015, pp. 1--5.

\bibitem{treust2017joint}
M.~Le~Treust, ``Joint empirical coordination of source and channel,''
  \emph{IEEE Transactions on Information Theory}, vol.~63, no.~8, pp.
  5087--5114, 2017.

\bibitem{haddadpour2017simulation}
F.~Haddadpour, M.~H. Yassaee, S.~Beigi, A.~Gohari, and M.~R. Aref,
  ``{S}imulation of a channel with another channel,'' \emph{{IEEE}
  {T}ransactions on {I}nformation {T}heory}, vol.~63, no.~5, pp. 2659--2677,
  2017.

\bibitem{Cervia2017}
G.~Cervia, L.~Luzzi, M.~Le~Treust, and M.~R. Bloch, ``Strong coordination of
  signals and actions over noisy channels,'' in \emph{2017 IEEE Int. Symp. Inf.
  Theory (ISIT)}, Jun. 2017, pp. 2835--2839.

\bibitem{cervia2018journal}
------, ``Strong coordination of signals and actions over noisy channels with
  two-sided state information,'' \emph{IEEE Transactions on Information
  Theory}, vol.~66, no.~8, pp. 4681--4708, 2020.

\bibitem{Cervia2019Fixed}
G.~Cervia, T.~J. Oechtering, and M.~Skoglund, ``Fixed-length strong
  coordination,'' in \emph{Proc. of IEEE Information Theory Workshop (ITW)},
  2019.

\bibitem{strassen1962asymptotische}
V.~Strassen, ``Asymptotische {A}bsch\"{a}tzungen in {S}hannon's
  {I}nformationstheorie,'' in \emph{Transactions of the Third Prague Conference
  on Information Theory etc, 1962. Czechoslovak Academy of Sciences, Prague},
  1962, pp. 689--723.

\bibitem{kontoyiannis1997second}
I.~Kontoyiannis, ``Second-order noiseless source coding theorems,'' \emph{IEEE
  Transactions on Information Theory}, vol.~43, no.~4, pp. 1339--1341, 1997.

\bibitem{baron2004quickly}
D.~Baron, M.~A. Khojastepour, and R.~G. Baraniuk, ``How quickly can we approach
  channel capacity?'' in \emph{Conference Record of the Thirty-Eighth Asilomar
  Conference on Signals, Systems and Computers, 2004.}, vol.~1.\hskip 1em plus
  0.5em minus 0.4em\relax IEEE, 2004, pp. 1096--1100.

\bibitem{hayashi2008second}
M.~Hayashi, ``Second-order asymptotics in fixed-length source coding and
  intrinsic randomness,'' \emph{IEEE Transactions on Information Theory},
  vol.~54, no.~10, pp. 4619--4637, 2008.

\bibitem{hayashi2009information}
------, ``Information spectrum approach to second-order coding rate in channel
  coding,'' \emph{IEEE Transactions on Information Theory}, vol.~55, no.~11,
  pp. 4947--4966, 2009.

\bibitem{polyanskiy2010channel}
Y.~Polyanskiy, H.~V. Poor, and S.~Verd{\'u}, ``Channel coding rate in the
  finite blocklength regime,'' \emph{IEEE Transactions on Information Theory},
  vol.~56, no.~5, p. 2307, 2010.

\bibitem{verdu2012non}
S.~Verd{\'u}, ``Non-asymptotic achievability bounds in multiuser information
  theory,'' in \emph{2012 50th Annual Allerton Conference on Communication,
  Control, and Computing (Allerton)}.\hskip 1em plus 0.5em minus 0.4em\relax
  IEEE, 2012, pp. 1--8.

\bibitem{jazi2012simpler}
E.~M. Jazi and J.~N. Laneman, ``Simpler achievable rate regions for multiaccess
  with finite blocklength,'' in \emph{2012 IEEE International Symposium on
  Information Theory Proceedings}.\hskip 1em plus 0.5em minus 0.4em\relax IEEE,
  2012, pp. 36--40.

\bibitem{kostina2012fixed}
V.~Kostina and S.~Verd{\'u}, ``Fixed-length lossy compression in the finite
  blocklength regime,'' \emph{IEEE Transactions on Information Theory},
  vol.~58, no.~6, pp. 3309--3338, 2012.

\bibitem{kostina2013lossyit}
------, ``Lossy joint source-channel coding in the finite blocklength regime,''
  \emph{IEEE Transactions on Information Theory}, vol.~59, no.~5, pp.
  2545--2575, 2013.

\bibitem{tan2013dispersions}
V.~Y. Tan and O.~Kosut, ``On the dispersions of three network information
  theory problems,'' \emph{IEEE Transactions on Information Theory}, vol.~60,
  no.~2, pp. 881--903, 2013.

\bibitem{Watanabe2015Nonasymptotic}
S.~{Watanabe}, S.~{Kuzuoka}, and V.~Y.~F. {Tan}, ``Nonasymptotic and
  second-order achievability bounds for coding with side-information,''
  \emph{IEEE Transactions on Information Theory}, vol.~61, no.~4, pp.
  1574--1605, 2015.

\bibitem{nomura2014second}
R.~Nomura \emph{et~al.}, ``Second-order slepian-wolf coding theorems for
  non-mixed and mixed sources,'' \emph{IEEE transactions on information
  theory}, vol.~60, no.~9, pp. 5553--5572, 2014.

\bibitem{blahut1974hypothesis}
R.~Blahut, ``Hypothesis testing and information theory,'' \emph{IEEE
  Transactions on Information Theory}, vol.~20, no.~4, pp. 405--417, 1974.

\bibitem{campo2012converse}
A.~T. Campo, G.~Vazquez-Vilar, A.~G. i~F{\`a}bregas, and A.~Martinez,
  ``Converse bounds for finite-length joint source-channel coding,'' in
  \emph{2012 50th Annual Allerton Conference on Communication, Control, and
  Computing (Allerton)}.\hskip 1em plus 0.5em minus 0.4em\relax IEEE, 2012, pp.
  302--307.

\bibitem{yassaee2013technique}
M.~H. Yassaee, M.~R. Aref, and A.~Gohari, ``A technique for deriving one-shot
  achievability results in network information theory,'' in \emph{Proc. of IEEE
  International Symposium on Information Theory (ISIT)}, 2013, pp. 1287--1291.

\bibitem{yassaee2013non}
------, ``Non-asymptotic output statistics of random binning and its
  applications,'' in \emph{2013 IEEE International Symposium on Information
  Theory}.\hskip 1em plus 0.5em minus 0.4em\relax IEEE, 2013, pp. 1849--1853.

\bibitem{vazquez2013meta}
G.~Vazquez-Vilar, A.~T. Campo, A.~G. i~F{\`a}bregas, and A.~Martinez, ``The
  meta-converse bound is tight,'' in \emph{2013 IEEE International Symposium on
  Information Theory}.\hskip 1em plus 0.5em minus 0.4em\relax IEEE, 2013, pp.
  1730--1733.

\bibitem{Vazquez2013Bayesian}
G.~{Vazquez-Vilar}, A.~{Tauste Campo}, A.~{Guill\'en i F\`abregas}, and
  A.~{Martinez}, ``Bayesian $m$-ary hypothesis testing: The meta-converse and
  verd\'u-han bounds are tight,'' \emph{IEEE Transactions on Information
  Theory}, vol.~62, no.~5, pp. 2324--2333, 2016.

\bibitem{kostina2013lossy}
V.~Kostina, ``Lossy data compression: Non-asymptotic fundamental limits,''
  \emph{Ph. D. dissertation}, 2013.

\bibitem{yassaee2014achievability}
M.~H. Yassaee, M.~R. Aref, and A.~Gohari, ``{Achievability Proof via Output
  Statistics of Random Binning},'' \emph{{IEEE Trans. Inf. Theory}}, vol.~60,
  no.~11, pp. 6760--6786, Nov. 2014.

\bibitem{Lindvall1992coupling}
T.~Lindvall, \emph{Lectures on the Coupling Method}.\hskip 1em plus 0.5em minus
  0.4em\relax John Wiley \& Sons, Inc., 1992. Reprint: Dover paperback edition,
  2002.

\bibitem{erokhin1958varepsilon}
V.~Erokhin, ``$\varepsilon$-entropy of a discrete random variable,''
  \emph{Theory of Probability \& Its Applications}, vol.~3, no.~1, pp. 97--100,
  1958.

\bibitem{polyanskiy2014lecture}
Y.~Polyanskiy and Y.~Wu, ``Lecture notes on information theory,'' \emph{Lecture
  Notes for ECE563 (UIUC) and}, vol.~6, no. 2012-2016, p.~7, 2014.

\bibitem{scarlett2013mismatched}
J.~Scarlett, A.~Martinez, and A.~G. i~F{\`a}bregas, ``Mismatched decoding:
  Finite-length bounds, error exponents and approximations,'' \emph{arXiv
  preprint arXiv:1303.6166}, 2012.

\bibitem{martinez2009bit}
A.~Martinez, A.~G. i~Fabregas, G.~Caire, and F.~M. Willems, ``Bit-interleaved
  coded modulation revisited: A mismatched decoding perspective,'' \emph{IEEE
  Transactions on Information Theory}, vol.~55, no.~6, pp. 2756--2765, 2009.

\bibitem{yassaee2013nonarxiv}
M.~H. Yassaee, M.~R. Aref, and A.~Gohari, ``Non-asymptotic output statistics of
  random binning and its applications,'' \emph{arXiv preprint arXiv:1303.0695},
  2013.

\bibitem{shannon1959coding}
C.~E. Shannon, ``Coding theorems for a discrete source with a fidelity
  criterion,'' \emph{IRE Nat. Conv. Rec}, vol.~4, no. 142-163, p.~1, 1959.

\bibitem{elgamal2011nit}
A.~El~Gamal and Y.~H. Kim, \emph{Network information theory}.\hskip 1em plus
  0.5em minus 0.4em\relax Cambridge University Press, 2011.

\end{thebibliography}
\end{small}

\end{document}